\UseRawInputEncoding

\documentclass[conference,onecolumn]{IEEEtran}
\IEEEoverridecommandlockouts
 \usepackage{url} 

\usepackage{xspace}
\usepackage[utf8]{inputenc} 
\usepackage{cite}
\usepackage{algorithm}
\usepackage{textcomp}
\usepackage{algpseudocode}    
\usepackage{siunitx} 
\usepackage{multirow} 

\usepackage{comment}
\usepackage{tikz}
\usepackage{amsmath, amssymb, graphicx, subcaption}

\usepackage{textcomp}
\usepackage{xcolor}
\usepackage{float} 
\usepackage{graphicx} 

\usepackage{booktabs}
\usepackage[utf8]{inputenc}
\newcommand{\linebreakand}{%
  \end{@IEEEauthorhalign}
  \hfill\mbox{}\par
  \mbox{}\hfill\begin{@IEEEauthorhalign}
}
\usepackage{array}
\usepackage{geometry}
\def\BibTeX{{\rm B\kern-.05em{\sc i\kern-.025em b}\kern-.08em
    T\kern-.1667em\lower.7ex\hbox{E}\kern-.125emX}}
\begin{document}

\title{ROC Analysis with Covariate Adjustment Using Neural Network Models: Evaluating the Role of Age in the Physical Activity–Mortality Association}

\author{\IEEEauthorblockN{Ziad Akram Ali Hammouri}

\IEEEauthorblockA{\textbf{Corresponding author}\\
\textit{Dept. of Computer Science} \\
\textit{Al-Balqa Applied University}\\
Aqaba, Jordan \\
z.hammouri@bau.edu.jo}
  \and
  \IEEEauthorblockN{Yating Zou}
  \IEEEauthorblockA{Dept.\ of Biostatistics\\
   University of North Carolina\\
    Chapell Hill, USA\\
   yating@live.unc.edu}
  \and
  \IEEEauthorblockN{Rahul Ghosal}
  \IEEEauthorblockA{Dept.\ of Epidemiology and Biostatistics\\
    University of South Carolina\\
    Columbia, USA\\
    rghosal@mailbox.sc.edu}
  \and
  \IEEEauthorblockN{Juan C.\ Vidal}
  \IEEEauthorblockA{Research Center on Intelligent Technologies\\
    University of Santiago de Compostela\\
    Santiago, Spain\\
    juan.vidal@usc.es}
  \and
  \IEEEauthorblockN{Marcos Matabuena \textbf{Corresponding author}}
  \IEEEauthorblockA{Dept.\ of Biostatistics\\
    Harvard University\\
    Boston, USA\\
    mmatabuena@hsph.harvard.edu}
}

\maketitle

\begin{abstract}
The receiver operating characteristic (ROC) curve and its summary measure, the Area Under the Curve (AUC), are well-established tools for evaluating the efficacy of biomarkers in biomedical studies.  Compared to the traditional ROC curve, the covariate-adjusted ROC curve allows for individual evaluation of the biomarker. However, the use of machine learning models has rarely been explored in this context, despite their potential to develop more powerful and sophisticated approaches for biomarker evaluation. The goal of this paper is to propose a framework for neural network-based covariate-adjusted ROC modeling that allows flexible and nonlinear evaluation of the effectiveness of a biomarker to discriminate between two reference populations. The finite-sample performance of our method is investigated through extensive simulation tests under varying dependency structures between biomarkers, covariates, and referenced populations. The methodology is further illustrated in a clinically case study that assesses daily physical activity - measured as total activity time (TAC), a proxy for daily step count - as a biomarker to predict mortality at three, five and eight years. Analyzes stratified by sex and adjusted for age and BMI reveal distinct covariate effects on mortality outcomes. These results underscore the importance of covariate-adjusted modeling in biomarker evaluation and highlight TAC’s potential as a functional capacity biomarker based on specific individual characteristics.


\end{abstract}
\begin{IEEEkeywords}
ROC analysis, Biomarkers, Diagnostic Performance, Machine learning, Neural Network, Physical Activity.
\end{IEEEkeywords}

\section{Introduction}

The receiver operating curve (ROC) is the cornerstone of diagnostic test evaluation, providing a rigorous framework to characterize the trade-off between sensitivity and specificity between different classification thresholds \cite{gonen2006receiver,perez2021visualizing,perez2020roc,fawcett2006introduction,martinez2022area,perez2018nsroc}. Applied extensively in medical diagnostics, machine learning, and statistical science, ROC analysis offers a principled criterion to quantify the discriminatory capacity of biomarkers or reference variables to distinguish between two populations, such as diseased and healthy individuals\cite{PMID:17668917,cai2008regression,zheng2006application,cai2004semi,cai2004semi1,
cai2002semiparametric}. Among its most widely used indices, the area under the curve (AUC), the integral of the ROC curve, summarizes the general precision of the diagnosis: an AUC of $0.5$ corresponds to random guessing, while an AUC of $1.0$ indicates perfect classification \cite{pepe2003statistical,obuchowski2018receiver,sadoon2024classification,krzanowski2009roc,nakas2023roc,zou2011statistical}.

However, classical ROC analysis captures only marginal associations between a biomarker and clinical status, neglecting analysis of individual-specific characteristics such as age, sex, or comorbidities, which can markedly influence diagnostic performance \cite{janes2009adjusting,inacio2022covariate,lee2023covariate}. Covariate-adjusted ROC methods address this limitation by incorporating individual-level factors, enabling refined diagnostic precision evaluations and supporting the choice of personalized decision thresholds \cite{pepe2003statistical,janes2009adjusting,ghosal2025impact,vickers2006decision,jiang2024analyzing}. Such methods are particularly relevant in clinical contexts where phenotypic heterogeneity strongly drives diagnostic outcomes, as shown in recent cardiology and endocrinology applications \cite{obuchowski2019statistical,vaid2023implications}. Conditional ROC curves, a special case of covariate adjustment, further sharpen inference by evaluating diagnostic accuracy at specific covariate values, thus facilitating individualized clinical decision rules \cite{pepe2000combining,inacio2022covariate}.

Despite these advances, significant challenges remain in the area. Many existing approaches rely on restrictive parametric assumptions \cite{cai2007model,perezvignette,martinez2024comparing,perez2021visualizing,diaz2020cumulative}, and others fail to capture multivariate nonlinear statistical associations \cite{fluss2005estimation,rodriguez2011roc}.  Recent contributions, including Bayesian nonparametric estimators \cite{inacio2022covariate}, offer greater flexibility, but are computationally intensive and not easily integrated into modern machine learning pipelines.

The goal of this paper is to introduce a covariate-adjusted ROC analysis framework based on feedforward neural networks (FNNs) under the lens of a semiparametric Gaussian scale and localization model. Using their ability to approximate complex nonlinear functions \cite{hinton1986learning,lecun2015deep,nair2010rectified}, FNNs provide a natural semiparametric mechanism to estimate conditional distributions within each clinical population based on conditional mean and variance estimation. Under classical smoothing conditions, the proposed approach yields efficient estimators of conditional ROC curves. Crucially, when covariates lie on a low-dimensional manifold or underlying regression functions follow hierarchical structures \cite{kohler2021rate}, as is common in medical imaging and high-dimensional clinical data \cite{bartlett2021deep
}, the method remains effective, in contrast to classical smoothing techniques (e.g., local polynomical regression algorithms, Nadaraya-Watson estimators), which deteriorate their performance very fast in such regimes with a moderate increase of the dimension of predictors.

To illustrate the clinical relevance of this framework, we examine the prognostic value of total activity count (TAC), a surrogate biomarker for daily step count, in predicting mortality over three, five, and eight-year horizons. Adjusted for age, body mass index, and sex, our analysis demonstrates substantial covariate-driven shifts in the TAC–mortality association, highlighting the need for covariate-based ROC evaluation. More specifically, Figure \ref{fig:tac2_age_all} displays the bivariate density of age and TAC, stratified by sex and clinical status (mortality in the three-, five- and eight-year horizons). The figure reveals marked age-dependent changes among people who died, providing clear evidence of an interaction between TAC and age that underscores the importance of covariate adjustment when evaluating the clinical value of TAC for mortality prediction tasks. 

Physical activity is widely recognized as the cornerstone of functional capacity and one of the most effective nonpharmacological determinants of chronic disease outcomes \cite{matabuena2022physical}. However, clinical guidelines remain largely population-based and not sufficiently tailored to individuals. Our results show that the predictive value of TAC strongly depends on specific covariate contexts, underscoring its potential as a personalized biomarker of functional capacity.

This perspective has two key implications. First, it underscores the value of physical-activity metrics as biomarkers in research on aging and chronic disease. Second, it opens new possibilities for precision public health, in which recommendations are tailored to individual characteristics rather than broad, population-level guidelines, based on each person’s specific health impact.

\begin{figure}[ht!]
\centering
\begin{subfigure}{0.32\textwidth}
    \includegraphics[width=\textwidth]{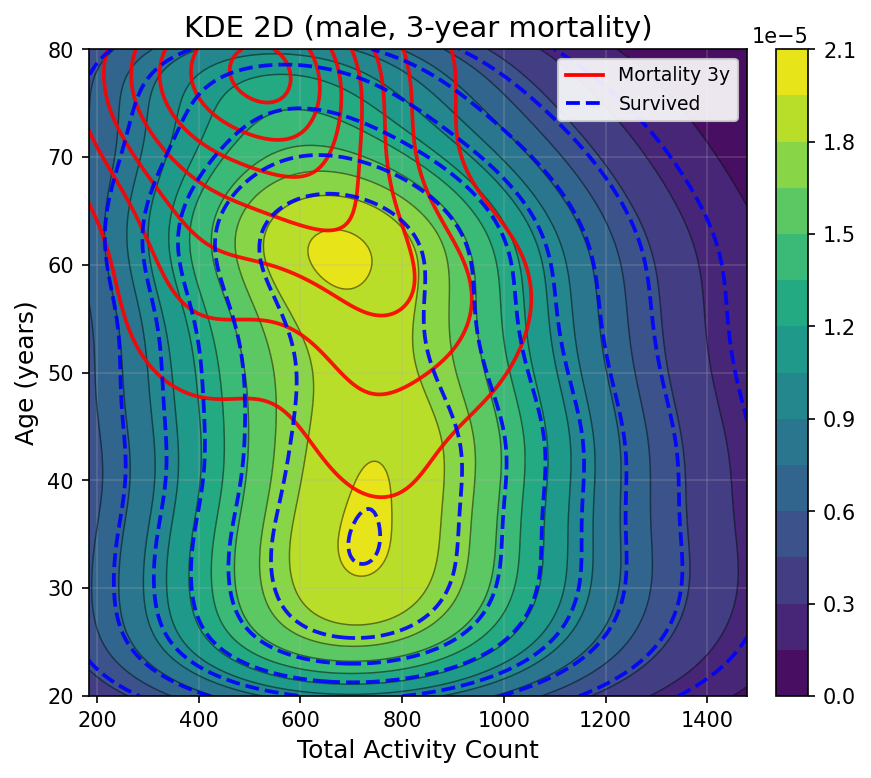}
    \caption{Males, 3-year mortality}
    \label{fig:tac2_age_m3}
\end{subfigure}
\hfill
\begin{subfigure}{0.32\textwidth}
    \includegraphics[width=\textwidth]{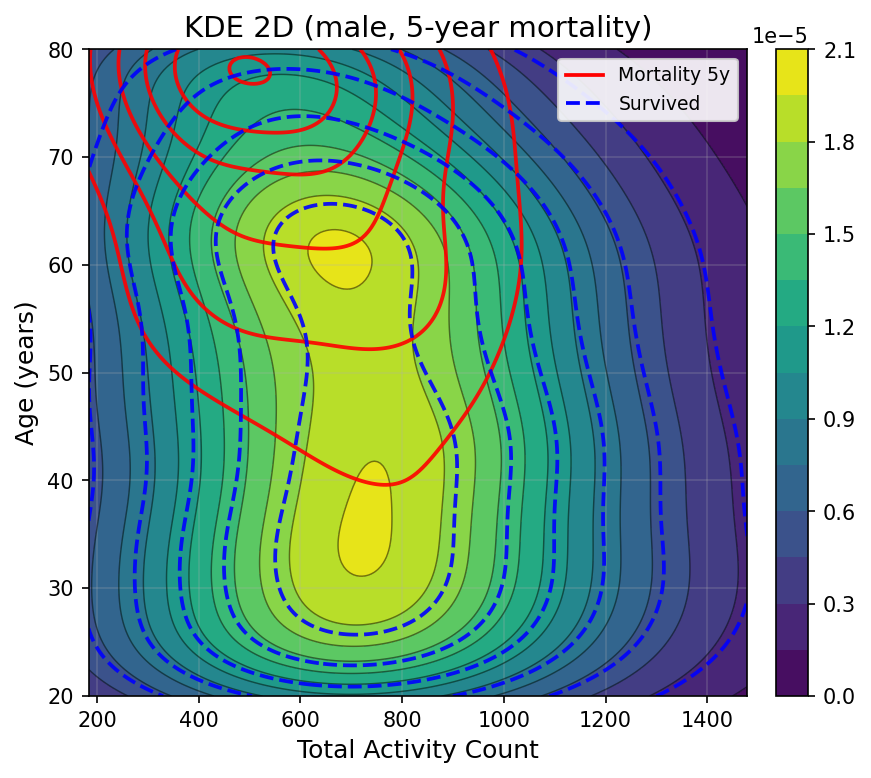}
    \caption{Males, 5-year mortality}
    \label{fig:tac2_age_m5}
\end{subfigure}
\hfill
\begin{subfigure}{0.32\textwidth}
    \includegraphics[width=\textwidth]{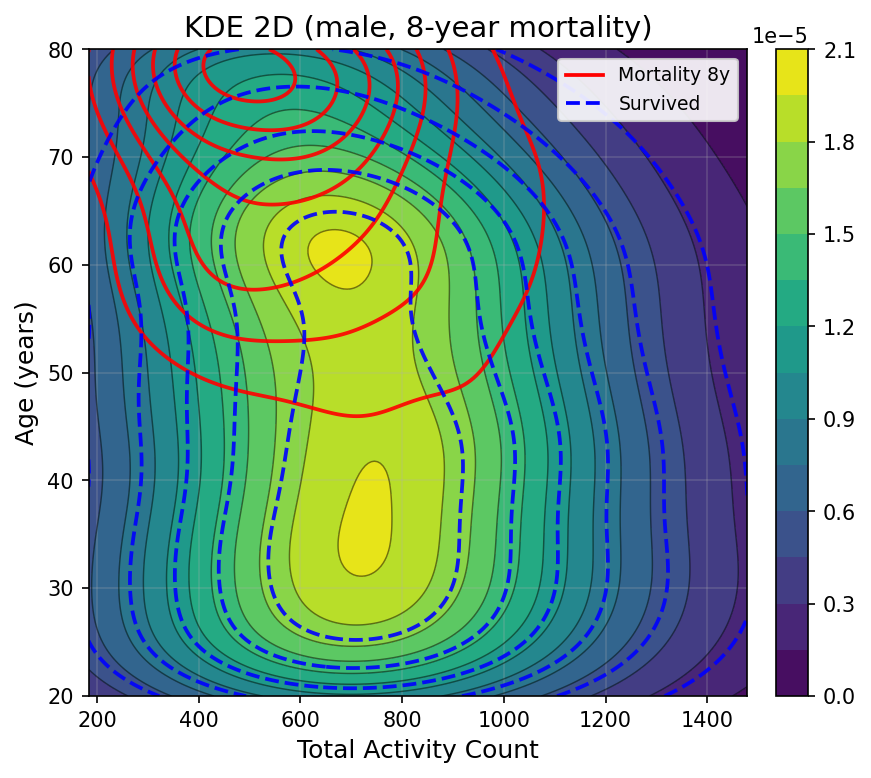}
    \caption{Males, 8-year mortality}
    \label{fig:tac2_age_m8}
\end{subfigure}

\vspace{0.5cm}

\begin{subfigure}{0.32\textwidth}
    \includegraphics[width=\textwidth]{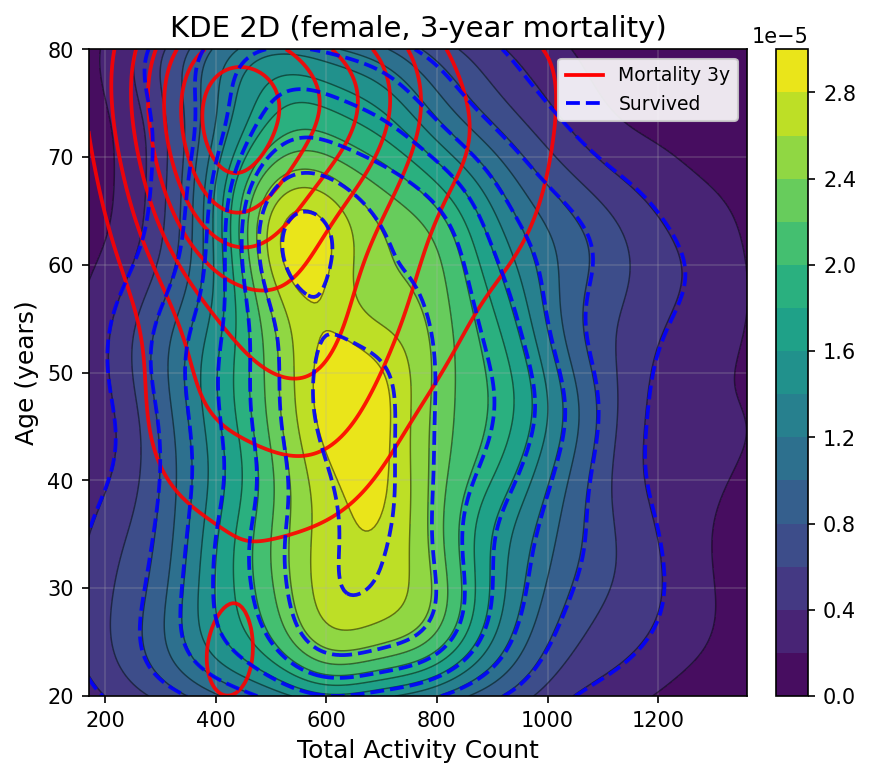}
    \caption{Females, 3-year mortality}
    \label{fig:tac2_age_f3}
\end{subfigure}
\hfill
\begin{subfigure}{0.32\textwidth}
    \includegraphics[width=\textwidth]{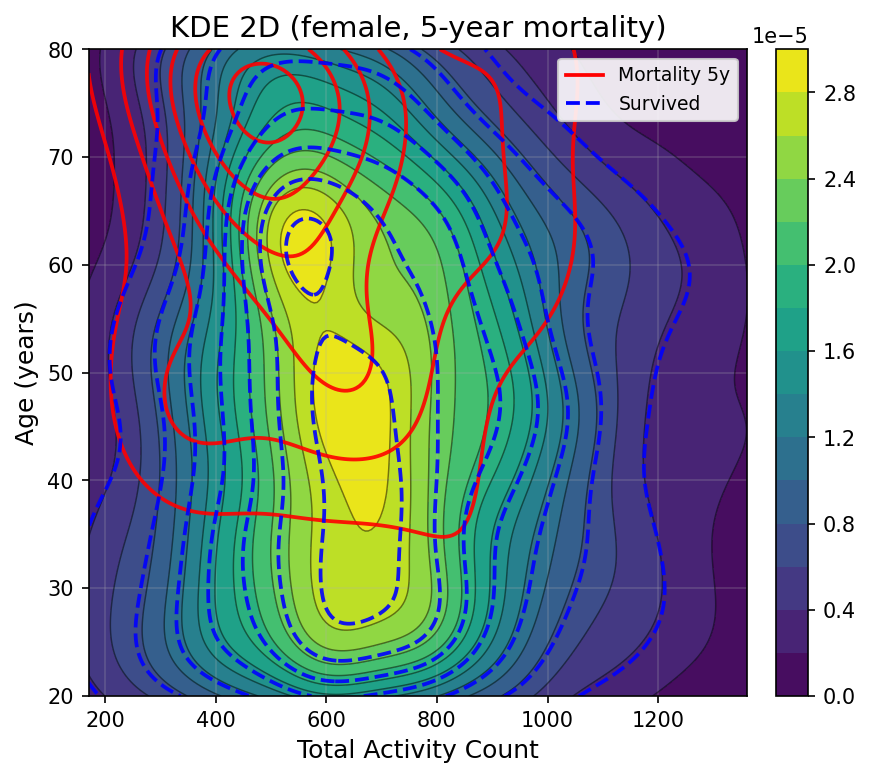}
    \caption{Females, 5-year mortality}
    \label{fig:tac2_age_f5}
\end{subfigure}
\hfill
\begin{subfigure}{0.32\textwidth}
    \includegraphics[width=\textwidth]{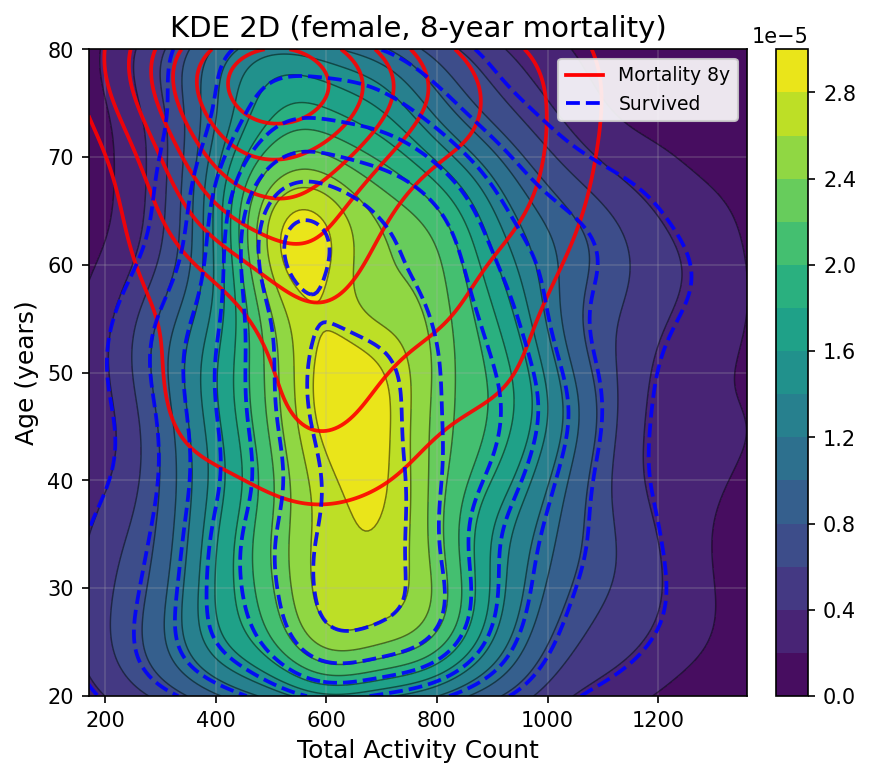}
    \caption{Females, 8-year mortality}
    \label{fig:tac2_age_f8}
\end{subfigure}

\caption{Bivariate kernel density estimates of total activity count versus age, stratified by sex and mortality follow-up period. Solid contours represent decedents and dashed contours represent survivors.}
\label{fig:tac2_age_all}
\end{figure}

\subsection{Outline}

The structure of the paper is as follows. In Section~\ref{sec:Related Work}, we briefly review the related work on covariate-adjusted ROC (AROC) analysis. Section~\ref{sec:models} introduces the mathematical framework and our proposed two-stage feedforward neural network (FNN) methodology. In Section~\ref{sec:nhanes}, we apply our method to the NHANES biomedical data set to evaluate the total activity count (TAC) as a covariate-adjusted biomarker of mortality. Finally, Section~\ref{sec:conclusion} provides a discussion of the implications, limitations, and directions for future research. The supplemental material contains simulation studies that confirm the satisfactory empirical performance of our proposal in different and other completementary results of the case study.

\section{Related Work}\label{sec:Related Work}

Traditional ROC analysis evaluates diagnostic performance without accounting for covariates, which can bias inference in heterogeneous populations \cite{pepe2003statistical, fawcett2006introduction}. To address this limitation, covariate-adjusted ROC (AROC) methods incorporate individual-level covariates, enabling more accurate assessments. In the following, we review key advances in the AROC methodology in recent years and discuss their limitations, focusing on both statistical and machine learning contributions that are directly linked to our proposed framework and the case study.

\subsection{Statistical Methods for Covariate-Adjusted ROC}

The early work of Alonzo \cite{alonzo2002distribution} introduced a ROC framework in the presence of covariates using the approaches of binary regression models. Janes and Pepe \cite{janes2006adjusting, janes2009adjusting} formalized the estimation of covariate-adjusted ROC by proposing semiparametric and nonparametric approaches, including location-scale regression models for cases and controls.  The mentioned approaches can be sensitive to model misspecification due to the semiparametric model of the regression function selected.

Rodriguez-alvarez \cite{rodriguez2011roc} extended this methodology by using kernel smoothing to estimate covariate-specific ROC curves in a fully nonparametric framework. These techniques are implemented in R packages such as \texttt{AROC}, \texttt{npROCRegression}, and \texttt{ROCnReg} \cite{rodriguez2021rocnreg}, which support both frequentist and Bayesian approaches. A conceptually related approach involves placement values, introduced by Pepe \cite{pepe2000interpretation}. Here, a marker's value is transformed into its percentile within the covariate-specific control distribution. This transformation enables covariate adjustment by standardizing test scores, and ROC curves are derived as the distribution of placement values among cases \cite{pepe2004combining, inacio2022covariate}.

More recently, Inácio de Carvalho et al.~\cite{inacio2022covariate} proposed a Bayesian nonparametric estimator for the AROC using a Dirichlet process mixture model and Bayesian bootstrap. This method flexibly captures complex and non-linear covariate effects and was applied to cardiovascular risk data with promising results. However, the approach is computationally intensive, making it prohibitive for large national health studies and modern electronic health record (EHR) datasets.

\subsection{Machine Learning in Diagnostic Evaluation and ROC software}

Machine learning methods, including ensemble approaches such as Random Forests and gradient boosting machines, have gained traction in classification tasks due to their ability to model complex, nonlinear relationships \cite{fernandez2014we,sadoon2024classification}. However, in regression settings, these methods can be less flexible, often capturing primarily additive functional relationships. Moreover, many machine learning approaches are not commonly considered for estimating the ROC curve in the presence of covariates.

Feedforward neural networks (FNNs), in particular, provide powerful function approximation capabilities through their layered architectures and nonlinear activation functions \cite{hinton1986learning, lecun2015deep, nair2010rectified}. Although FNNs have been successfully applied to regression and classification tasks in healthcare \cite{steelefeed}, their use in covariate adjusted ROC analysis remains limited.

With respect to the software in R, available libraries such as \texttt{pROC} and \texttt{ROCR} support pooled ROC analysis but do not allow for covariate adjustment. Packages like \texttt{AROC} and \texttt{npROCRegression} enable linear and certain nonlinear covariate adjustments, but are not designed to integrate with machine learning pipelines and generally support only a limited number of predictors.


\subsection{Contribution of This Work}

Our framework bridges covariate-adjusted ROC (aROC) methodology and deep learning by using feedforward neural networks (FNNs) to model complex, non-linear covariate effects. Unlike traditional aROC approaches that rely on parametric forms or kernel smoothing, our method leverages the expressiveness of neural networks to provide robust, flexible covariate adjustment while avoiding practical challenges such as bandwidth/tuning-parameter selection. Compared with Bayesian nonparametric models \cite{inacio2022covariate}, our approach is more computationally efficient and scales easily to high-dimensional settings. For example, in the analyzed case study, running the Bayesian alternative only in one example was computationally prohibitive on available hardware within several hours.

In the Supplementary Material, we validate the model in nine synthetic scenarios designed to mimic realistic medical data, including linear, non-linear, and interaction effects. Relative to a semiparametric benchmark and our implementation of ROC random forests with covariates, the FNN approach achieves superior finite-sample accuracy. The PyTorch implementation facilitates reproducibility, scalability, and easy integration into broader machine learning workflows. In general, this contribution addresses a gap in the diagnostic evaluation literature by providing a modern, scalable tool for personalized medicine and biomarker evaluation.

\section{Methodology}
\label{sec:models}
This study develops a framework for covariate-adjusted ROC analysis. In this section, we begin with a review of the foundational concepts of ROC curves in the presence of covariates, and then introduce our proposal for the two-stage NN ROC algorithm.

\subsection{Foundations of ROC Curve Analysis}

ROC curves provide a framework for evaluating the performance of a continuous diagnostic biomarker, denoted by $Y$, in discriminating between two populations: a diseased group ($D=1$) and a nondiseased group ($D=0$).  

For a given threshold $c \in \mathbb{R}$, the test result is classified as positive if $Y > c$ and negative otherwise. This leads to the definitions of the true positive rate (TPR) and the false positive rate (FPR).  
\[
\text{TPR}(c) = P(Y > c \mid D=1), 
\qquad 
\text{FPR}(c) = P(Y > c \mid D=0).
\]  

Equivalently, if $F_1$ and $F_0$ denote the conditional cumulative distribution functions of $Y$ given $D=1$ and $D=0$, then
\[
\text{TPR}(c) = 1 - F_1(c), 
\qquad 
\text{FPR}(c) = 1 - F_0(c).
\]  

The ROC curve is formally defined as the mapping

\[
\text{ROC}(p) = 1 - F_1\!\left(F_0^{-1}(1-p)\right), \quad p \in [0,1],
\]
which trace $\text{TPR}(c)$ against $\text{FPR}(c)$ across all possible thresholds $c$.  

In practice, the so-called pooled ROC curve, $\text{ROC}(p)$, is empirically estimated by comparing the sample distributions of $Y$ between groups without accounting for other covariates. If $\{Y_{1i}\}_{i=1}^{n_1}$ and $\{Y_{0j}\}_{j=1}^{n_0}$ denote the observed biomarker values for diseased and non-diseased individuals, then
\[
\widehat{\text{TPR}}(c) = \frac{1}{n_1}\sum_{i=1}^{n_1}\mathbf{1}(Y_{1i} > c), 
\qquad
\widehat{\text{FPR}}(c) = \frac{1}{n_0}\sum_{j=1}^{n_0}\mathbf{1}(Y_{0j} > c),
\]
and plotting these quantities as $c$ varies produces the empirical ROC curve.  

A widely used summary index of the ROC curve is the Area Under the Curve (AUC), defined as
\[
\text{AUC} = \int_{0}^{1} \text{ROC}(p)\, dp.
\]
This measure has a simple probabilistic interpretation:
\[
\text{AUC} = P(Y_1 > Y_0),
\]
\noindent where $Y_1 \sim  (Y \mid D=1)$ and $Y_0 \sim (Y \mid D=0)$. Intuitively, the AUC represents the probability that a randomly selected diseased subject has a higher biomarker value than a randomly selected non-diseased subject.  

  \subsection{The Need for Covariate-Adjusted ROC Analysis}

\noindent
A fundamental limitation of the pooled (marginal) ROC curve is its implicit assumption of population homogeneity. In practice, the distribution of the biomarker $Y$ and its discriminatory capacity can be substantially influenced by subject-specific covariates $\mathbf{X}$ (e.g., age, sex, body mass index). For example, physical activity levels (the biomarker) naturally decline with age; thus, a threshold indicative of elevated mortality risk for a 30-year-old may be entirely normal for an 80-year-old.  

\noindent
Ignoring such covariates can therefore lead to biased assessments of biomarker performance. To overcome this limitation, the \textit{covariate-adjusted} ROC (aROC) curve conditions the ROC analysis on a specific covariate profile $\mathbf{x}$, providing a personalized evaluation of diagnostic accuracy:
\[
\text{aROC}(p \mid \mathbf{x})
= P\!\left( Y_1 > F_{0}^{-1}(1-p \mid \mathbf{x}) \,\big|\, \mathbf{X}=\mathbf{x} \right),
\]
\noindent
where $Y_1$ denotes the biomarker in the diseased group, $F_{0}(\cdot \mid \mathbf{x})$ is the conditional CDF in the non-diseased group, and $p \in [0,1]$ is the false positive rate.  

\noindent
Equivalently, for a threshold $c$, the covariate-specific true- and false-positive rates are
\[
\text{TPR}(c \mid \mathbf{x}) = P(Y_1 > c \mid \mathbf{X}=\mathbf{x}) = 1 - F_{1}(c \mid \mathbf{x}), 
\qquad
\text{FPR}(c \mid \mathbf{x}) = P(Y_0 > c \mid \mathbf{X}=\mathbf{x}) = 1 - F_{0}(c \mid \mathbf{x}),
\]
\noindent
where $F_{1}(\cdot \mid \mathbf{x})$ is the conditional distribution of $Y$ in diseased subjects. The aROC curve plots $\text{TPR}(c \mid \mathbf{x})$ against $\text{FPR}(c \mid \mathbf{x})$ across thresholds, conditional on $\mathbf{X}=\mathbf{x}$.  

\noindent
Traditional estimators of the aROC often rely on parametric assumptions (for example, location-scale families) or semiparametric regression models. Although useful, these approaches may not capture complex, non-linear relationships between covariates and biomarker distributions, motivating more flexible and robust methodologies.

\subsection{Proposed Framework: A Two-Stage Neural Network Approach}

\noindent
To address these limitations, we propose a flexible and non-parametric framework based on feedforward neural networks (FNNs). The key idea is to estimate the covariate-specific conditional distributions of the biomarker and then construct the aROC curve from these estimates. The procedure is summarized in Algorithm~\ref{alg:fnn_aroc_mathematical} (See Supplemental Material).  

\subsubsection{ROC Curves under Covariate-Specific Location–Scale Models}

\noindent
Suppose that we observe a random sample $\mathcal{D}_n = \{(\mathbf{X}_i, Y_i, D_i)\}_{i=1}^n$, where $\mathbf{X}_i \in \mathbb{R}^p$ are covariates, $Y_i\in \mathbb{R}$ is a continuous biomarker, and $D_i \in \{0,1\}$ denotes the disease status. For $d \in \{0,1\}$ (non-diseased, diseased), assume that conditional on $\mathbf{x}$ the biomarker follows a Gaussian location-scale model:
\[
Y \mid (\mathbf{X}=\mathbf{x}, D=d) \;\sim\; \mu_d(\mathbf{x}) + \sigma_d(\mathbf{x}) \, \varepsilon_d,\quad d \in \{0,1\},
\]
\noindent
where $\mu_d(\mathbf{x})$ and $\sigma_d^2(\mathbf{x})$ are the conditional mean and variance functions, and $\varepsilon_d\sim \mathcal{N}  (0,1)$ is a standard Gaussian random variable.  

\noindent
In this formulation, the aROC for a fixed covariate profile $\mathbf{x}$ can be expressed as
\[
\text{aROC}(p \mid \mathbf{x})
= 1 - \Phi\!\left( b(\mathbf{x}) \, \Phi^{-1}(1-p) - a(\mathbf{x}) \right),
\]
\noindent
where $\Phi(\cdot)$ is the standard normal CDF, and
\[
a(\mathbf{x}) = \frac{\mu_1(\mathbf{x}) - \mu_0(\mathbf{x})}{\sigma_1(\mathbf{x})},
\qquad
b(\mathbf{x}) = \frac{\sigma_0(\mathbf{x})}{\sigma_1(\mathbf{x})}.
\]
\noindent
The corresponding covariate-specific AUC is
\[
\text{AUC}(\mathbf{x}) = \int_0^1 \text{aROC}(p \mid \mathbf{x}) \, dp.
\]

\subsubsection{Model Estimation via Neural Networks}
\label{subsec:model-estimation}

\noindent
The main challenge is to flexibly estimate $\mu_d(\mathbf{x})$ and $\sigma_d^2(\mathbf{x})$ from the data. We adopt a two-stage estimation strategy using FNNs.  

\paragraph{Stage 1: Conditional Mean Estimation.}
\noindent
Model the conditional mean for subgroup $d$ as
\[
\hat{\mu}_d(\mathbf{x}; \boldsymbol{\theta}_d)
= f^{(L)} \circ f^{(L-1)} \circ \dots \circ f^{(1)}(\mathbf{x}),
\]
\noindent
where $\boldsymbol{\theta}_d$ are the network parameters. Each layer is
\[
f^{(\ell)}(\mathbf{h}^{(\ell-1)})
= \phi^{(\ell)}\!\left(\mathbf{W}^{(\ell)} \mathbf{h}^{(\ell-1)} + \mathbf{b}^{(\ell)}\right),
\]
\noindent
with $\mathbf{h}^{(0)} = \mathbf{x}$, $\mathbf{W}^{(\ell)}$ and $\mathbf{b}^{(\ell)}$ the weight matrix and bias vector, and $\phi^{(\ell)}$ a nonlinear activation (e.g., ReLU). The final layer is linear to produce $\hat{\mu}_d(\mathbf{x})$. Parameters are estimated by minimizing the mean squared error:
\[
\hat{\boldsymbol{\theta}}_d
= \arg\min_{\boldsymbol{\theta}_d} \sum_{i \in \mathcal{D}_d} \big( Y_i - \hat{\mu}_d(\mathbf{X}_i; \boldsymbol{\theta}_d) \big)^2.
\]

\paragraph{Stage 2: Conditional Variance Estimation.}
\noindent
Estimate the variance function $\sigma_d^2(\mathbf{x}; \boldsymbol{\phi}_d)$ by regressing squared residuals, $\hat{r}_i = \big(Y_i - \hat{\mu}_d(\mathbf{X}_i)\big)^2$, on the covariates using a second FNN:

\[
\hat{\sigma}_d^2(\mathbf{x}; \boldsymbol{\phi}_d) = \operatorname{softplus}\!\big(g(\mathbf{x}; \boldsymbol{\phi}_d)\big)
\]

\noindent
with parameters obtained via
\[
\hat{\boldsymbol{\phi}}_d
= \arg\min_{\boldsymbol{\phi}_d} \sum_{i \in \mathcal{D}_d} \big( \hat{r}_i - \hat{\sigma}_d^2(\mathbf{X}_i; \boldsymbol{\phi}_d) \big)^2.
\]

\noindent
Substituting $\hat{\mu}_d(\mathbf{x})$ and $\hat{\sigma}_d(\mathbf{x})$ into the location-scale formulation produces the estimated aROC and AUC surfaces across the covariate profiles.

\subsection*{Model and Computational Details}

We implement the Feed-Forward Neural Network (FNN) as a Multi-Layer Perceptron (MLP) in PyTorch for regression. The architecture comprises an input layer, a stack of hidden layers, and a single neuron output layer with linear activation to predict continuous outcomes. Each hidden layer applies an affine transformation followed by a Rectified Linear Unit (ReLU) nonlinearity. To mitigate overfitting, we insert dropout after every hidden layer (default rate $0.2$) and apply weight decay $L_2$ through the \texttt{Adam} optimizer (default $1\times 10^{-4}$). The depth and width of the hidden stack, specified by the \texttt{hidden} hyperparameter \_layers (e.g., $[128,128,128]$ or $[64,32]$), determine the representation capacity: larger models capture richer nonlinearities between covariates $\mathbf{x}_i$ and biomarker values $y_i$, but can overfit without regularization. ReLU promotes efficient optimization and helps to avoid vanishing gradients in deeper networks.

Training seeks parameters $\theta$ that minimize the mean squared error (MSE) over a mini-batch $\mathcal{B}$:
\[
\mathcal{L}(\theta) \;=\; \frac{1}{|\mathcal{B}|}\sum_{(i)\in \mathcal{B}}\bigl(y_i - f_\theta(\mathbf{x}_i)\bigr)^2 \;+\; \lambda \,\|\theta\|_2^2,
\]
where $f_\theta$ denotes the network mapping and $\lambda$ is the weight-decay coefficient. Unless otherwise stated, we use \texttt{Adam} with default momentum parameters, learning rate $\eta=10^{-3}$, and mini-batch size $32$. Early stopping based on validation MSE halts training when no improvement is observed for several epochs, preventing unnecessary updates and overfitting.

Model selection and hyperparameter tuning follow a two-stage, data-driven protocol. First, we split the data set into $80\%$ training/validation and $20\%$ hold-out tests. Within the $80\%$ portion, we perform $k$-fold cross-validation with $k=5$ to evaluate candidate configurations. The following hyperparameters are exposed as command-line arguments to enable a grid or randomized search: number and width of hidden layers (\texttt{hidden\_layers}), dropout rate, learning rate, and weight decay. For each fold, we initialize a new model, train the training folds, and compute MSE validation for the held-out fold. The MSE of MSE for average validation in $5$ folds provides a robust estimate of generalization for each candidate. Optionally, we repeat the procedure on multiple random seeds to assess the robustness.

After selecting the optimal configuration, we retrain a final model on the full training/validation split ($80\%$ of the data) using the chosen hyperparameters and early-stopping settings. We then evaluate this model once on the $20\%$ hold-out test set to obtain an unbiased estimate of the out-of-sample performance. This final model is subsequently used for all downstream analyzes, including out-of-sample prediction and covariate-adjusted ROC estimation. The procedure prioritizes generalization rather than in-sample fit, thereby reinforcing the reliability and reproducibility of the reported results.

\paragraph{Random Forest configuration.}

For comparison, we train a Random Forest (RF) regressor (scikit-learn) with \texttt{n\_estimators=500}, \texttt{max\_depth=3}, \texttt{min\_samples\_split=20}, \texttt{min\_samples\_leaf=20} and \texttt{max\_features=2}, using the squared error criterion. All other parameters remain at their default values. These settings prioritize generalization—shallow trees and stricter minimum-sample thresholds that reduce variance and mitigate overfitting—while retaining the capacity to capture nonlinear effects. This configuration provides a stable baseline against which we compare the Feedforward Neural Network (FNN).

\section{Physical Activity Case Study: NHANES 2011--2014}\label{sec:nhanes}

\subsection{Background and Aims}

We applied an individualized, covariate- adjusted ROC framework to accelerometry data from the National Health and Nutrition Examination Survey (NHANES, 2011--2014) to evaluate \emph{Total Activity Count} (TAC) as a biomarker for all-cause mortality. The TAC was calculated from monitor independent movement summary units at minute level (MIMS) recorded by ActiGraph GT3X+ devices worn for seven consecutive days. Mortality status at 3-, 5- and 8-year horizons was obtained by link to the National Death Index.

Many prior studies using logistic and Cox regression have shown that higher activity levels are associated with a lower risk of mortality. However, stratified analyzes (e.g., by age group) can conflate biomarker effects with individual characteristics. To isolate the discriminative value of TAC while accounting for covariates such as age, sex, and BMI, we employ a covariate-adjusted ROC analysis.

The purpose of this section is to perform different analyzes to address the following two goals: (i) quantify the value of TAC as a biomarker for short-term (3-year), medium-term (5-year) and long-term (8-year) mortality; and (ii) assess how age, BMI, and sex modify this value, addressing two questions: \emph{Is physical activity more effective for promoting longevity in younger versus older individuals?} and \emph{Is the effect comparable between women and men?}. For this purpose, we use the proposed ROC--NN algorithm along with classical additive ROC methods and a machine learning comparator in which the neural network regression component is replaced by a random forest, allowing for a direct empirical comparison on the same prediction task.

\subsection{Results}

The analytical cohort comprised $n=5{,}006$ adults aged 20--80 years with valid accelerometry data and complete covariates (age, sex, and body mass index [BMI]). Table~\ref{tab:desc_stats_side_by_side} summarizes participant characteristics; continuous variables are reported as mean~($\pm$~SD) and categorical variables as counts~(\%). As expected, the deceased were on average nearly two decades older and exhibited substantially lower total activity counts (TAC), while BMI distributions were broadly similar between groups. These contrasts underscore the importance of covariate adjustment in assessing the prognostic value of TAC.

Kernel density plots (Figures~\ref{fig:tac2_age_all}--\ref{fig:age_bmi_all}) corroborate these patterns: (i) survivors maintained higher TAC throughout the age spectrum, while decedents clustered in the lowest age-specific TAC percentiles; (ii) survivors more frequently occupied the high TAC region, whereas BMI showed minimal separation between groups; and (iii) mortality density was concentrated in the 60--80 year range. 

\begin{table}[H]
\centering
\caption{Descriptive Statistics for the Cohort, Stratified by Sex.}
\label{tab:desc_stats_side_by_side}

\begin{tabular}{@{}lcccccc@{}}
\toprule
\multirow{2}{*}{\textbf{Variable}} 
  & \multicolumn{3}{c}{\textbf{Female Cohort}} 
  & \multicolumn{3}{c}{\textbf{Male Cohort}} \\
\cmidrule(lr){2-4} \cmidrule(lr){5-7}
  & \textbf{Total} & \textbf{Deceased} & \textbf{Survivors} 
  & \textbf{Total} & \textbf{Deceased} & \textbf{Survivors} \\
\midrule

\textbf{Age} 
  & 48.23 (±17.09) & 67.40 (±13.49) & 46.80 (±16.46)
  & 48.92 (±17.40) & 67.41 (±14.51) & 47.08 (±16.57) \\

\textbf{BMI} 
  & 29.54 (±7.54) & 29.58 (±7.39) & 29.54 (±7.56)
  & 28.56 (±5.87) & 28.51 (±5.99) & 28.57 (±5.86) \\

\textbf{TAC} 
  & 680.75 (±249.32) & 599.24 (±221.51) & 686.84 (±250.26)
  & 735.04 (±920.60) & 610.01 (±250.23) & 747.47 (±961.19) \\
\midrule

\textbf{Sex} & \multicolumn{3}{c}{2385 (100.0\%), 166 Deceased (7.0\%)} 
             & \multicolumn{3}{c}{2621 (100.0\%), 237 Deceased (9.0\%)} \\
\bottomrule
\end{tabular}

\end{table}

We compared a feedforward neural network (FNN) with two common alternatives, a Random Forest and a semiparametric additive GAM / ROC model, to evaluate how well total activity counts (TAC) discriminate mortality risk after adjusting for age, sex and BMI in a reasonably large cohort (n=5,006). The FNN uses standard regularization (dropout and weight decay), early stopping, and cross-validated hyperparameter tuning, and consistently uncovers an epidemiologically coherent pattern: The discriminative ability of TAC increases with age. In early adulthood, adjusted AUCs are modest $(\approx 0.58\text{--}0.62)$, increase in midlife, and reach $\approx 0.75\text{--}0.79$ in older groups (e.g., around 60 years of age in some strata). By prediction horizon, men show the strongest discrimination at 3 years and a step down at 5 years with a further decline at 8 years; women display moderate discrimination at 3 and 8 years (typically around 0.60) and a more favorable profile at 5 years, where AUCs can approach 0.70. Across both sexes, performance generally attenuates by 8 years, consistent with increasing heterogeneity and competing risks over longer follow-up. Age emerges as the dominant modifier of the prognostic value of TAC, with a clear monotone improvement from young to old adults; BMI shows only small non-monotone differences after adjustment, indicating a limited modification of the effect by adiposity. In general, FNN AUCs range from approximately 0.57 to 0.79 across age, sex, and horizon strata. Against this benchmark, the Random Forest achieves comparable discrimination in several strata but exhibits signs of miscalibration (over- or under-confident risk estimates), while the semiparametric GAM/ROC model generally yields lower AUCs, likely reflecting restricted flexibility to capture age-dependent nonlinearities and interactions. Results should be interpreted with caution in strata with fewer events, particularly younger ages and the female 3-year horizon, where seemingly favorable short-term associations are less biologically plausible and may reflect model misspecification or sampling variability. Taken together, the evidence indicates a sex effect and supports that higher physical activity is more strongly protective - and more effectively discriminatory - in older ages, with the FNN providing the most coherent and robust summary across different time horizons.

In general, ROC analysis shows that TAC is a clinically significant biomarker, but a biomarker whose value depends on age and sex. In older adults, TAC provides stronger prognostic information than in younger adults, supporting its use in age-targeted risk stratification. The lack of strong modification by BMI suggests that clinicians should encourage activity across weight categories. Crucially, reporting a single pooled AUC obscures this heterogeneity; Personalized and covariate-conditional  provide new information on the prediction capacity of the TAC variable to detect mortality in three, five, and eight years of age with change in age and gender distribution. 
The ROC functions for different combinations of variables are shown in the Supplementary Material (Figures \ref{fig:females_3year}--\ref{fig:males_8year}).

\begin{figure}[H]
    \centering
    \begin{subfigure}{\textwidth}
        \centering
        \includegraphics[width=0.32\textwidth]{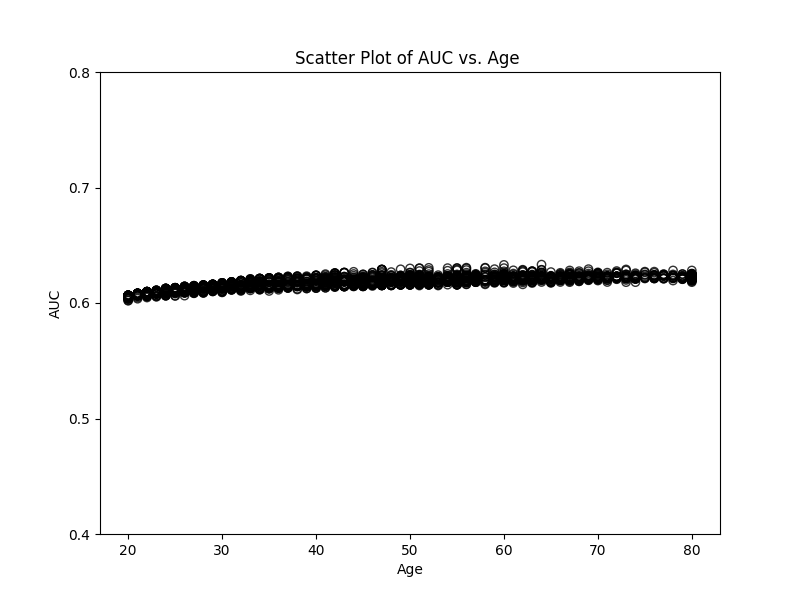}
        \includegraphics[width=0.32\textwidth]{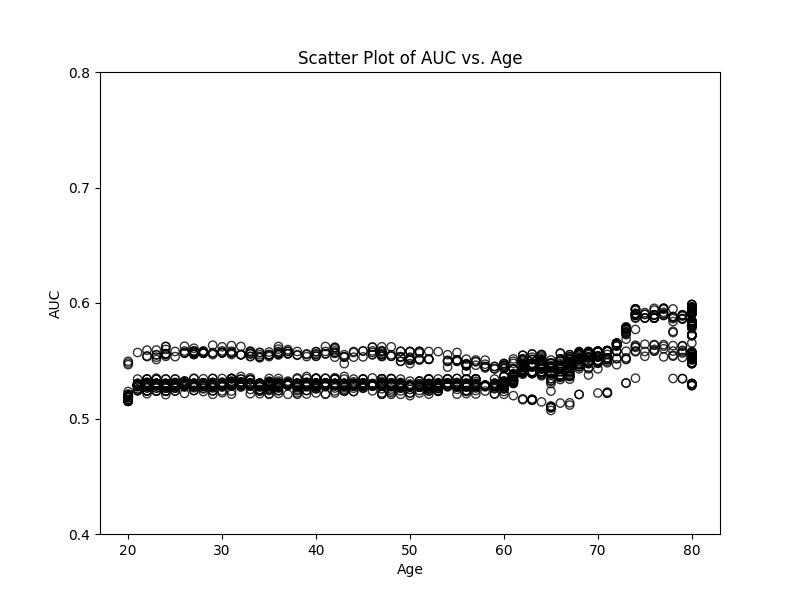}
        \includegraphics[width=0.32\textwidth]{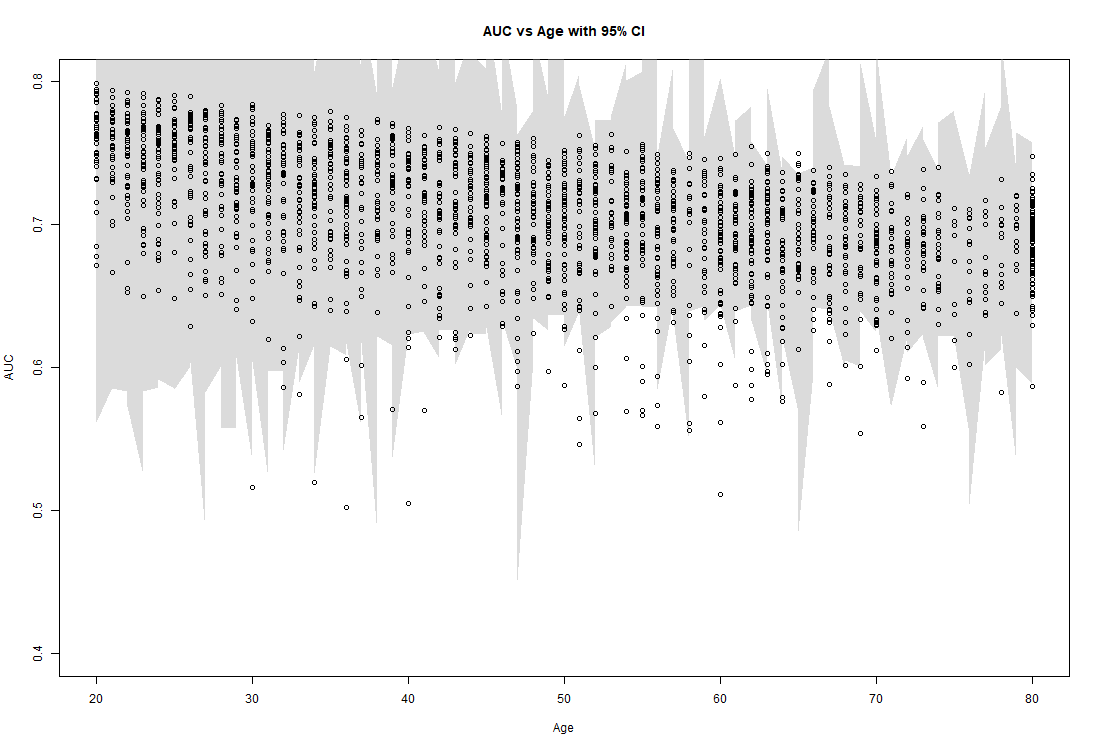}
        \caption{Female. \textbf{Left:} FNN model. \textbf{Center:} Random Forest model. \textbf{Right:} Semiparametric R model.}
        \label{fig:3_year_auc_f}
    \end{subfigure}
    \par\bigskip 
    \begin{subfigure}{\textwidth}
        \centering
        \includegraphics[width=0.32\textwidth]{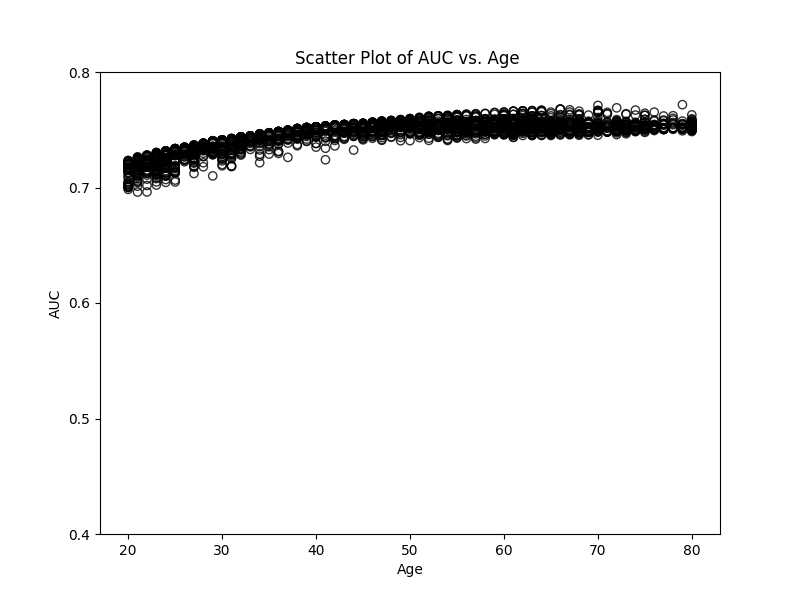}
        \includegraphics[width=0.32\textwidth]{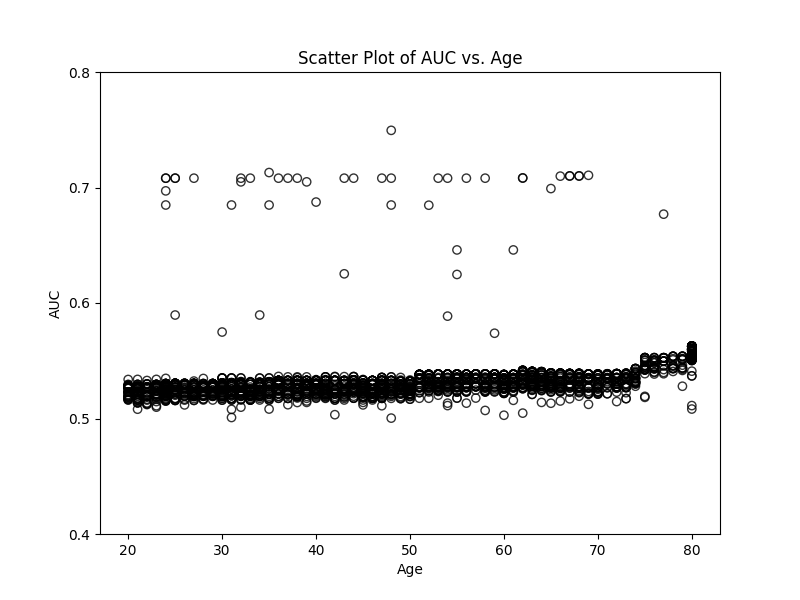}
        \includegraphics[width=0.32\textwidth]{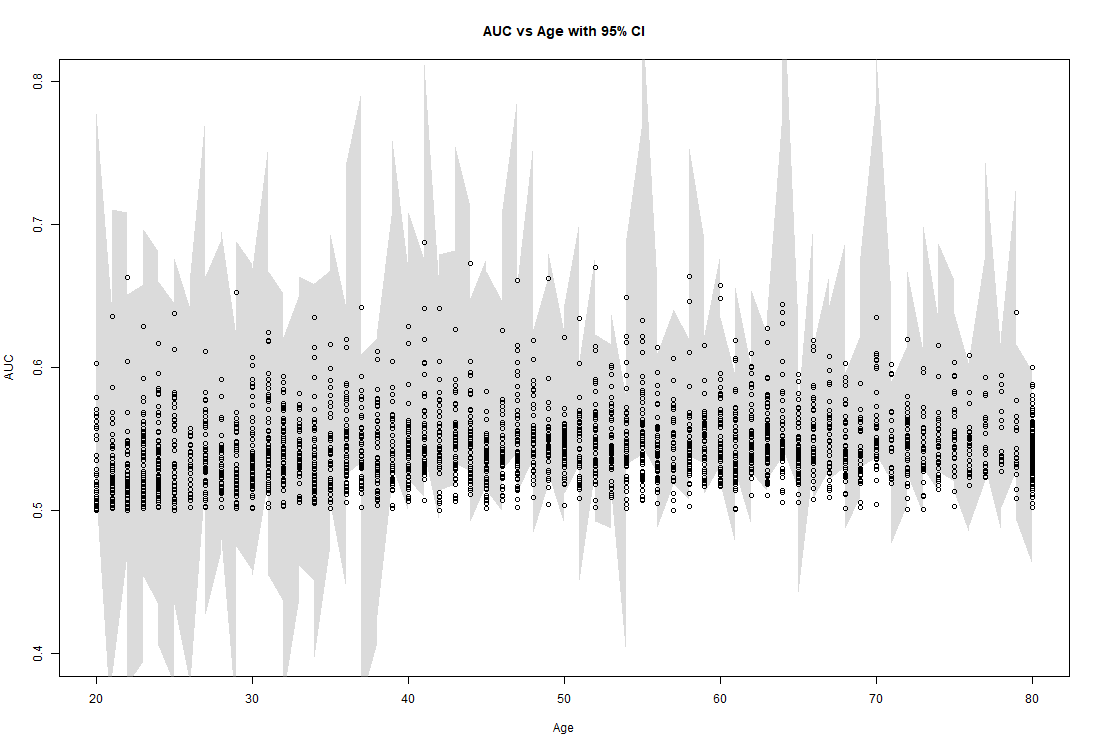}
        \caption{Male. \textbf{Left:} FNN model. \textbf{Center:} Random Forest model. \textbf{Right:} Semiparametric R model.}
        \label{fig:3_year_auc_m}
    \end{subfigure}
    \caption{Covariate-adjusted AUC for 3-year mortality prediction by age and sex. The FNN model (left panels) reveals a clear monotonic increase in AUC with age. The Random Forest model (center panels) shows high volatility with no clear trend, while the R model (right panels) suggests a decreasing relationship for females and a flat pattern for males.}
    \label{fig:3_year_auc}
\end{figure}

\begin{figure}[H]
    \centering
    \begin{subfigure}{\textwidth}
        \centering
        \includegraphics[width=0.32\textwidth]{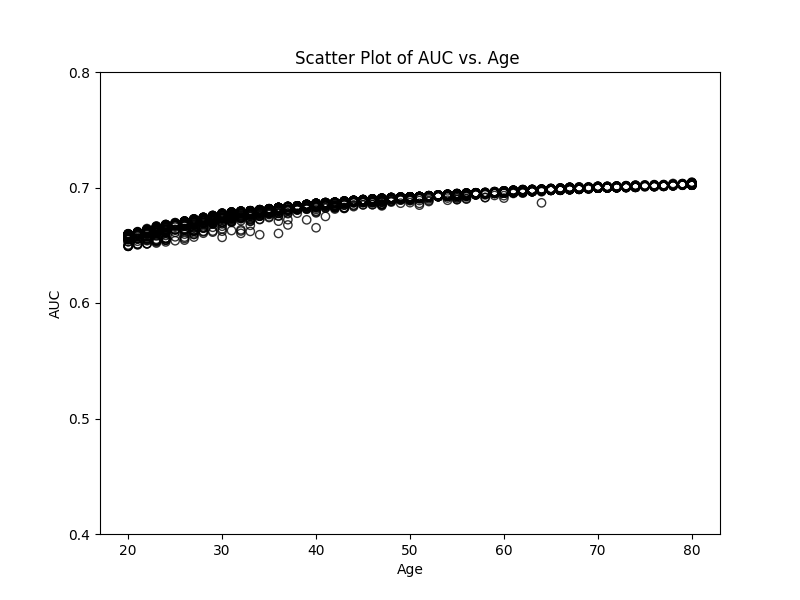}
        \includegraphics[width=0.32\textwidth]{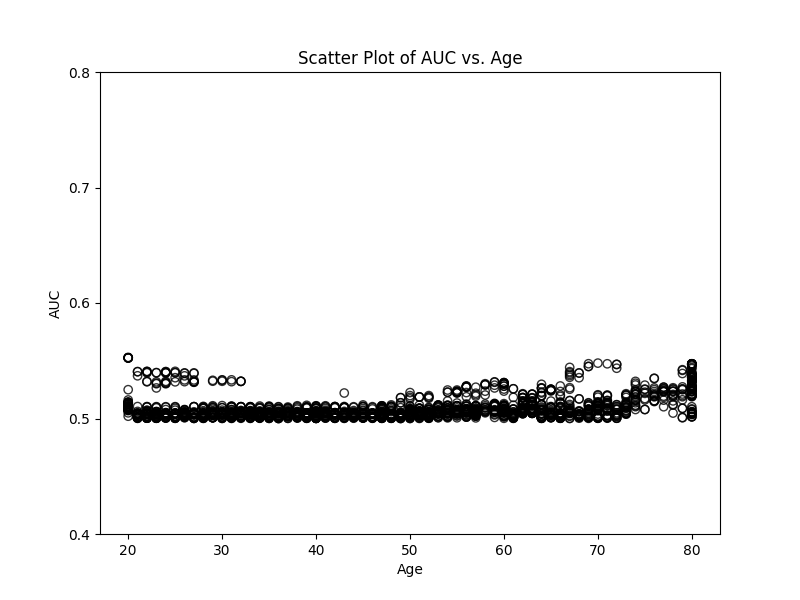}
        \includegraphics[width=0.32\textwidth]{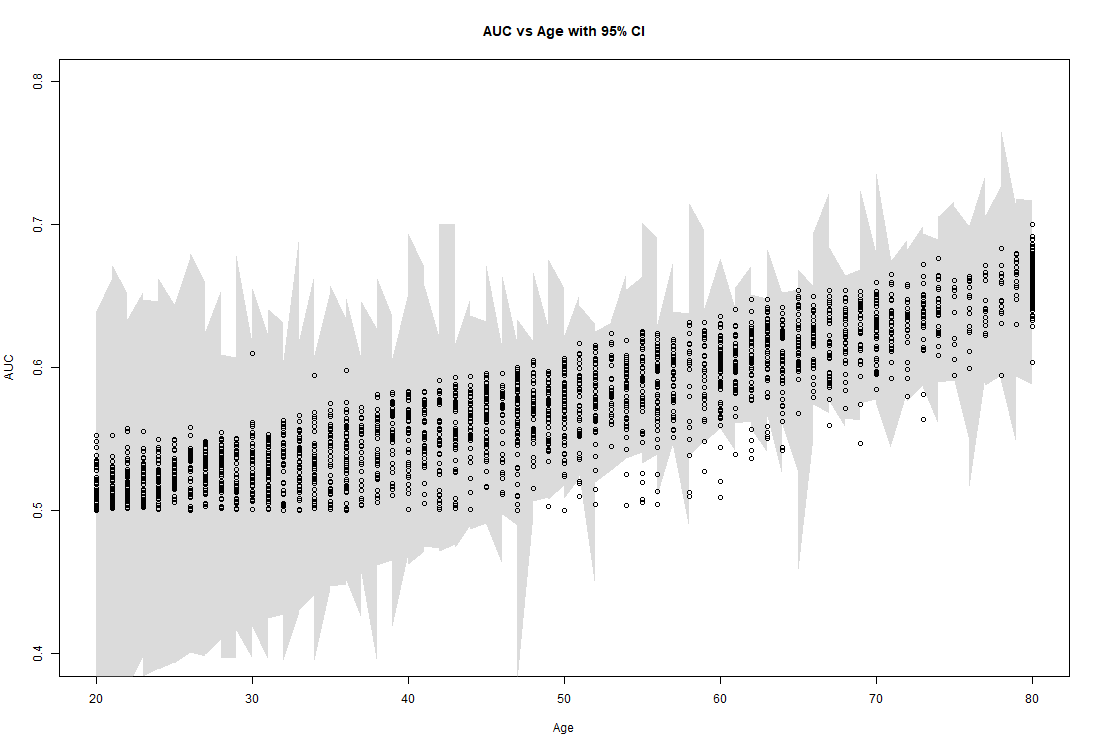}
        \caption{Female. \textbf{Left:} FNN model. \textbf{Center:} Random Forest model. \textbf{Right:} Semiparametric R model.}
        \label{fig:5_year_auc_f}
    \end{subfigure}
    \par\bigskip
    \begin{subfigure}{\textwidth}
        \centering
        \includegraphics[width=0.32\textwidth]{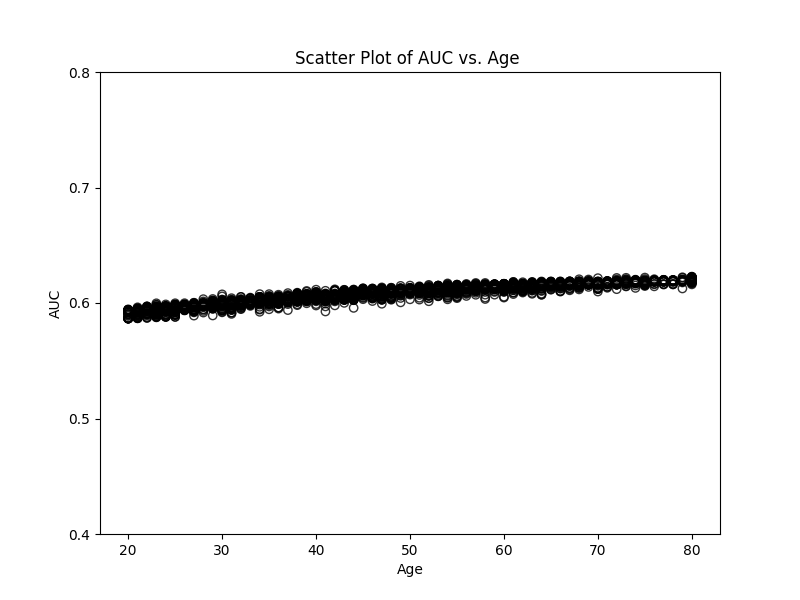}
        \includegraphics[width=0.32\textwidth]{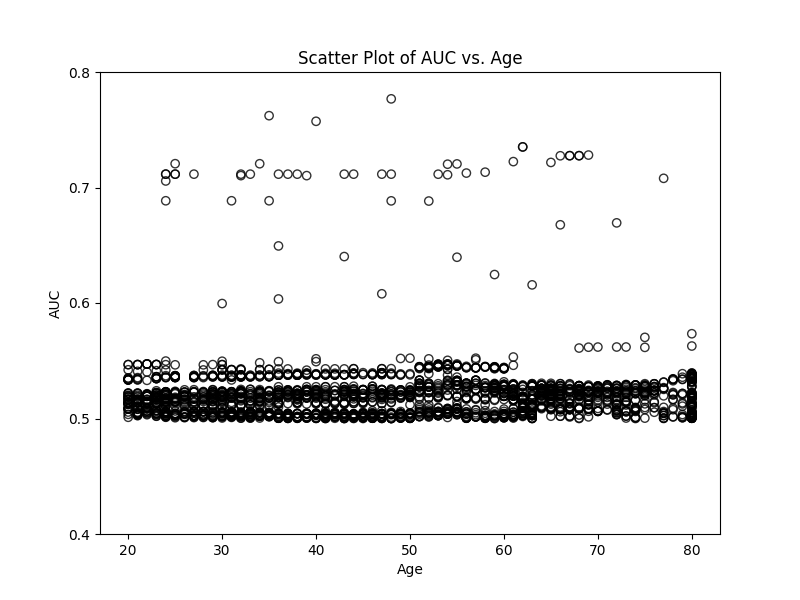}
        \includegraphics[width=0.32\textwidth]{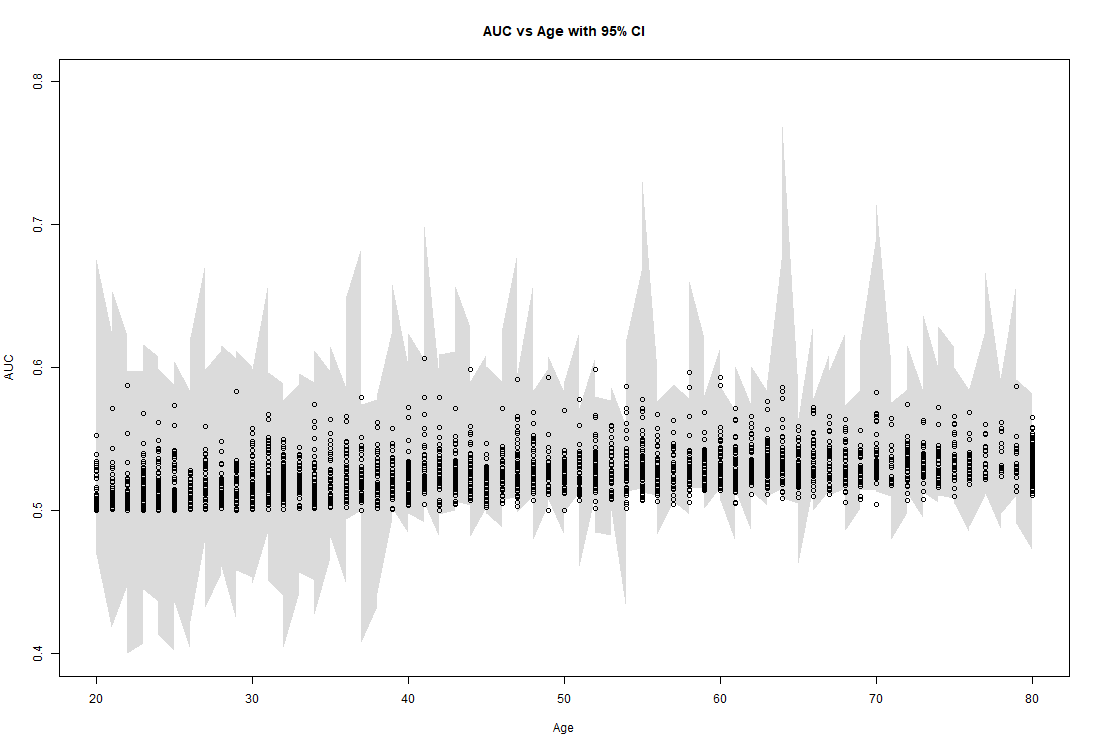}
        \caption{Male. \textbf{Left:} FNN model. \textbf{Center:} Random Forest model. \textbf{Right:} Semiparametric R model.}
        \label{fig:5_year_auc_m}
    \end{subfigure}
    \caption{Covariate-adjusted AUC for 5-year mortality prediction by age and sex. The trends observed in the 3-year analysis are strengthened over a 5-year horizon. The FNN model consistently shows that the predictive power of TAC grows with age for both males and females.}
    \label{fig:5_year_auc}
\end{figure}

\begin{figure}[H]
    \centering
    \begin{subfigure}{\textwidth}
        \centering
        \includegraphics[width=0.32\textwidth]{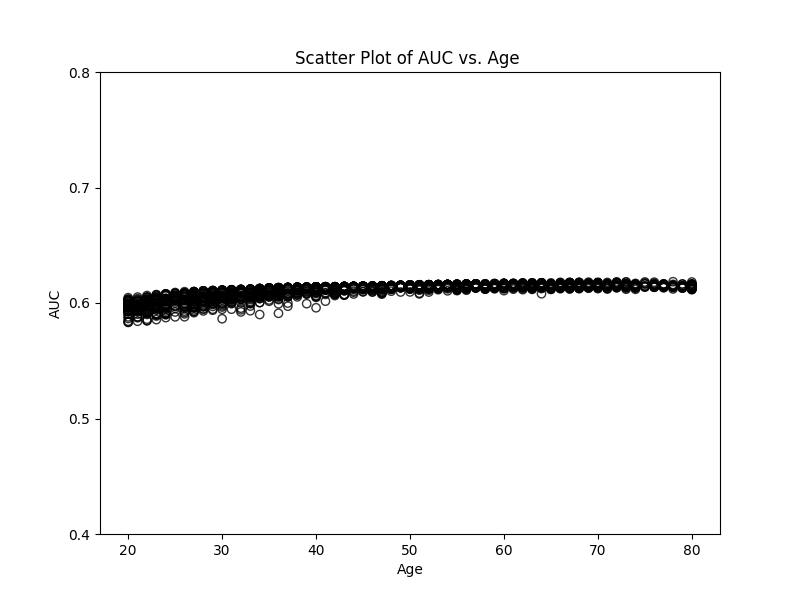}
        \includegraphics[width=0.32\textwidth]{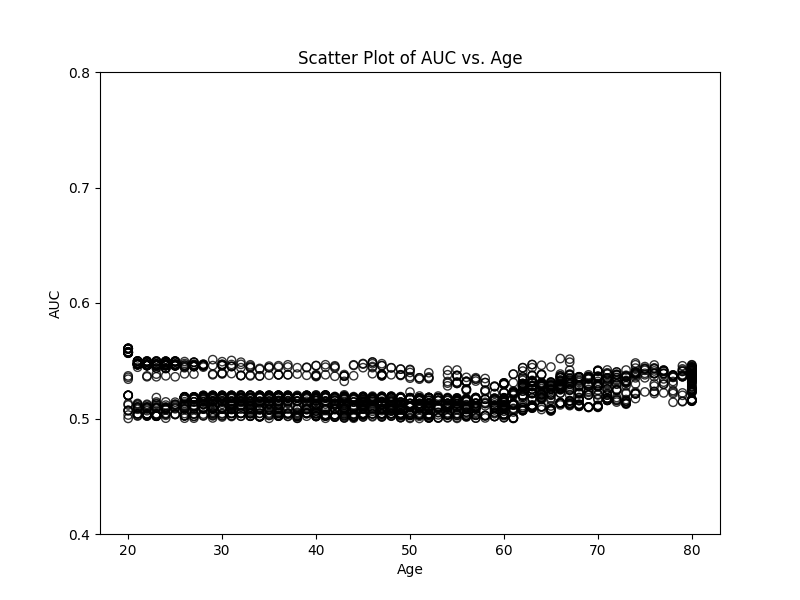}
        \includegraphics[width=0.32\textwidth]{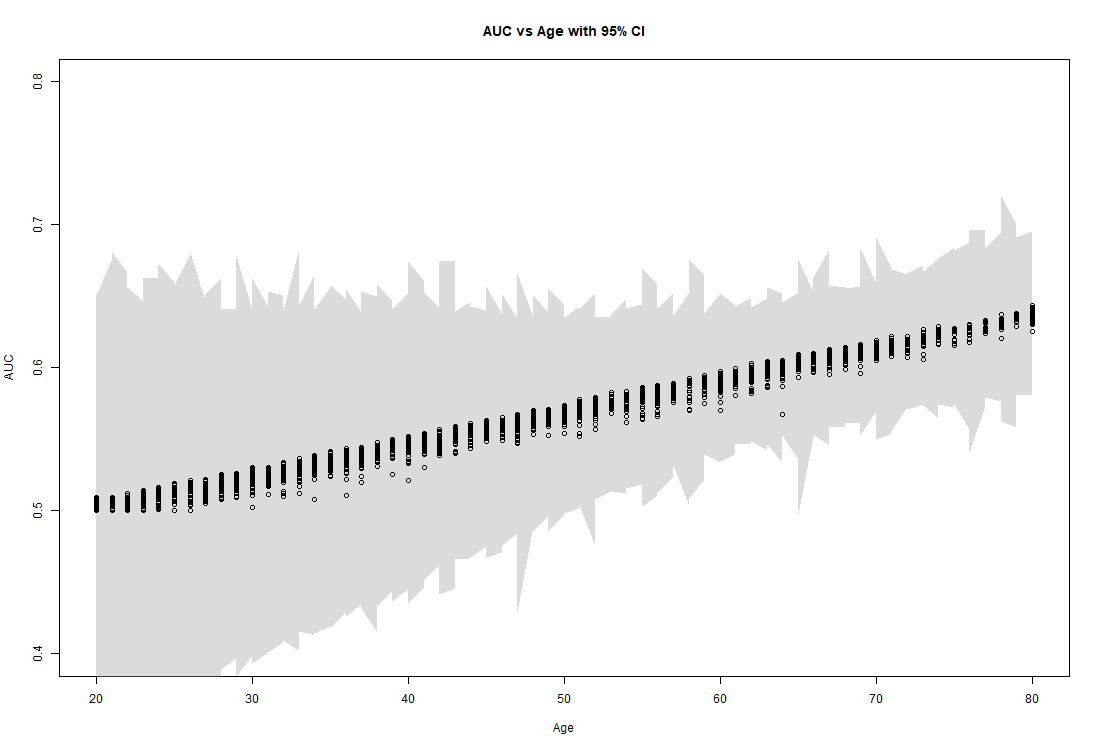}
        \caption{Female. \textbf{Left:} FNN model. \textbf{Center:} Random Forest model. \textbf{Right:} Semiparametric R model.}
        \label{fig:8_year_auc_f}
    \end{subfigure}
    \par\bigskip
    \begin{subfigure}{\textwidth}
        \centering
        \includegraphics[width=0.32\textwidth]{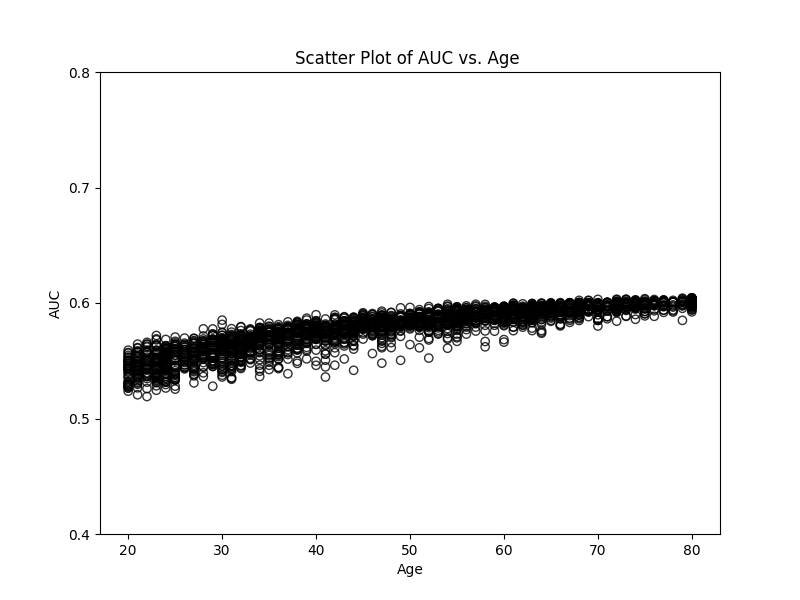}
        \includegraphics[width=0.32\textwidth]{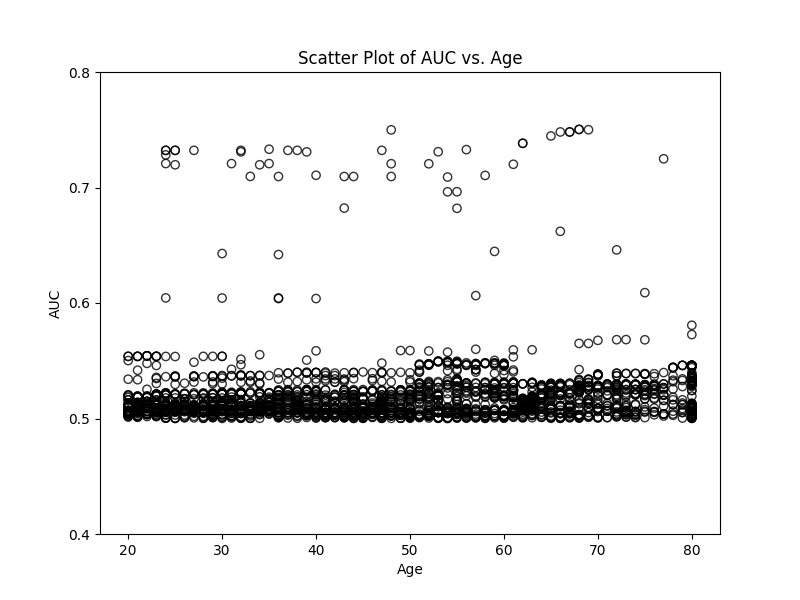}
        \includegraphics[width=0.32\textwidth]{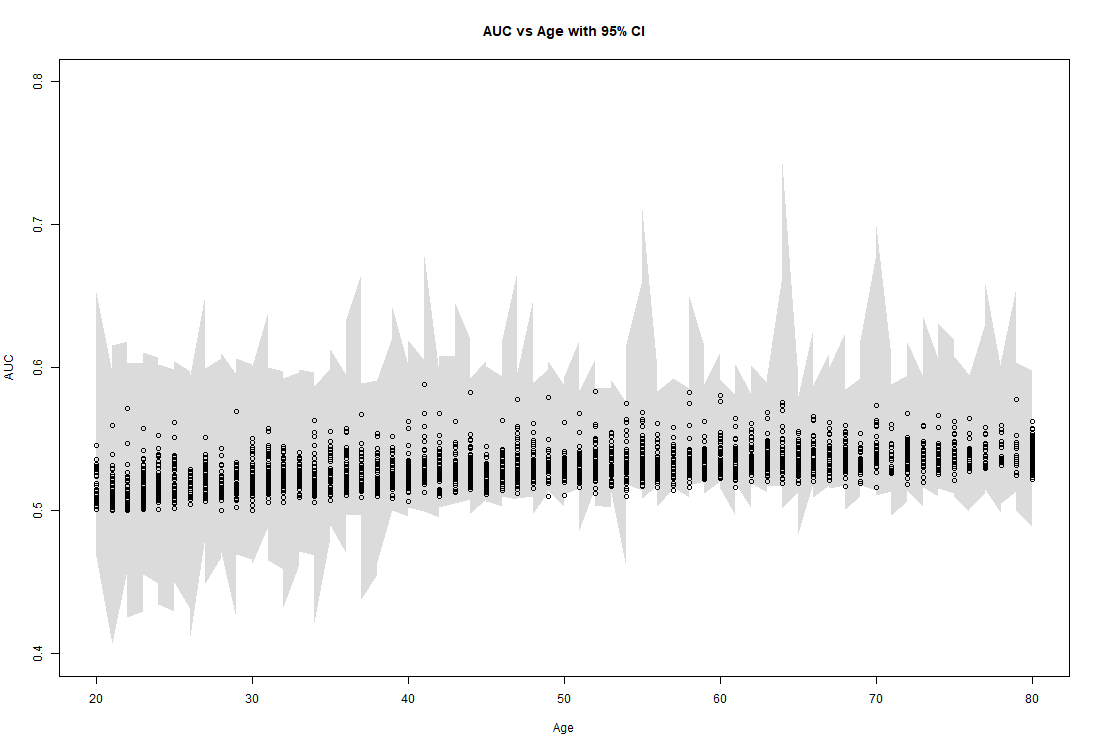}
        \caption{Male. \textbf{Left:} FNN model. \textbf{Center:} Random Forest model. \textbf{Right:} Semiparametric R model.}
        \label{fig:8_year_auc_m}
    \end{subfigure}
    \caption{Covariate-adjusted AUC for 8-year mortality prediction by age and sex. The FNN's finding of an age-dependent increase in AUC remains robust over the 8-year period, confirming the stability of this biological relationship.}
    \label{fig:8_year_auc}
\end{figure}

\begin{table}[h!]
\centering
\caption{Summary of Model Performance (AUC) by Key Stratifying Factors}
\label{tab:auc_summary}
\begin{tabular}{llc}
\toprule
\textbf{Stratifying Factor} & \textbf{Category} & \textbf{AUC Range} \\
\midrule
\multirow{2}{*}{Sex} & Male   & 0.57--0.79 \\
                     & Female & 0.59--0.67 \\
\midrule
\multirow{3}{*}{Time Horizon} & 3-year & 0.62--0.79 \\
                              & 5-year & 0.60--0.67 \\
                              & 8-year & 0.57--0.65 \\
\midrule
\multirow{3}{*}{Age Group} & Young (20)  & 0.57--0.72 \\
                           & Middle (40) & 0.60--0.77 \\
                           & Old (60)    & 0.62--0.79 \\
\midrule
\multirow{2}{*}{BMI} & 20 & 0.59--0.77 \\
                     & 40 & 0.57--0.79 \\
\bottomrule
\end{tabular}
\end{table}

\section{Discussion}\label{sec:conclusion}
 We introduce a flexible framework for covariate‐adjusted ROC analysis based on feedforward neural networks (FNNs). The methodological contribution is to relax the functional form constraints imposed by classical parametric and semiparametric approaches, enabling the regression function underlying the covariate-specific ROC curve to capture complex, nonlinear relationships among biomarkers, subject characteristics, and reference populations in clinical studies.

In nine simulation scenarios (see the Supplementary Materials), the proposed ROC FNN consistently outperformed a benchmark Random Forest in terms of lower mean squared error in approximately conditional mean and variance regression functions, and consequently in roc function approximations and greater in settings with nonlinear covariates and interaction effects. These results confirm satisfactory finite-sample performance for applications to biomedical data and nonlinear relationships, a common scenario in real-world biomedical data.

We evaluated practical utility using NHANES to study the covariate-specific discriminatory value of physical activity for all-cause mortality. The analysis revealed that the discriminatory capacity of the total activity counts (TAC) is not constant across age but \emph{increases with age}. Thus, a given TAC level conveys greater prognostic information for older adults than for younger adults, an effect that pooled AUC analyses may obscure. This pattern was identified by the proposed FNN framework, whereas the benchmark methods either did not detect the trend (Random Forest) or suggested inconsistent patterns (semiparametric ROC model) surely due to lack of flexibility.

Future work will extend the framework to time-to-event outcomes using a \emph{log-normal accelerated failure time} model and will consider biomarkers defined as mathematical functionals rather than scalars. We will also investigate \emph{Bayesian neural networks} to quantify uncertainty through posterior inference for the ROC curve and the derived metrics (e.g. AUC).

In summary, the FNN ROC based approach provides a robust and scalable tool for covariate-adjusted ROC analysis, supporting personalized evaluation of biomarker performance in the presence of individual-level covariates and facilitating clinial decision-making support.

\bibliography{references}
\newpage
\appendix

\section*{Pseudocode}

\begin{algorithm}[H]
\caption{Two-Stage Feedforward Neural Network for Covariate-Adjusted ROC Analysis}
\label{alg:fnn_aroc_mathematical}
\begin{algorithmic}[1]
\Require Dataset $\mathcal{D}_n=\{(\mathbf{X}_i, Y_i, D_i)\}_{i=1}^n$ with $\mathbf{X}_i\in\mathbb{R}^p$, $Y_i\in\mathbb{R}$, $D_i\in\{0,1\}$
\Ensure $\widehat{\mathrm{aROC}}(p \mid \mathbf{x})$ and $\widehat{\mathrm{AUC}}(\mathbf{x})$ for $\mathbf{x}\in\mathbf{X}_{\text{grid}}$
\State \textbf{Notation:} $\Phi$ is the standard normal CDF; $\Phi^{-1}$ is its quantile function.

\State \textbf{Split by disease status:}
\[
\mathcal{D}_0=\{(\mathbf{X}_i,Y_i):D_i=0\},\qquad
\mathcal{D}_1=\{(\mathbf{X}_i,Y_i):D_i=1\}.
\]

\Statex \textbf{Stage 1: Conditional Mean Estimation}
\For{each $d\in\{0,1\}$}
  \State Define a feedforward network $\mathrm{FNN}^{(\mu)}_{d}(\cdot;\theta)$ (final layer linear).
  \State Estimate $\hat{\theta}^{(\mu)}_{d}$ by
  \[
    \hat{\theta}^{(\mu)}_{d}
    = \arg\min_{\theta}\sum_{(\mathbf{X}_i,Y_i)\in\mathcal{D}_d}\!\big(Y_i-\mathrm{FNN}^{(\mu)}_{d}(\mathbf{X}_i;\theta)\big)^2.
  \]
  \State Set $\hat{\mu}_d(\mathbf{x})\gets \mathrm{FNN}^{(\mu)}_{d}(\mathbf{x};\hat{\theta}^{(\mu)}_{d})$.
\EndFor

\Statex \textbf{Stage 2: Conditional Variance Estimation}
\For{each $d\in\{0,1\}$}
  \State Compute squared residuals on $\mathcal{D}_d$:
  \[
    \hat{r}_i^2\gets\big(Y_i-\hat{\mu}_d(\mathbf{X}_i)\big)^2,\qquad (\mathbf{X}_i,Y_i)\in\mathcal{D}_d.
  \]
  \State Define $\mathrm{FNN}^{(\sigma^2)}_{d}(\cdot;\phi)$ with positive output, e.g.
  \[
    \hat{\sigma}_d^2(\mathbf{x};\phi)=\mathrm{softplus}\!\big(g_d(\mathbf{x};\phi)\big).
  \]
  \State Estimate $\hat{\phi}^{(\sigma^2)}_{d}$ by
  \[
    \hat{\phi}^{(\sigma^2)}_{d}
    = \arg\min_{\phi}\sum_{(\mathbf{X}_i,\hat{r}_i^2)\in\mathcal{D}_d}\!\big(\hat{r}_i^2-\hat{\sigma}_d^2(\mathbf{X}_i;\phi)\big)^2.
  \]
  \State Set $\hat{\sigma}^2_d(\mathbf{x})\gets \hat{\sigma}^2_d(\mathbf{x};\hat{\phi}^{(\sigma^2)}_{d})$ and $\hat{\sigma}_d(\mathbf{x})\gets\sqrt{\hat{\sigma}^2_d(\mathbf{x})}$.
\EndFor

\Statex \textbf{Stage 3: aROC and AUC Computation (Location–Scale Model)}
\For{each $\mathbf{x}\in\mathbf{X}_{\text{grid}}$}
  \State Location contrast: \(\displaystyle \hat{a}(\mathbf{x}) \gets \frac{\hat{\mu}_1(\mathbf{x})-\hat{\mu}_0(\mathbf{x})}{\hat{\sigma}_1(\mathbf{x})}\).
  \State Scale ratio: \(\displaystyle \hat{b}(\mathbf{x}) \gets \frac{\hat{\sigma}_0(\mathbf{x})}{\hat{\sigma}_1(\mathbf{x})}\).
  \State For any \(p\in[0,1]\), define
  \[
    \widehat{\mathrm{aROC}}(p\mid\mathbf{x})
    \gets 1-\Phi\!\big(\hat{b}(\mathbf{x})\,\Phi^{-1}(1-p)-\hat{a}(\mathbf{x})\big).
  \]
  \State Covariate-specific AUC (closed form under the model):
  \[
    \widehat{\mathrm{AUC}}(\mathbf{x})
    \gets \Phi\!\left(\frac{\hat{a}(\mathbf{x})}{\sqrt{1+\hat{b}^2(\mathbf{x})}}\right)
    \quad\text{(or numerically } \int_0^1 \widehat{\mathrm{aROC}}(p\mid\mathbf{x})\,dp\text{)}.
  \]
\EndFor

\State \textbf{Return} $\big\{\widehat{\mathrm{aROC}}(p\mid\mathbf{x}),\,\widehat{\mathrm{AUC}}(\mathbf{x}) : \mathbf{x}\in\mathbf{X}_{\text{grid}}\big\}$.
\end{algorithmic}
\end{algorithm}

\section*{Figures Case Study}

\subsection{Bivariate Density Function Estimators}

\begin{figure}[H]
\centering
\begin{subfigure}{0.32\textwidth}
    \includegraphics[width=\textwidth]{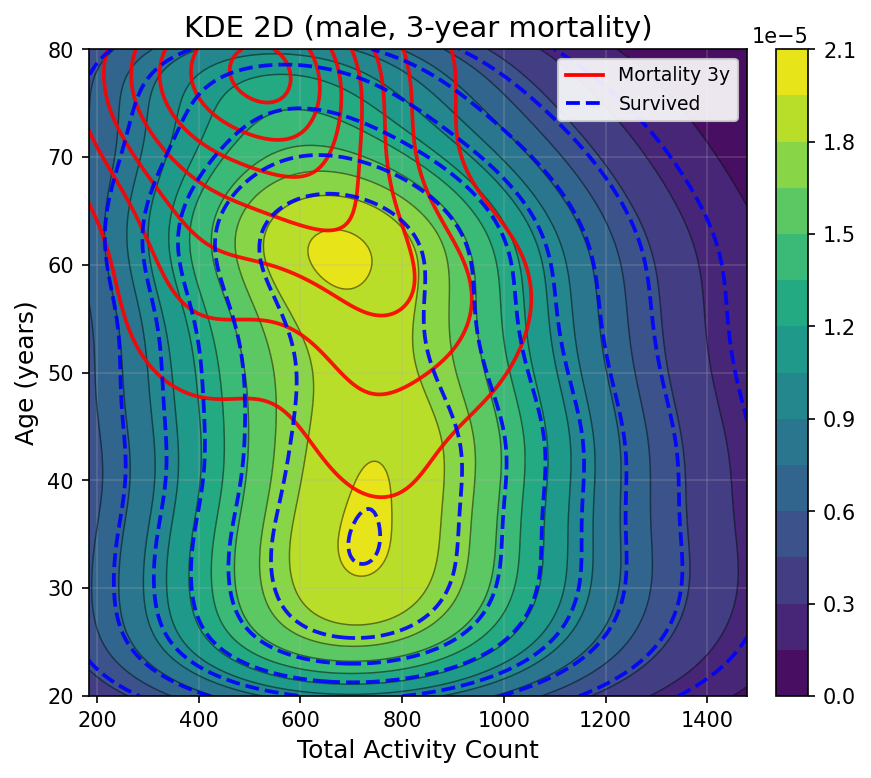}
    \caption{Males, 3-year mortality}
    \label{fig:tac2_age_m3}
\end{subfigure}
\hfill
\begin{subfigure}{0.32\textwidth}
    \includegraphics[width=\textwidth]{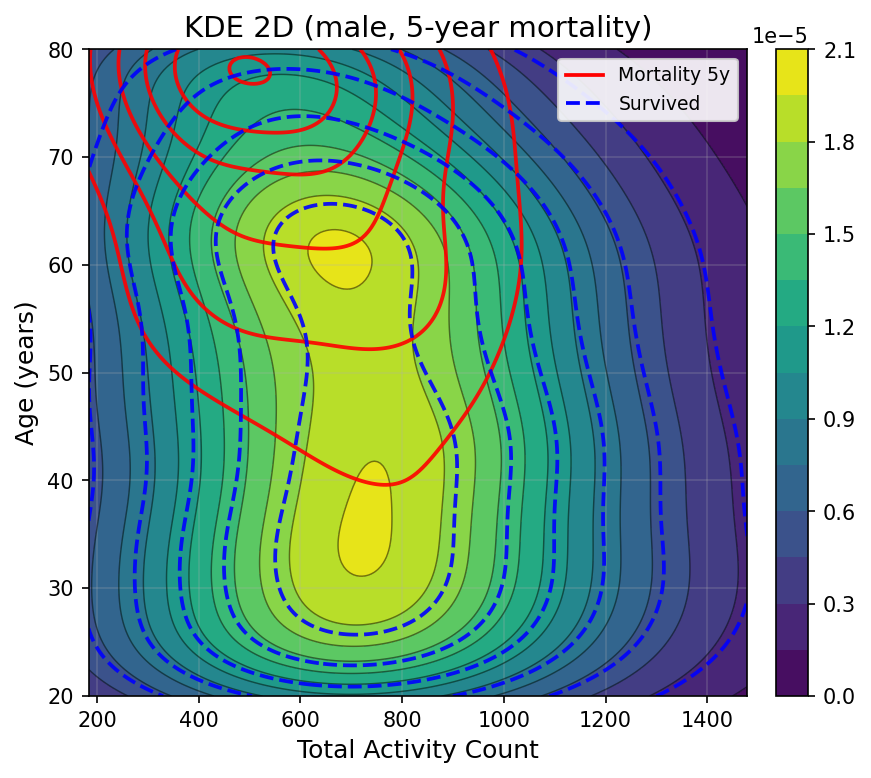}
    \caption{Males, 5-year mortality}
    \label{fig:tac2_age_m5}
\end{subfigure}
\hfill
\begin{subfigure}{0.32\textwidth}
    \includegraphics[width=\textwidth]{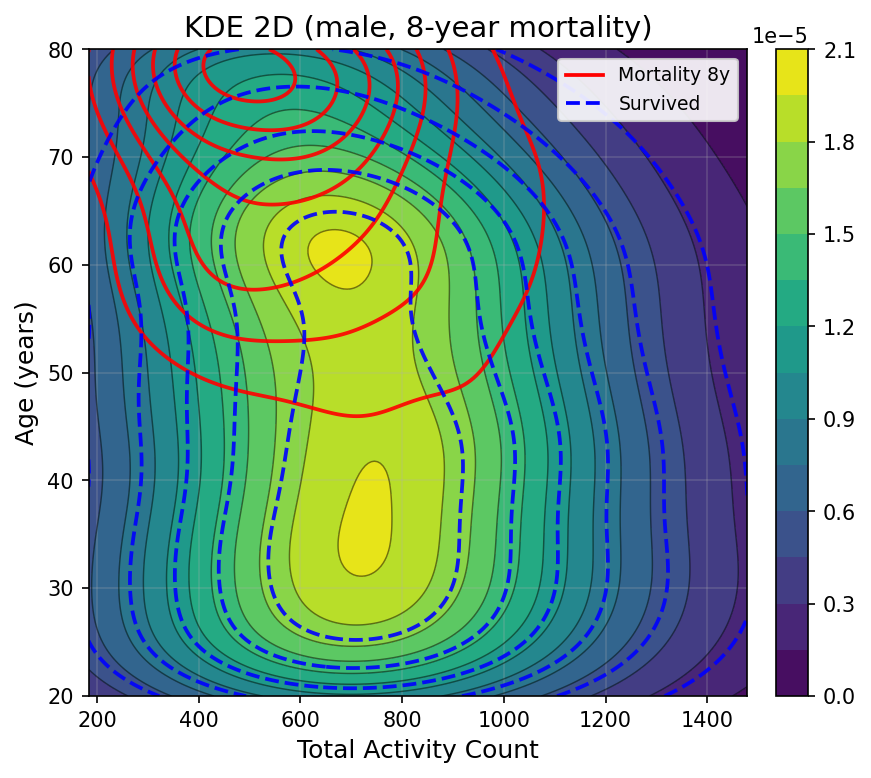}
    \caption{Males, 8-year mortality}
    \label{fig:tac2_age_m8}
\end{subfigure}

\vspace{0.5cm}

\begin{subfigure}{0.32\textwidth}
    \includegraphics[width=\textwidth]{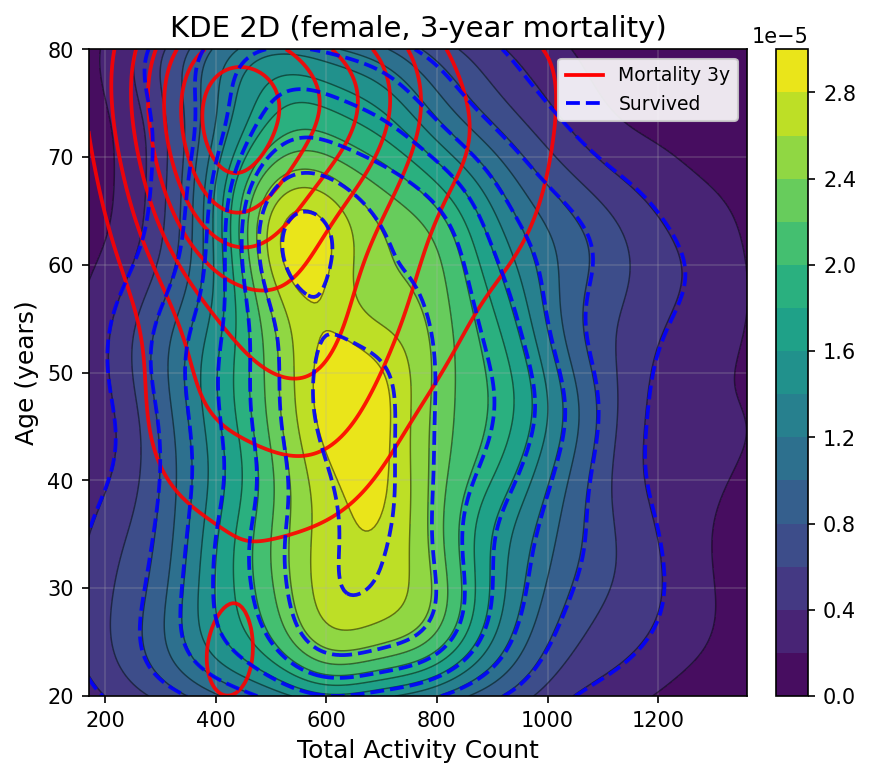}
    \caption{Females, 3-year mortality}
    \label{fig:tac2_age_f3}
\end{subfigure}
\hfill
\begin{subfigure}{0.32\textwidth}
    \includegraphics[width=\textwidth]{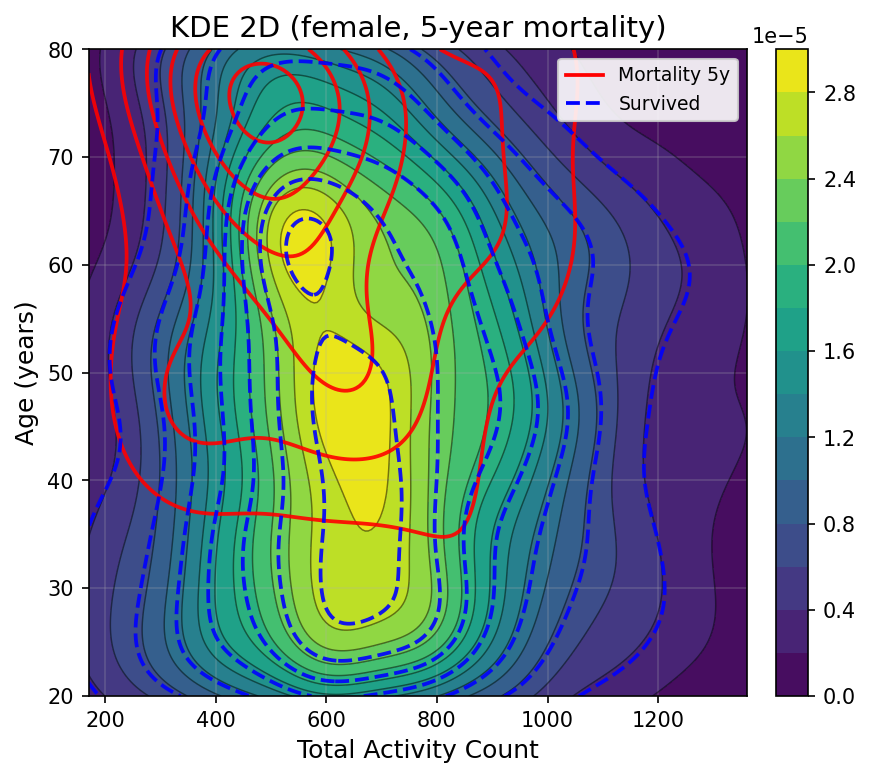}
    \caption{Females, 5-year mortality}
    \label{fig:tac2_age_f5}
\end{subfigure}
\hfill
\begin{subfigure}{0.32\textwidth}
    \includegraphics[width=\textwidth]{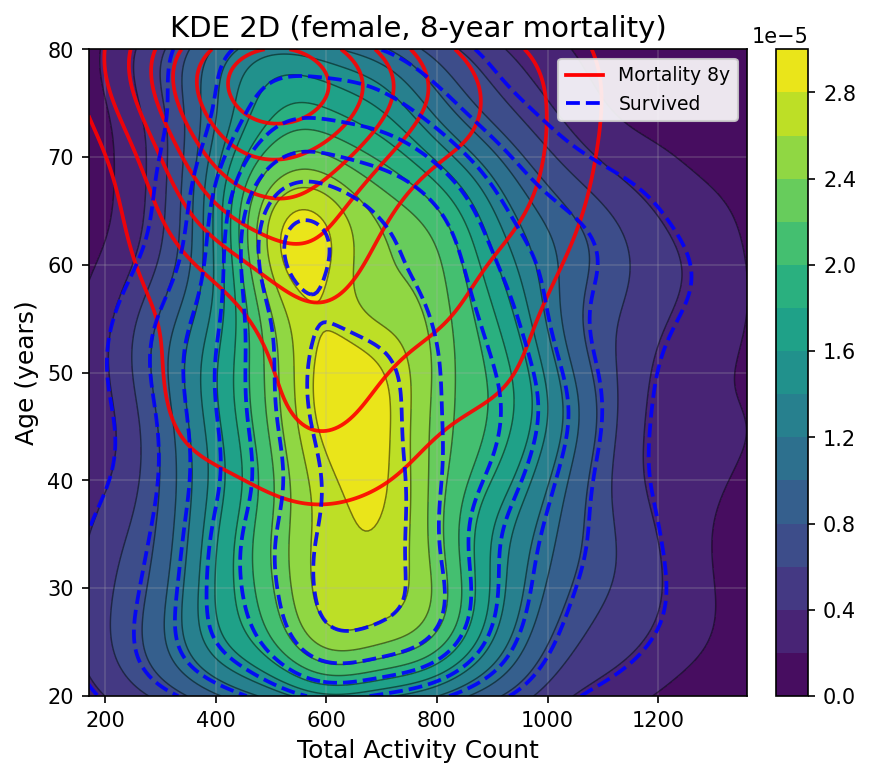}
    \caption{Females, 8-year mortality}
    \label{fig:tac2_age_f8}
\end{subfigure}

\caption{Bivariate kernel density estimates of total activity count versus age, stratified by sex and mortality follow-up period. Solid contours represent decedents and dashed contours represent survivors.}
\label{fig:tac2_age_all}
\end{figure}

\begin{figure}[H]
\centering
\begin{subfigure}{0.32\textwidth}
    \includegraphics[width=\textwidth]{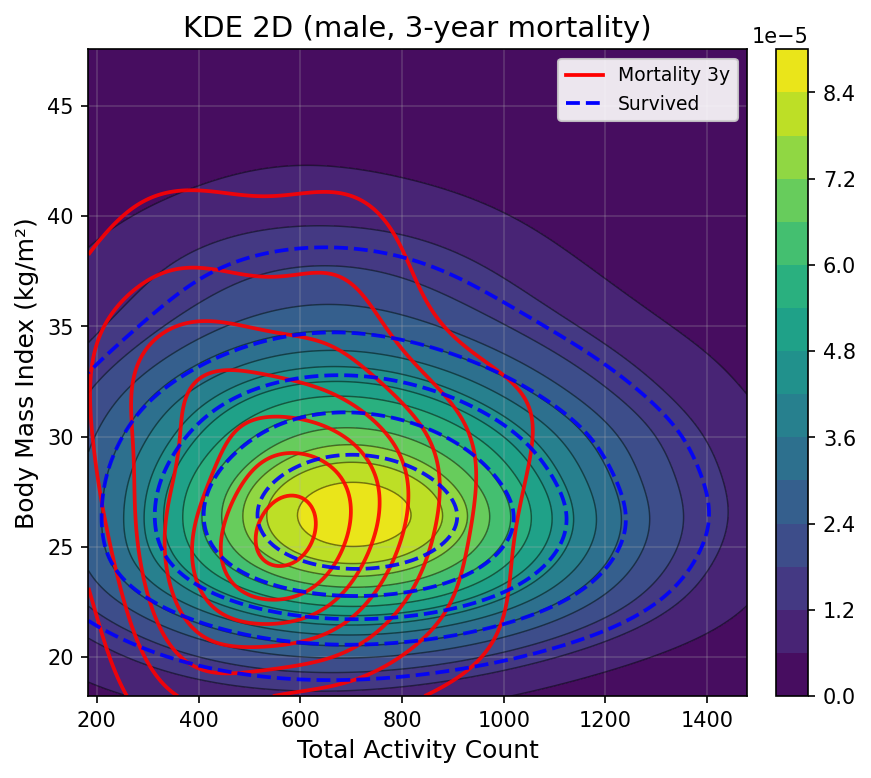}
    \caption{Males, 3-year mortality}
    \label{fig:tac2_bmi_m3}
\end{subfigure}
\hfill
\begin{subfigure}{0.32\textwidth}
    \includegraphics[width=\textwidth]{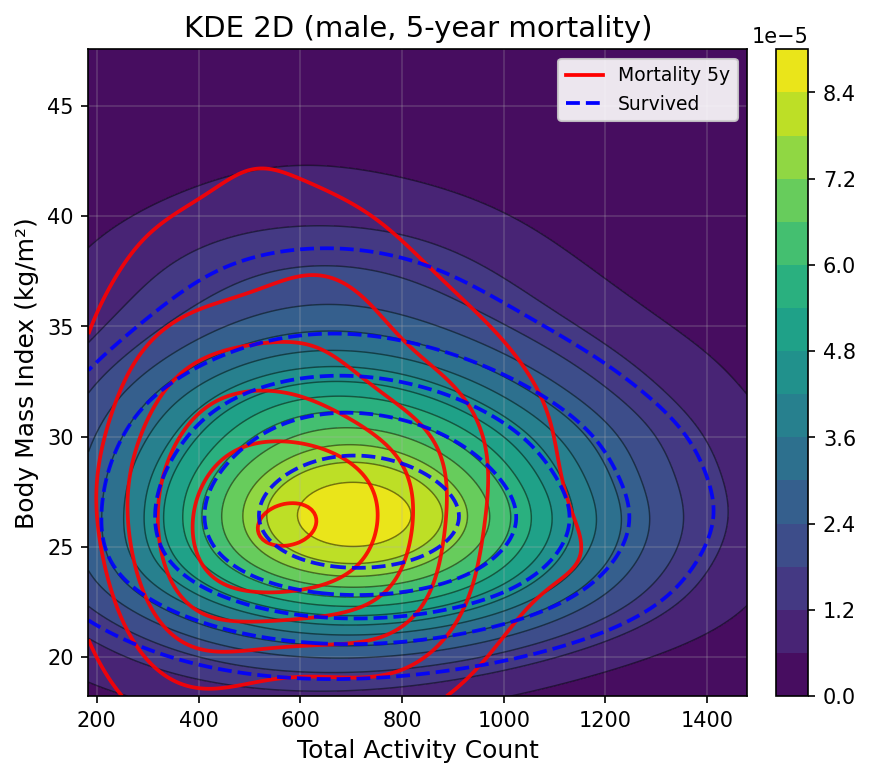}
    \caption{Males, 5-year mortality}
    \label{fig:tac2_bmi_m5}
\end{subfigure}
\hfill
\begin{subfigure}{0.32\textwidth}
    \includegraphics[width=\textwidth]{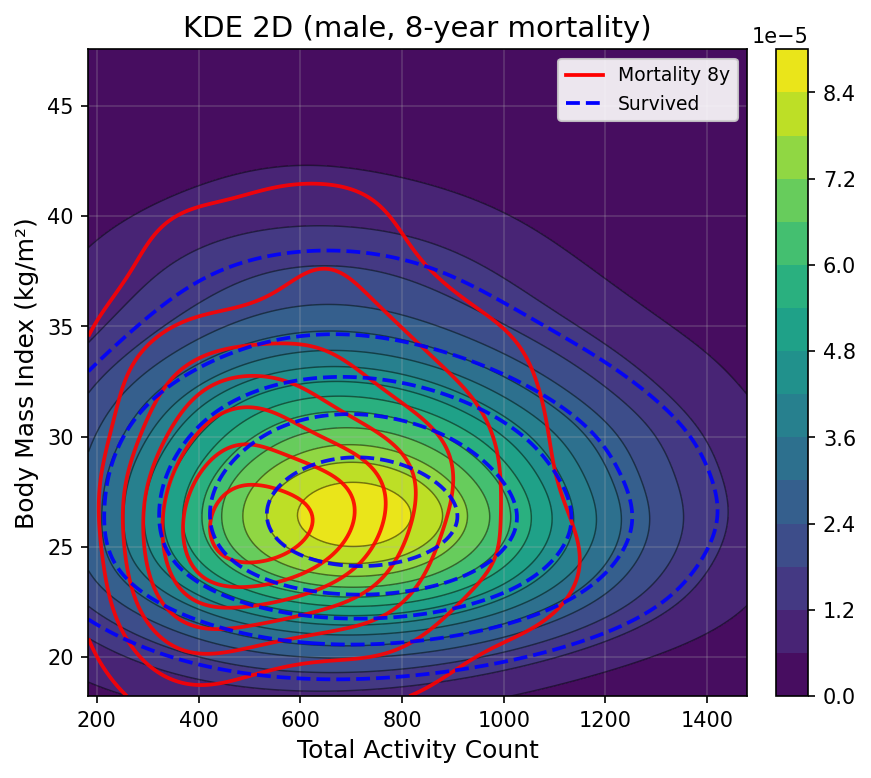}
    \caption{Males, 8-year mortality}
    \label{fig:tac2_bmi_m8}
\end{subfigure}

\vspace{0.5cm}

\begin{subfigure}{0.32\textwidth}
    \includegraphics[width=\textwidth]{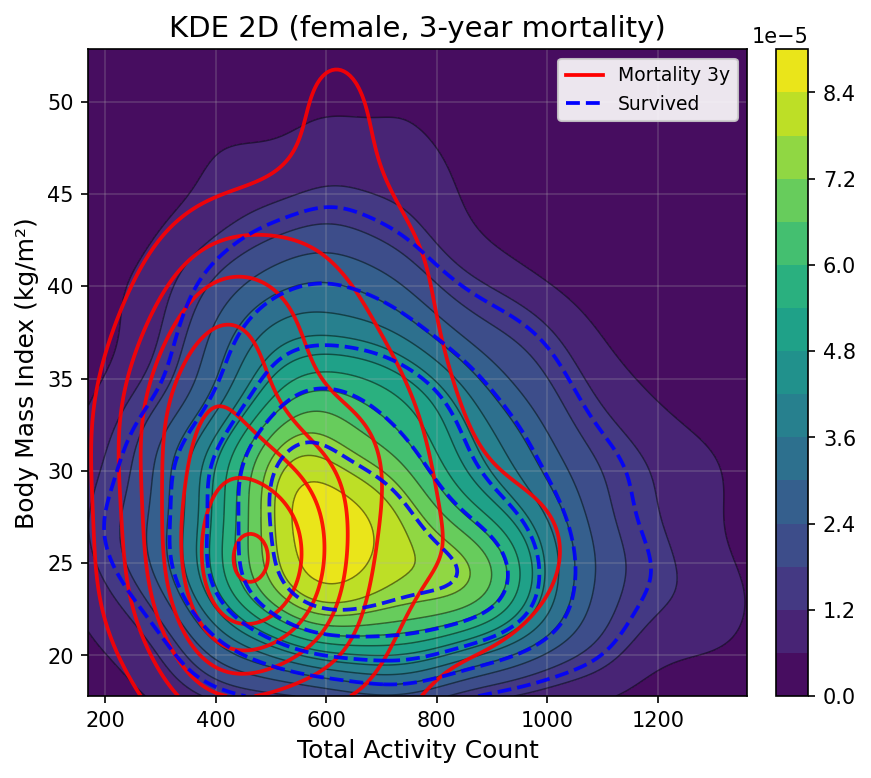}
    \caption{Females, 3-year mortality}
    \label{fig:tac2_bmi_f3}
\end{subfigure}
\hfill
\begin{subfigure}{0.32\textwidth}
    \includegraphics[width=\textwidth]{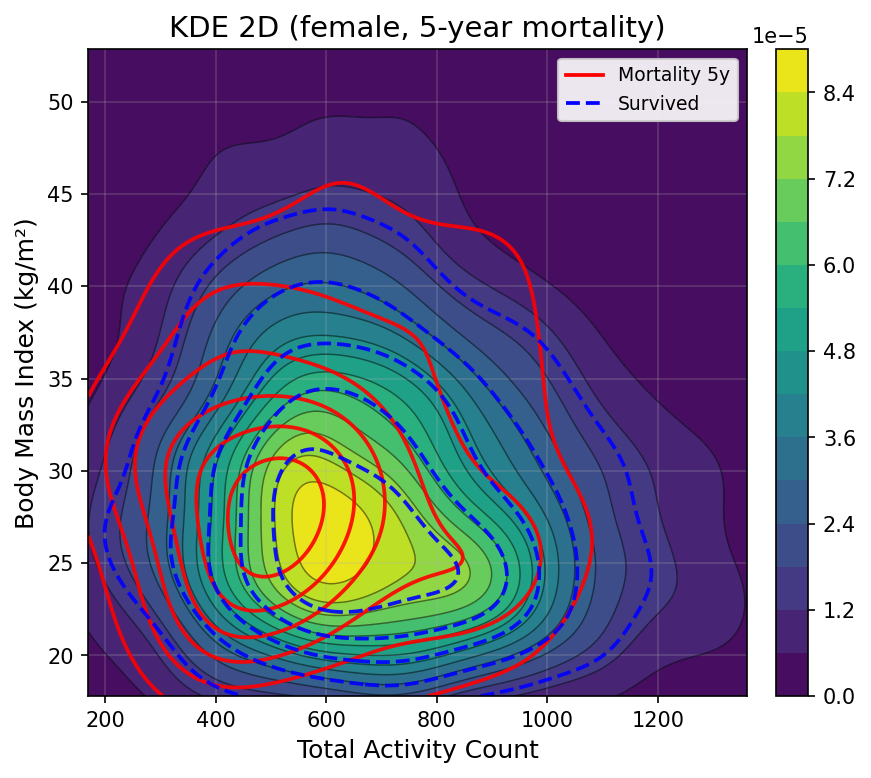}
    \caption{Females, 5-year mortality}
    \label{fig:tac2_bmi_f5}
\end{subfigure}
\hfill
\begin{subfigure}{0.32\textwidth}
    \includegraphics[width=\textwidth]{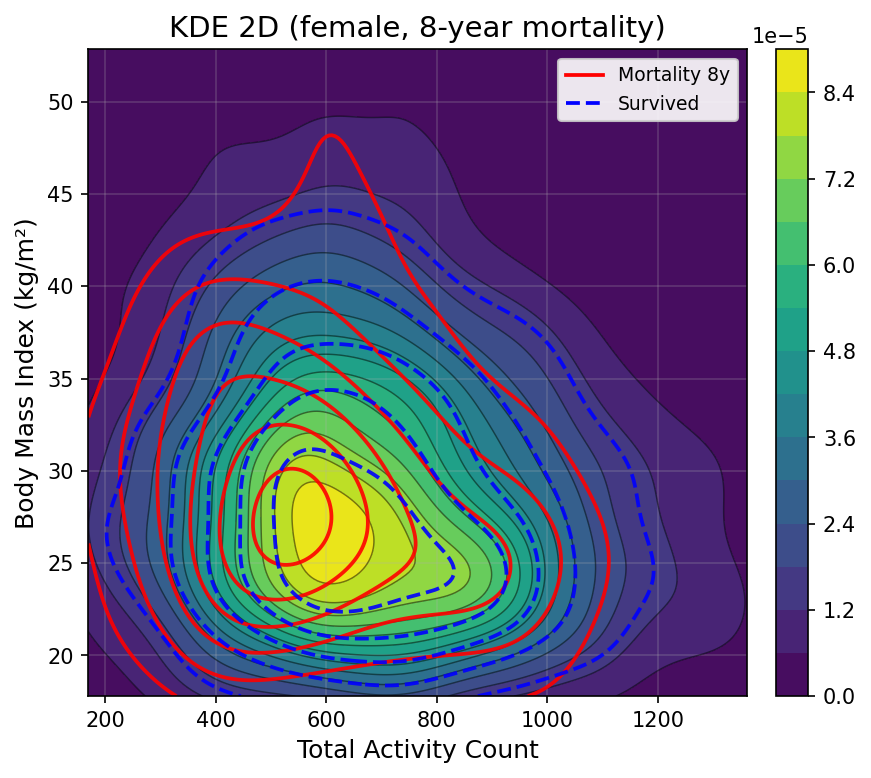}
    \caption{Females, 8-year mortality}
    \label{fig:tac2_bmi_f8}
\end{subfigure}

\caption{Bivariate kernel density estimates of total activity count versus body mass index (BMI), stratified by sex and mortality follow-up period. Solid contours represent decedents and dashed contours represent survivors.}
\label{fig:tac2_bmi_all}
\end{figure}

\begin{figure}[H]
\centering
\begin{subfigure}{0.32\textwidth}
    \includegraphics[width=\textwidth]{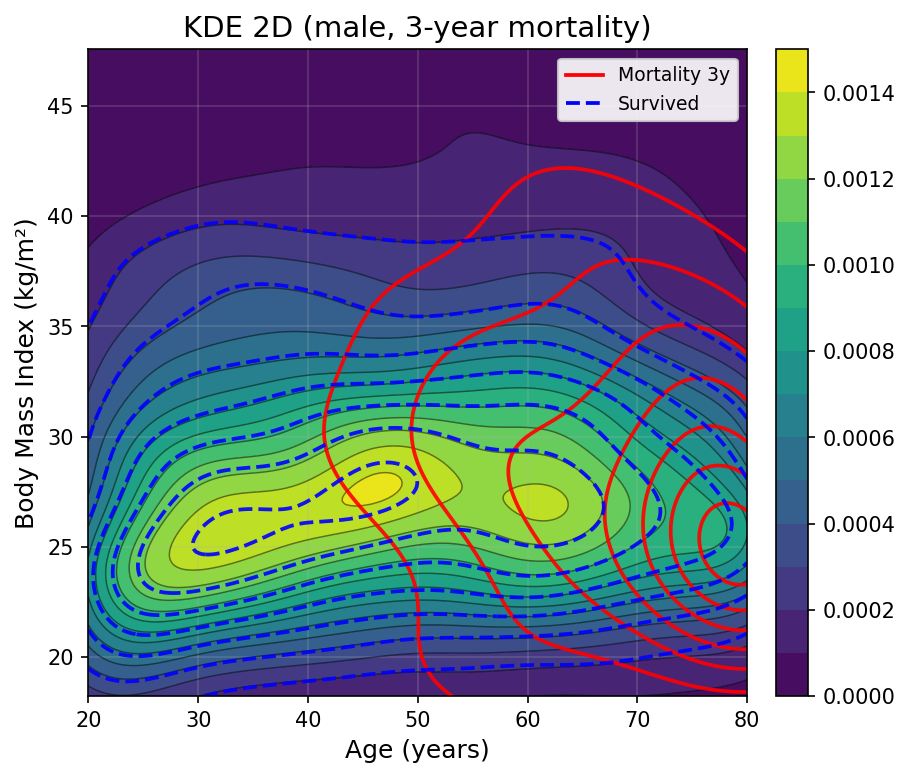}
    \caption{Males, 3-year mortality}
    \label{fig:age_bmi_m3}
\end{subfigure}
\hfill
\begin{subfigure}{0.32\textwidth}
    \includegraphics[width=\textwidth]{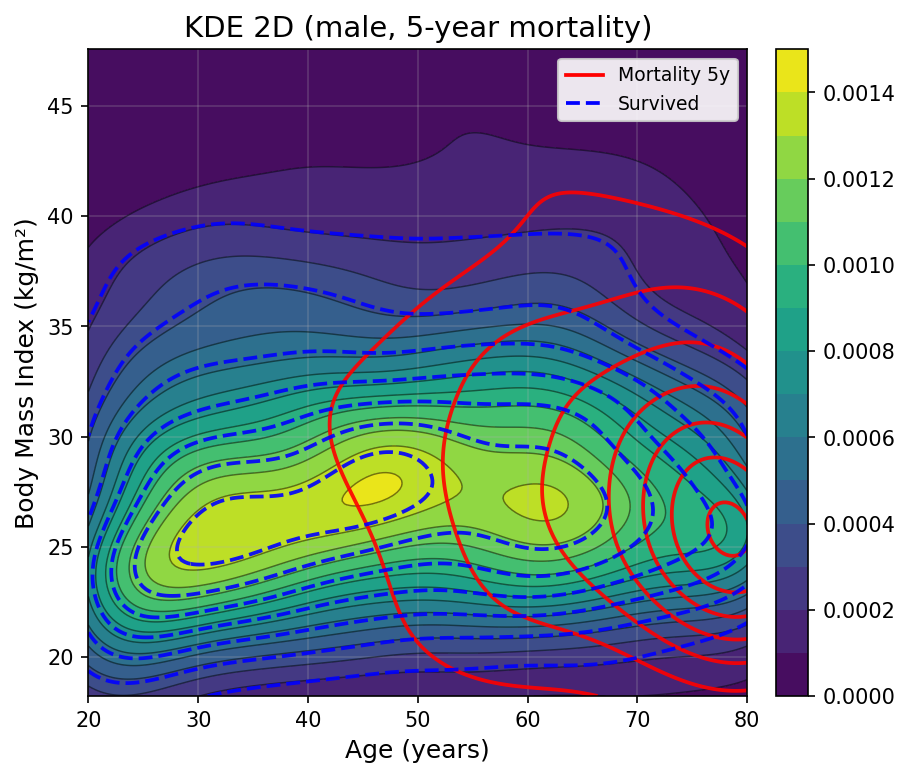}
    \caption{Males, 5-year mortality}
    \label{fig:age_bmi_m5}
\end{subfigure}
\hfill
\begin{subfigure}{0.32\textwidth}
    \includegraphics[width=\textwidth]{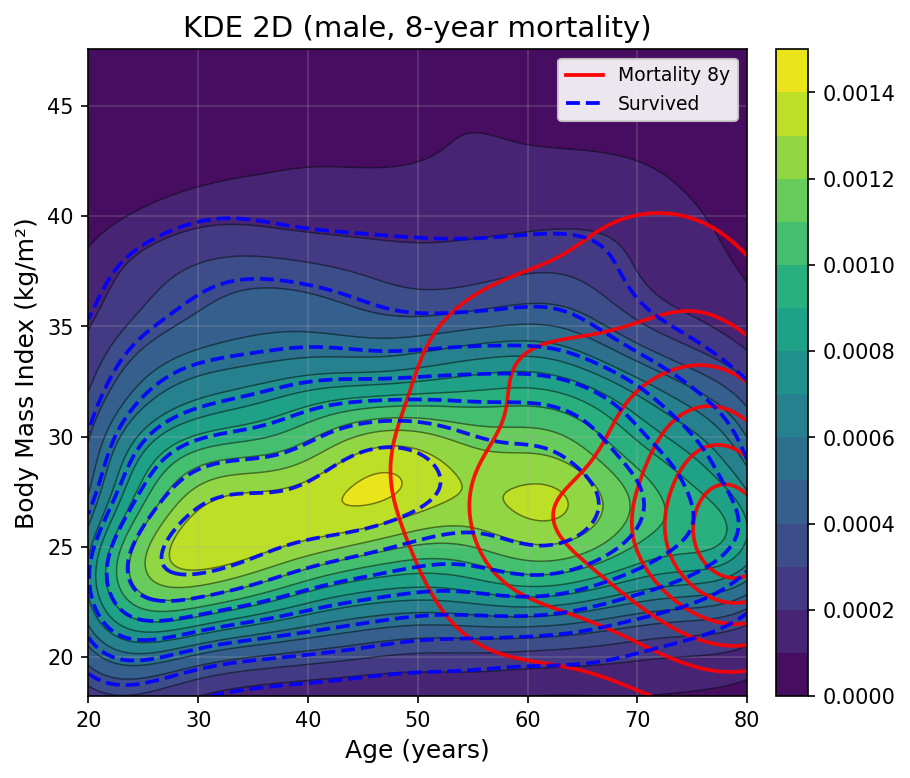}
    \caption{Males, 8-year mortality}
    \label{fig:age_bmi_m8}
\end{subfigure}

\vspace{0.5cm}

\begin{subfigure}{0.32\textwidth}
    \includegraphics[width=\textwidth]{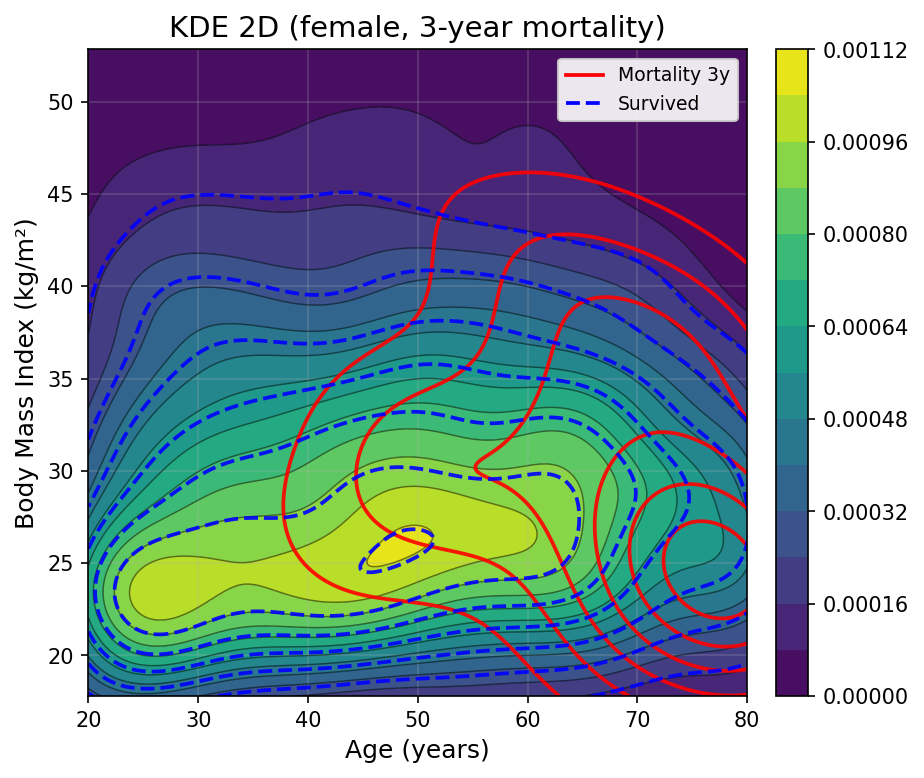}
    \caption{Females, 3-year mortality}
    \label{fig:age_bmi_f3}
\end{subfigure}
\hfill
\begin{subfigure}{0.32\textwidth}
    \includegraphics[width=\textwidth]{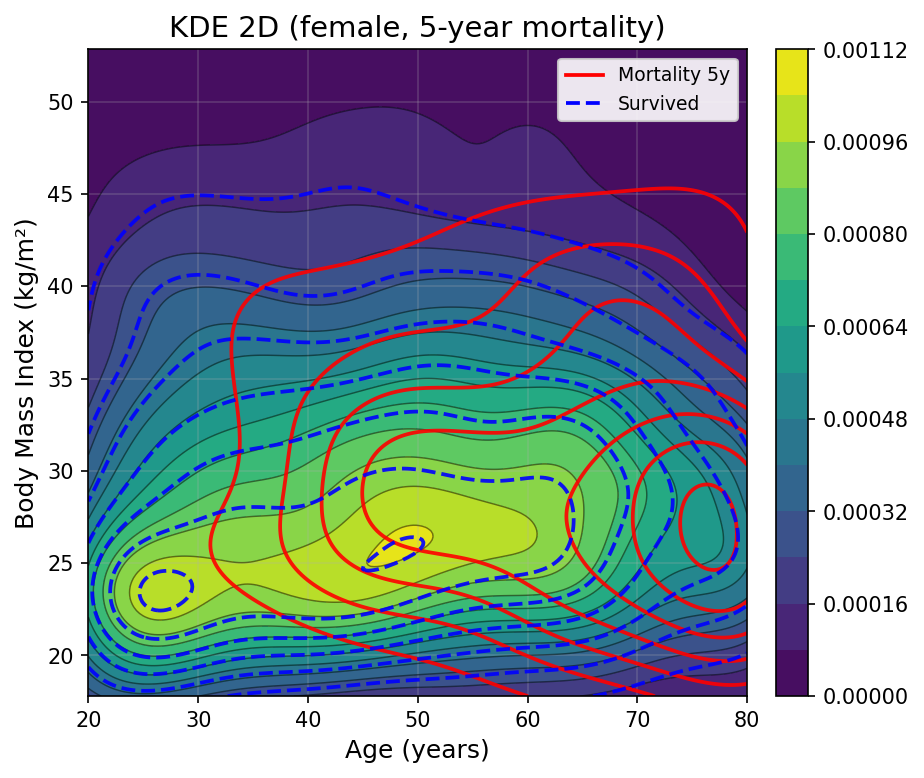}
    \caption{Females, 5-year mortality}
    \label{fig:age_bmi_f5}
\end{subfigure}
\hfill
\begin{subfigure}{0.32\textwidth}
    \includegraphics[width=\textwidth]{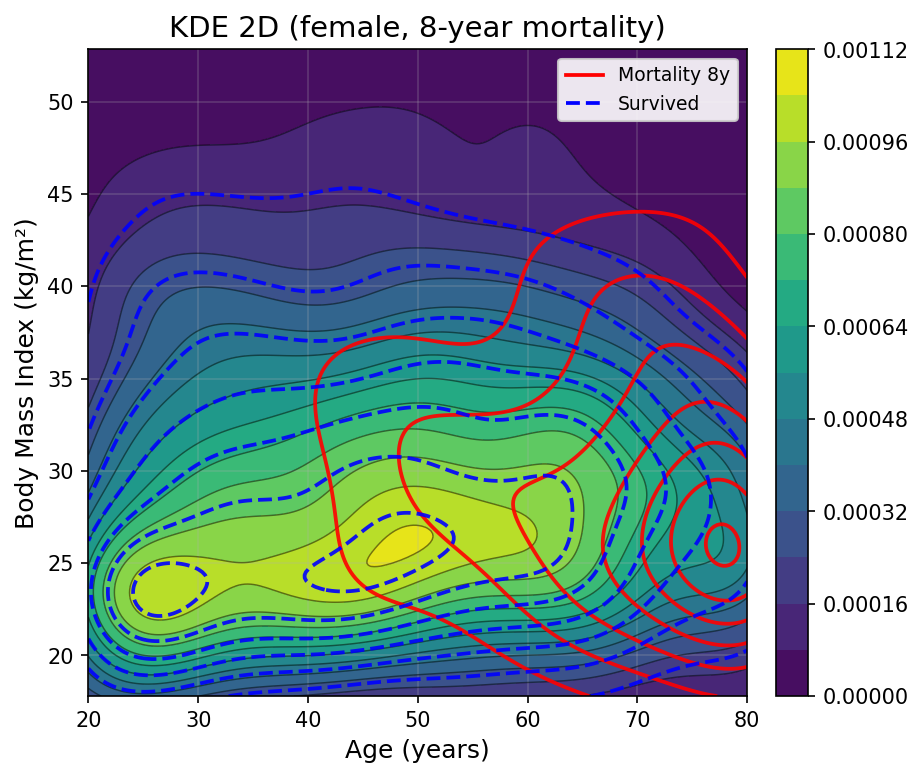}
    \caption{Females, 8-year mortality}
    \label{fig:age_bmi_f8}
\end{subfigure}

\caption{Bivariate kernel density estimates of age versus body mass index (BMI), stratified by sex and mortality follow-up period. Solid contours represent decedents and dashed contours represent survivors.}
\label{fig:age_bmi_all}
\end{figure}

\subsection{ROC Curves}

\begin{figure}[H]
    \centering
    \begin{subfigure}{0.3\textwidth}
        \includegraphics[width=\linewidth]{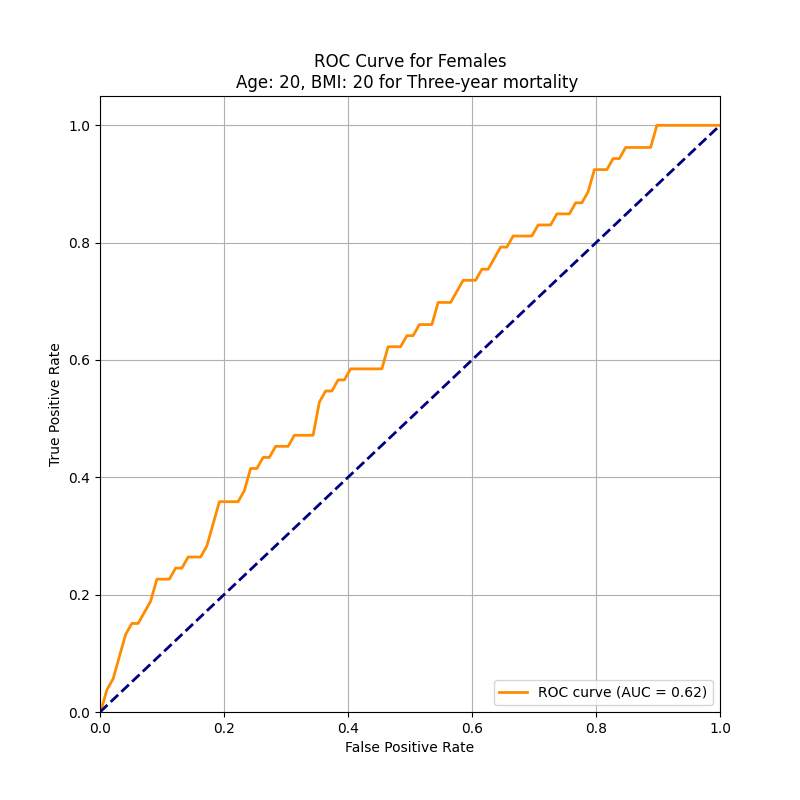}
        \caption{Age 20, BMI 20}
    \end{subfigure}
    \begin{subfigure}{0.3\textwidth}
        \includegraphics[width=\linewidth]{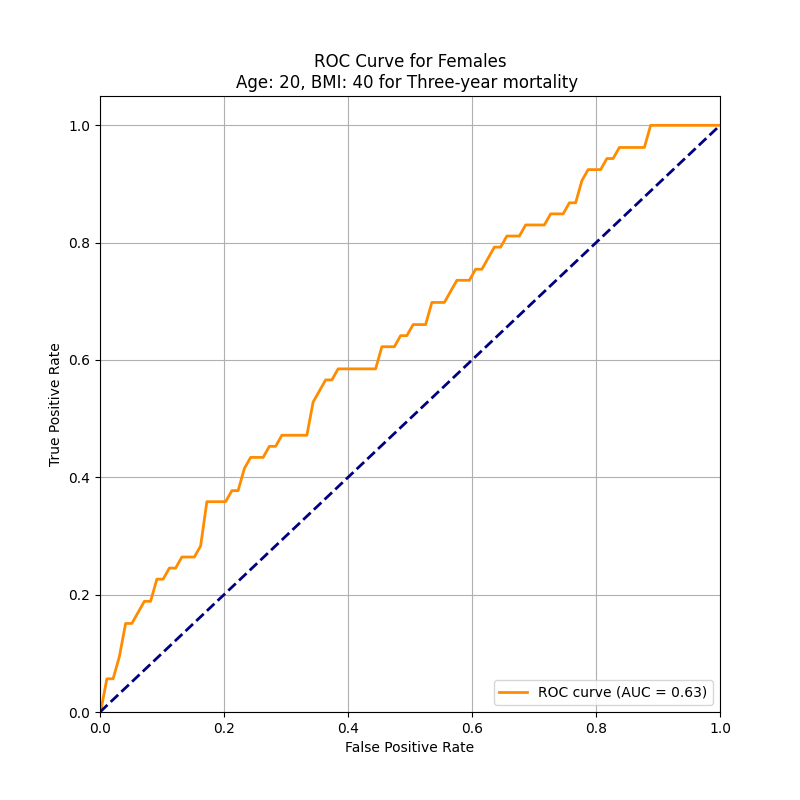}
        \caption{Age 20, BMI 40}
    \end{subfigure}
    \begin{subfigure}{0.3\textwidth}
        \includegraphics[width=\linewidth]{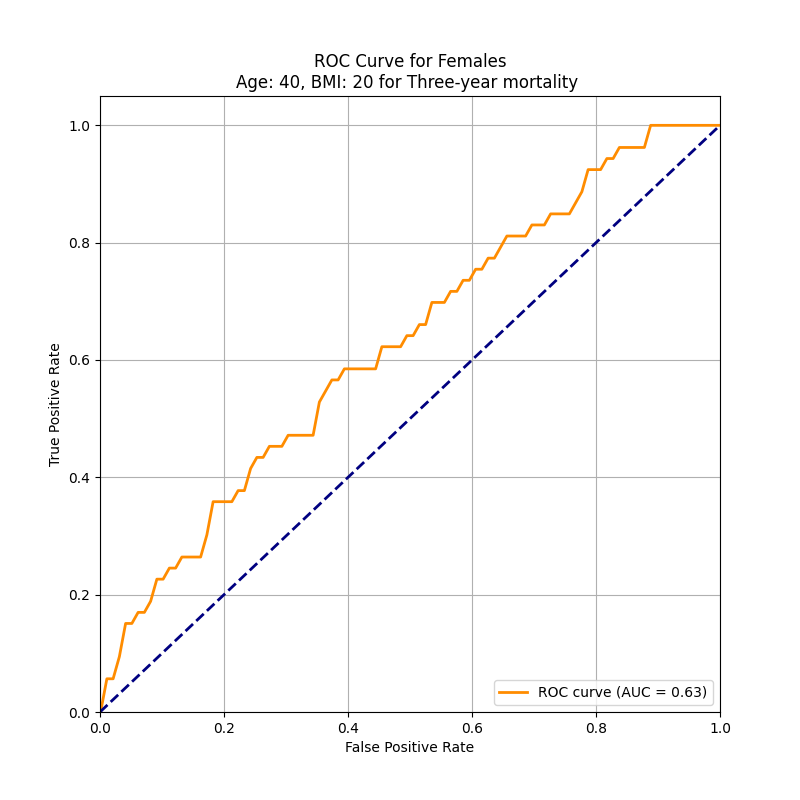}
        \caption{Age 40, BMI 20}
    \end{subfigure}

    \begin{subfigure}{0.3\textwidth}
        \includegraphics[width=\linewidth]{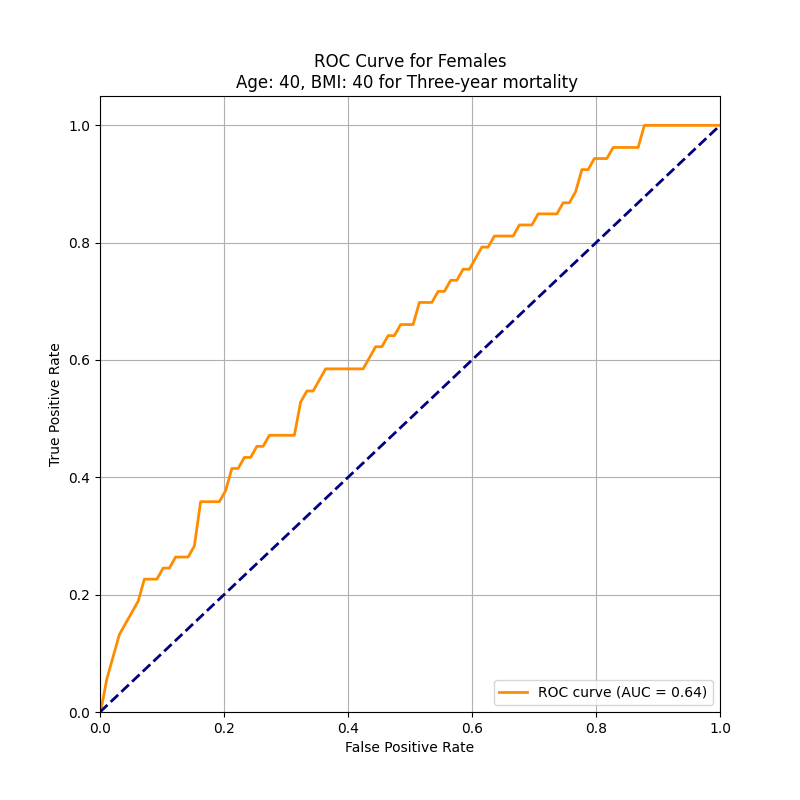}
        \caption{Age 40, BMI 40}
    \end{subfigure}
    \begin{subfigure}{0.3\textwidth}
        \includegraphics[width=\linewidth]{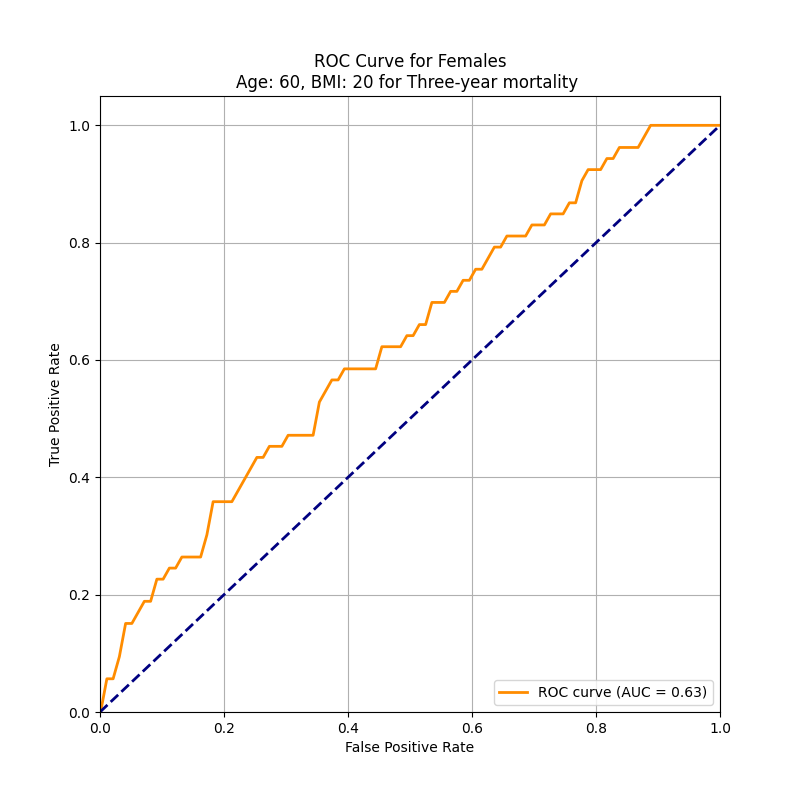}
        \caption{Age 60, BMI 20}
    \end{subfigure}
    \begin{subfigure}{0.3\textwidth}
        \includegraphics[width=\linewidth]{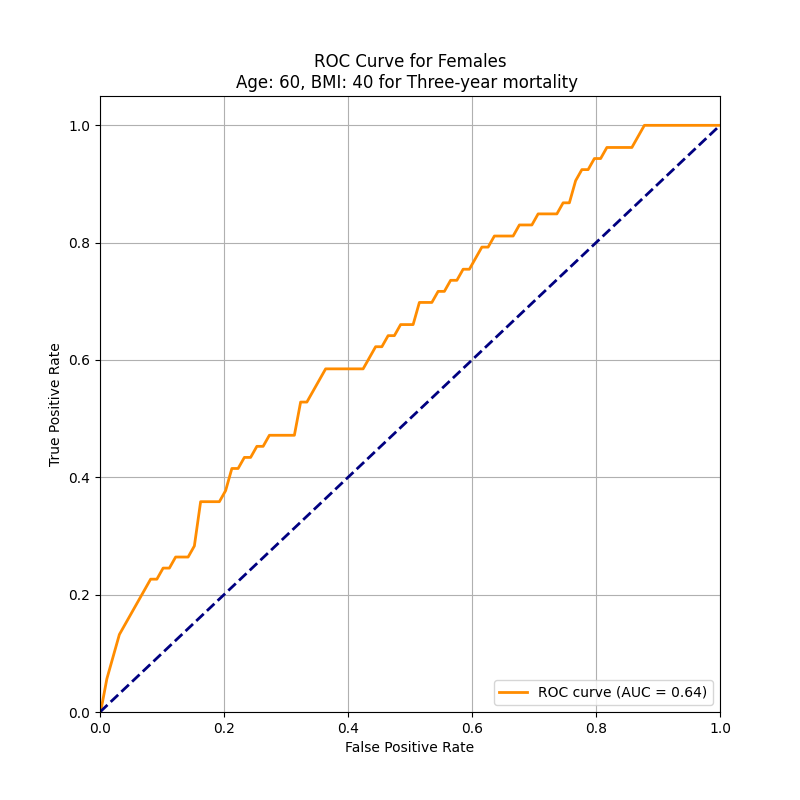}
        \caption{Age 60, BMI 40}
    \end{subfigure}
    \caption{ROC Curves for Females (Three-year mortality)}
    \label{fig:females_3year}
\end{figure}

\begin{figure}[H]
    \centering
    \begin{subfigure}{0.3\textwidth}
        \includegraphics[width=\linewidth]{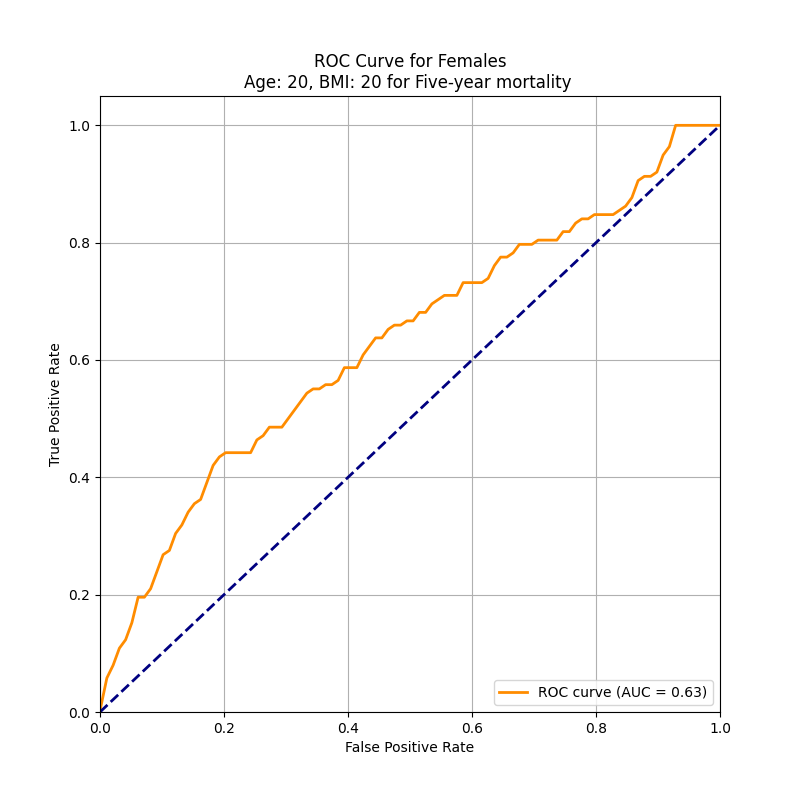}
        \caption{Age 20, BMI 20}
    \end{subfigure}
    \begin{subfigure}{0.3\textwidth}
        \includegraphics[width=\linewidth]{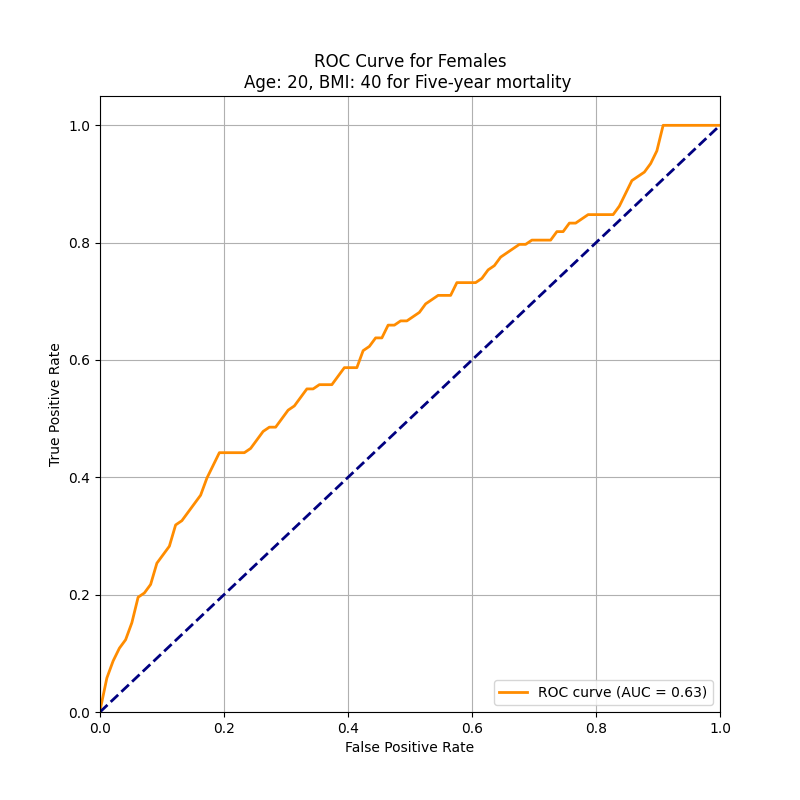}
        \caption{Age 20, BMI 40}
    \end{subfigure}
    \begin{subfigure}{0.3\textwidth}
        \includegraphics[width=\linewidth]{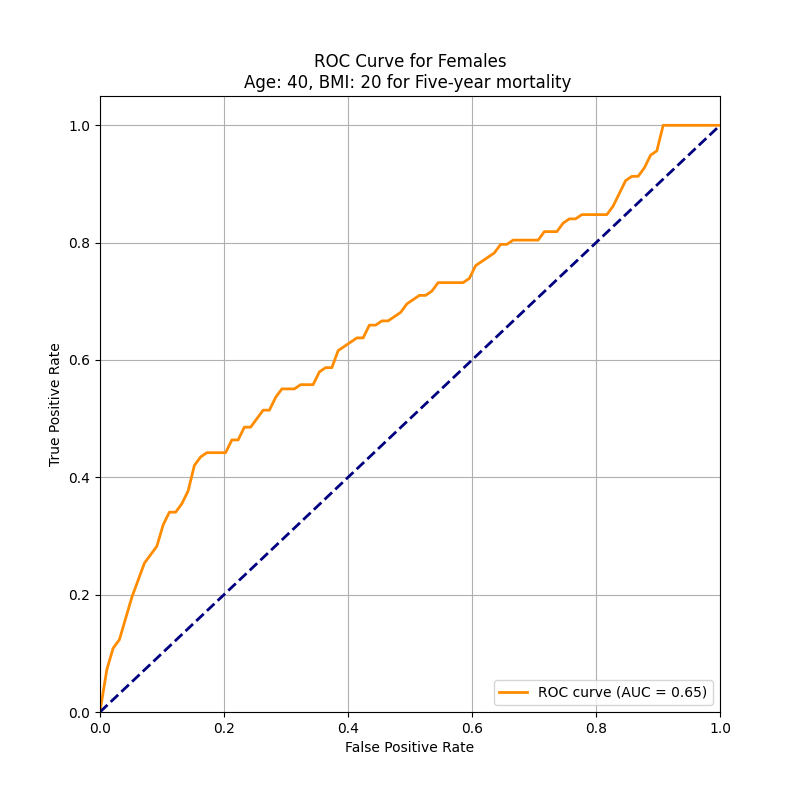}
        \caption{Age 40, BMI 20}
    \end{subfigure}

    \begin{subfigure}{0.3\textwidth}
        \includegraphics[width=\linewidth]{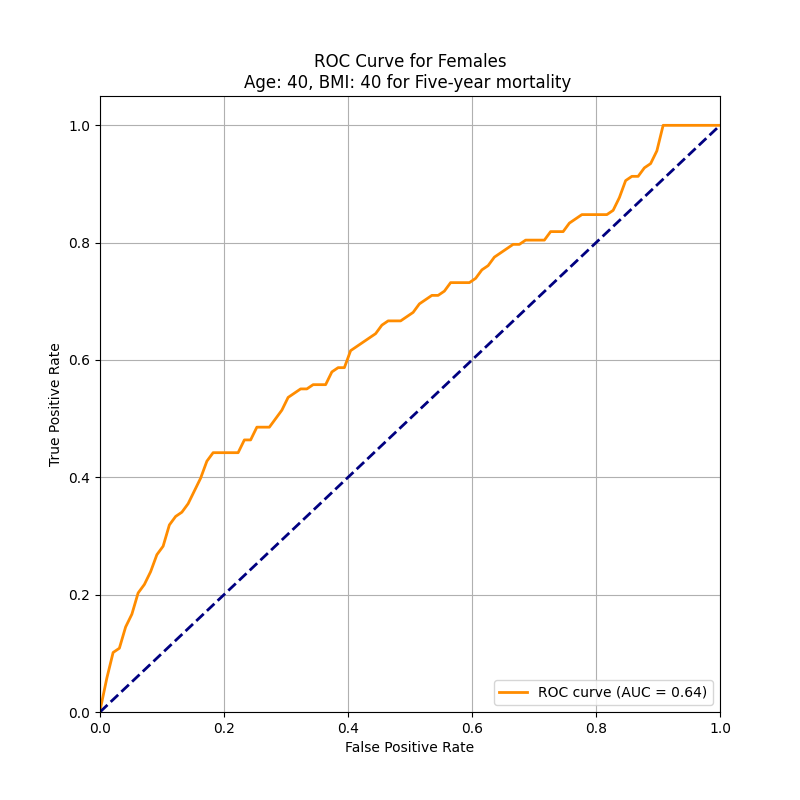}
        \caption{Age 40, BMI 40}
    \end{subfigure}
    \begin{subfigure}{0.3\textwidth}
        \includegraphics[width=\linewidth]{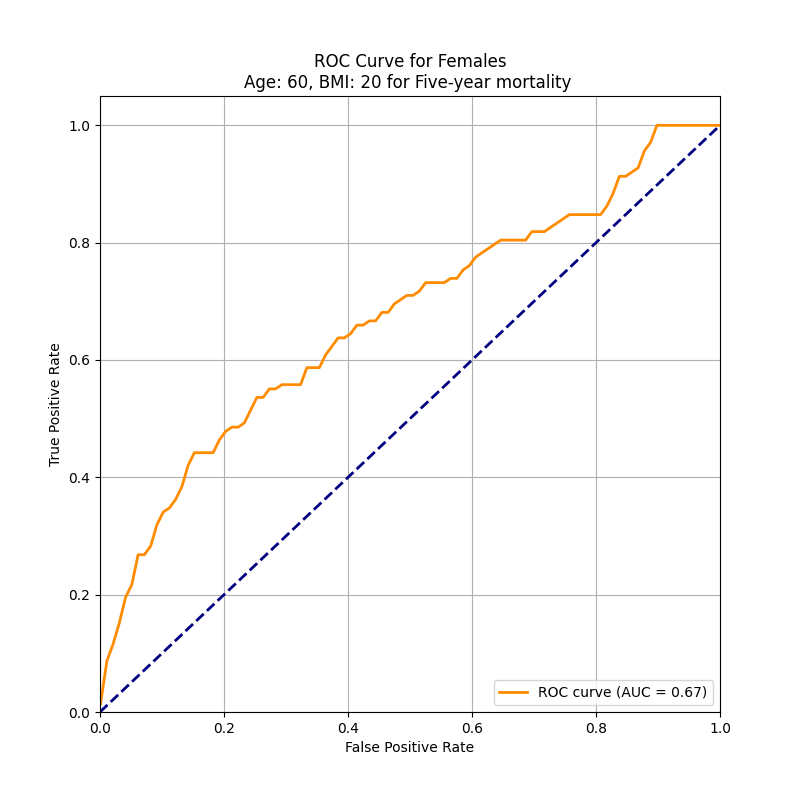}
        \caption{Age 60, BMI 20}
    \end{subfigure}
    \begin{subfigure}{0.3\textwidth}
        \includegraphics[width=\linewidth]{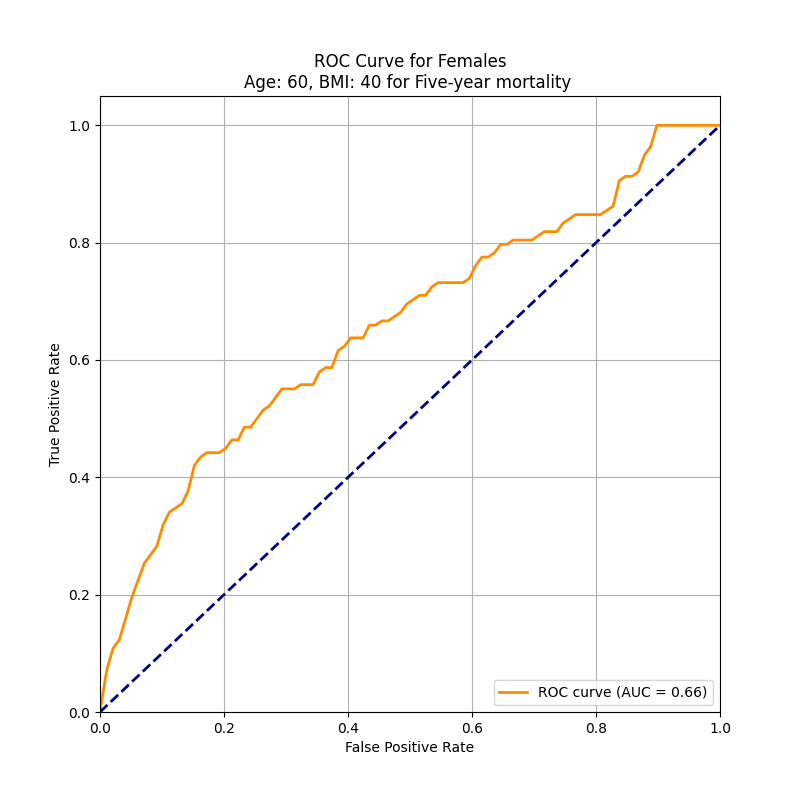}
        \caption{Age 60, BMI 40}
    \end{subfigure}
    \caption{ROC Curves for Females (Five-year mortality)}
     \label{fig:females_5year}
\end{figure}


\begin{figure}[H]
    \centering
    \begin{subfigure}{0.3\textwidth}
        \includegraphics[width=\linewidth]{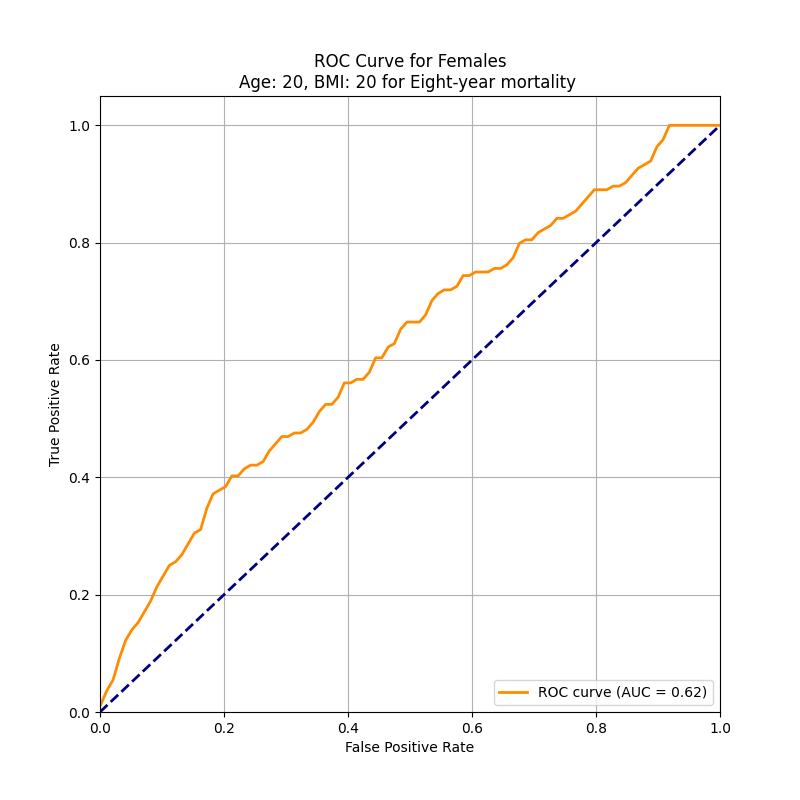}
        \caption{Age 20, BMI 20}
    \end{subfigure}
    \begin{subfigure}{0.3\textwidth}
        \includegraphics[width=\linewidth]{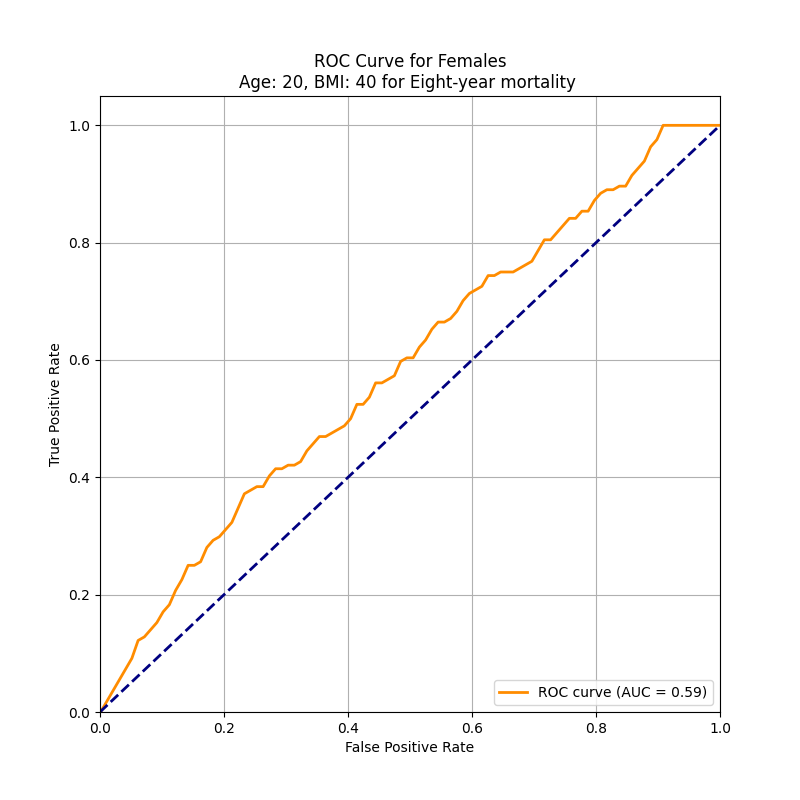}
        \caption{Age 20, BMI 40}
    \end{subfigure}
    \begin{subfigure}{0.3\textwidth}
        \includegraphics[width=\linewidth]{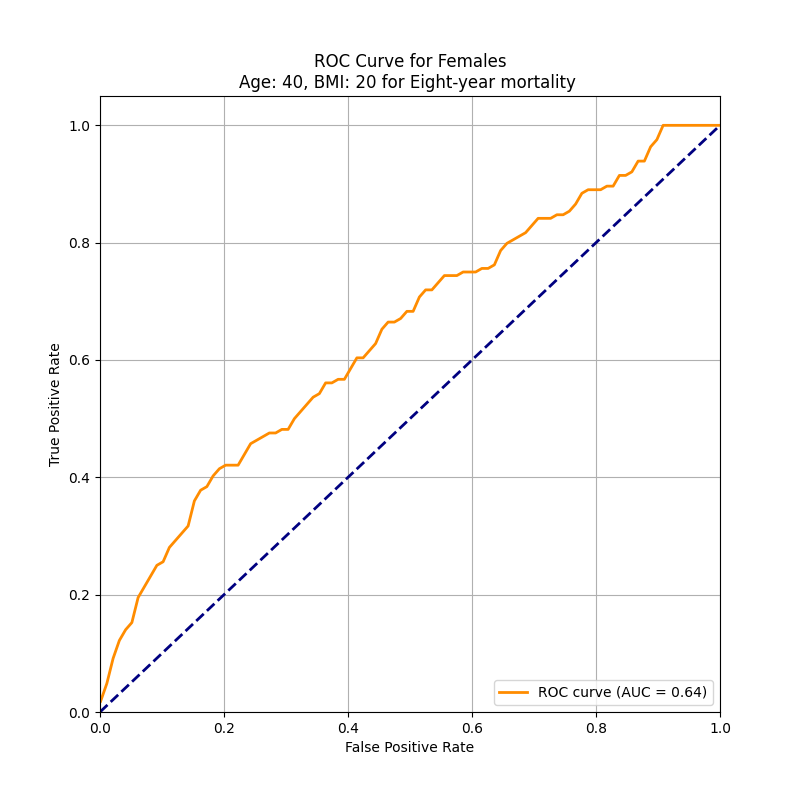}
        \caption{Age 40, BMI 20}
    \end{subfigure}

    \begin{subfigure}{0.3\textwidth}
        \includegraphics[width=\linewidth]{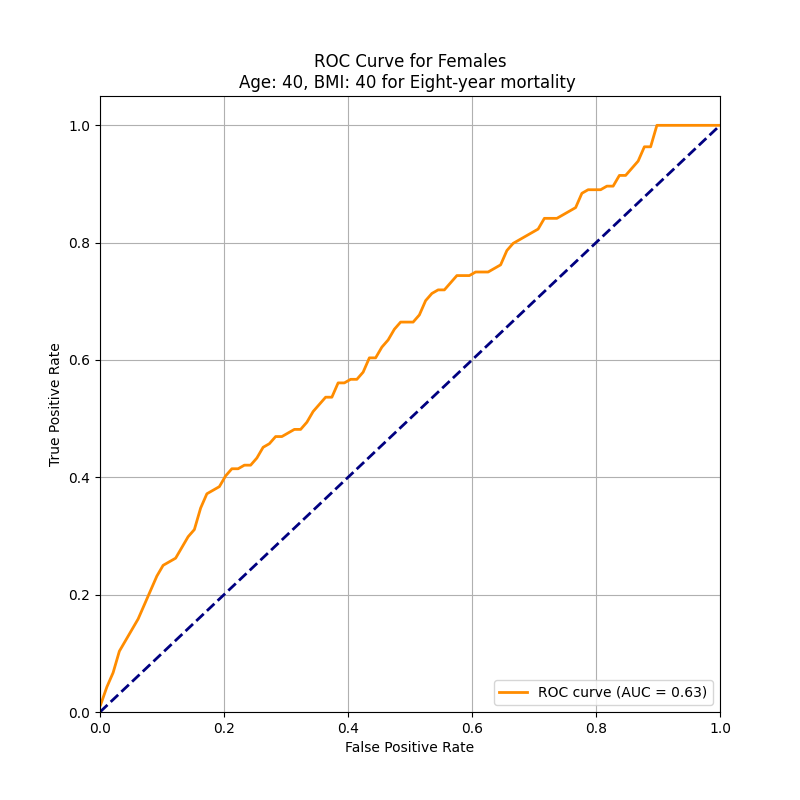}
        \caption{Age 40, BMI 40}
    \end{subfigure}
    \begin{subfigure}{0.3\textwidth}
        \includegraphics[width=\linewidth]{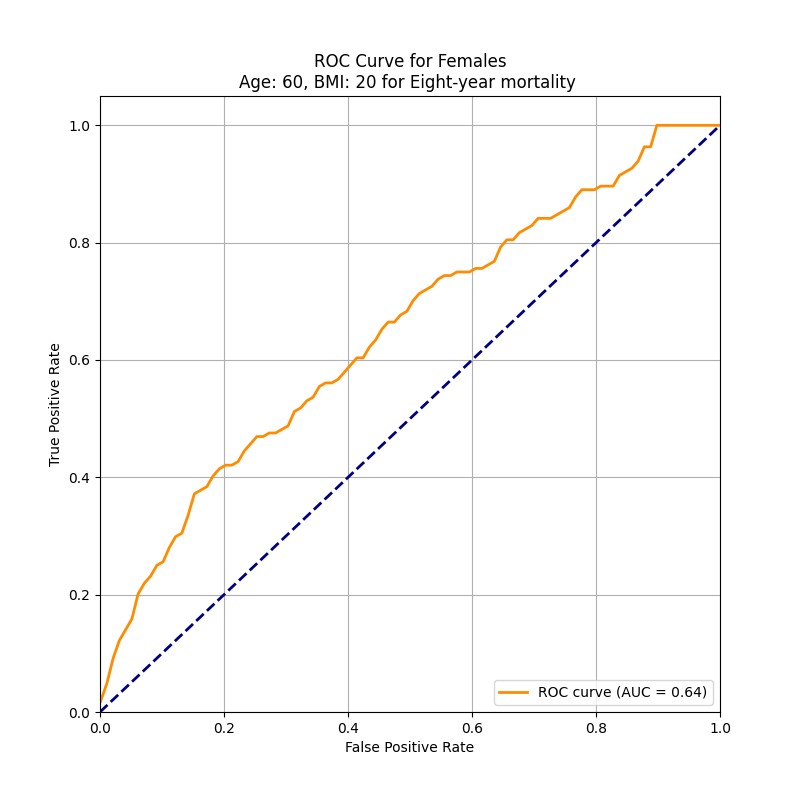}
        \caption{Age 60, BMI 20}
    \end{subfigure}
    \begin{subfigure}{0.3\textwidth}
        \includegraphics[width=\linewidth]{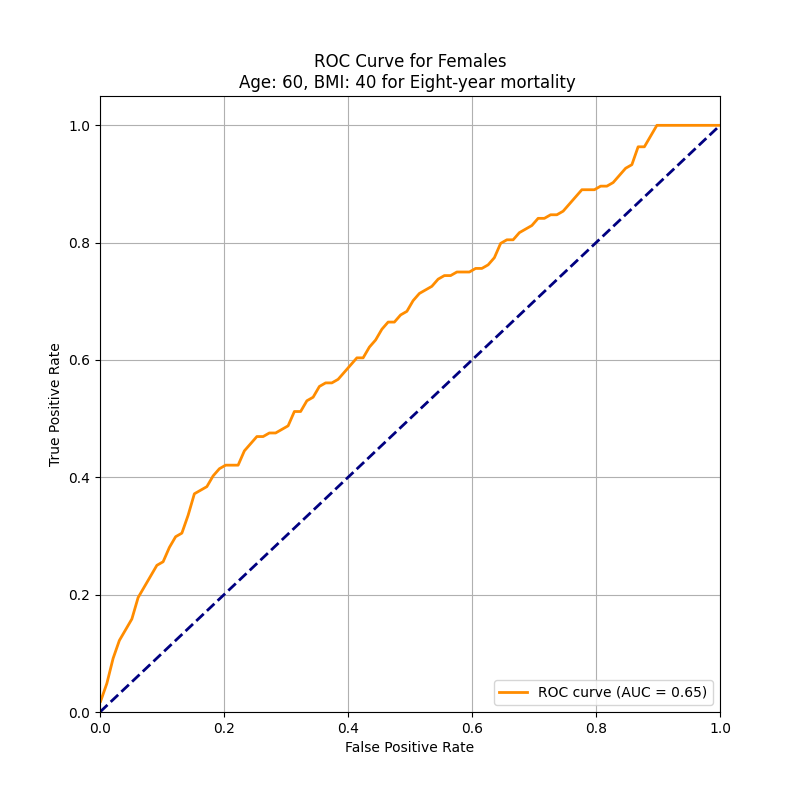}
        \caption{Age 60, BMI 40}
    \end{subfigure}
    \caption{ROC Curves for Females (Eight-year mortality)}
     \label{fig:females_8year}
\end{figure}


\begin{figure}[H]
    \centering
    \begin{subfigure}{0.3\textwidth}
        \includegraphics[width=\linewidth]{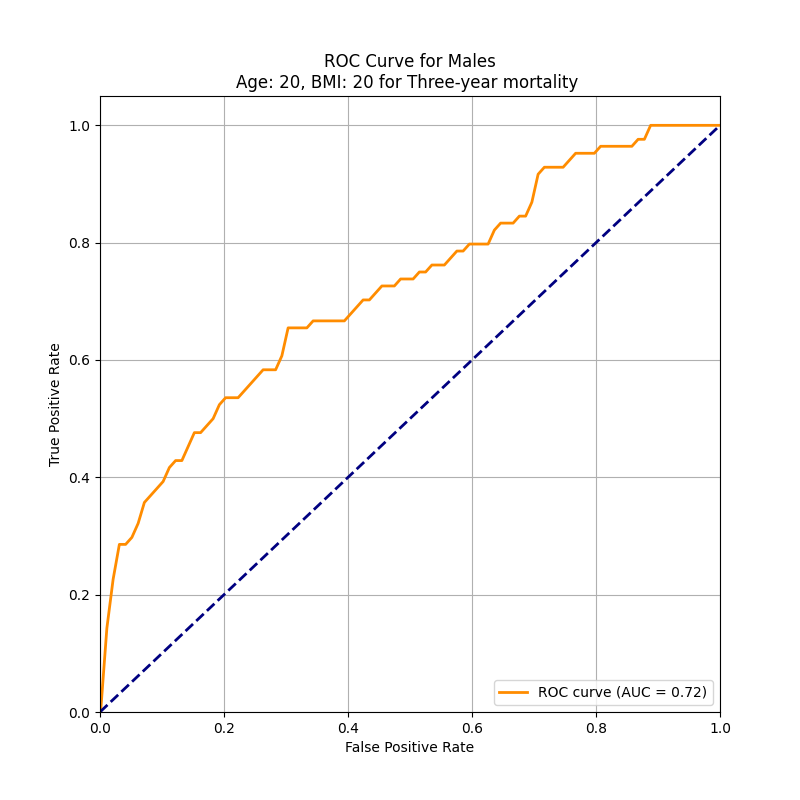}
        \caption{Age 20, BMI 20}
    \end{subfigure}
    \begin{subfigure}{0.3\textwidth}
        \includegraphics[width=\linewidth]{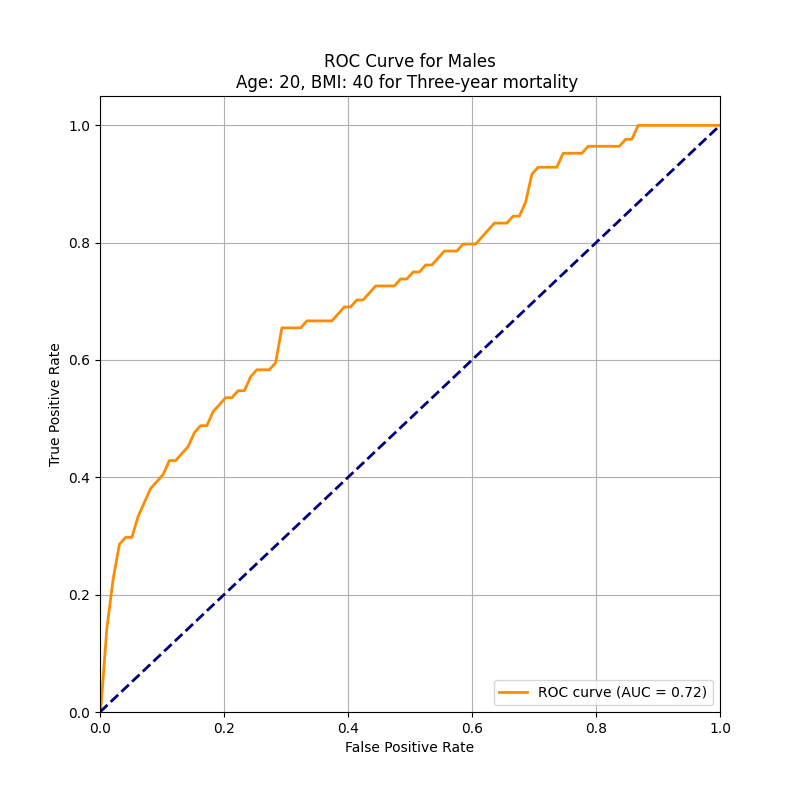}
        \caption{Age 20, BMI 40}
    \end{subfigure}
    \begin{subfigure}{0.3\textwidth}
        \includegraphics[width=\linewidth]{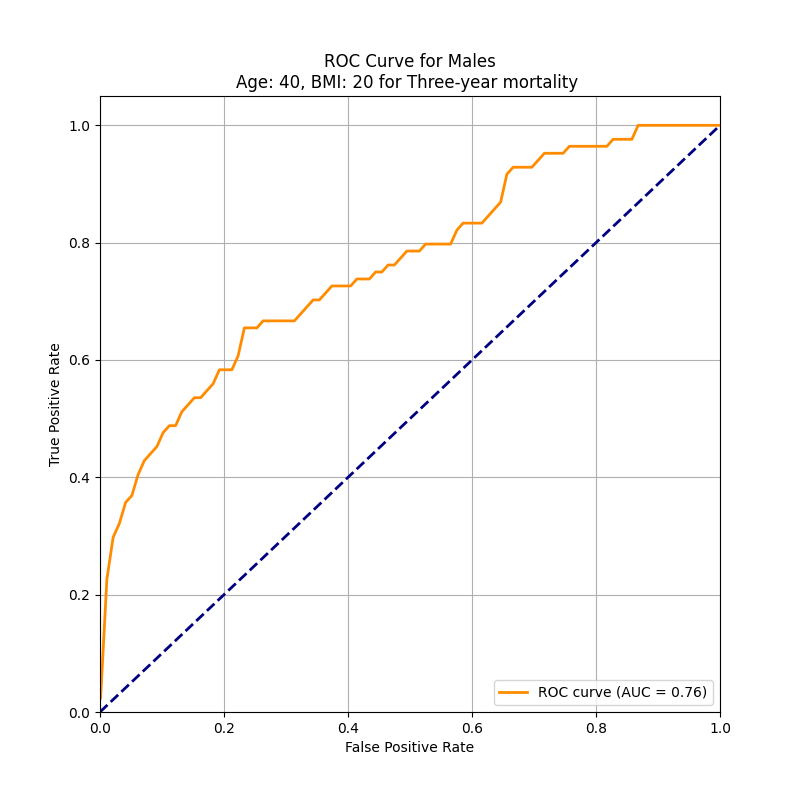}
        \caption{Age 40, BMI 20}
    \end{subfigure}

    \begin{subfigure}{0.3\textwidth}
        \includegraphics[width=\linewidth]{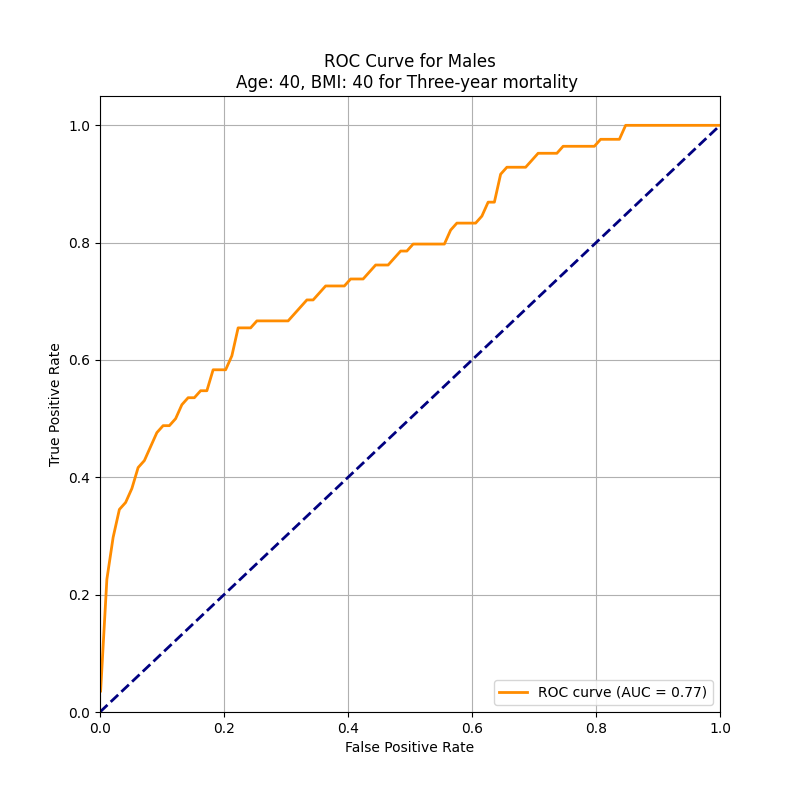}
        \caption{Age 40, BMI 40}
    \end{subfigure}
    \begin{subfigure}{0.3\textwidth}
        \includegraphics[width=\linewidth]{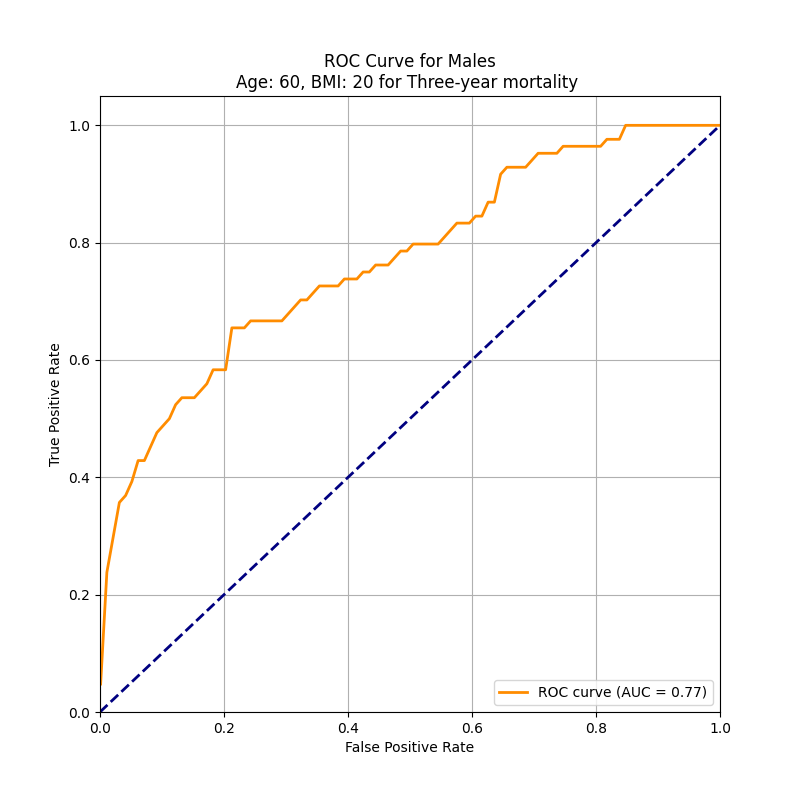}
        \caption{Age 60, BMI 20}
    \end{subfigure}
    \begin{subfigure}{0.3\textwidth}
        \includegraphics[width=\linewidth]{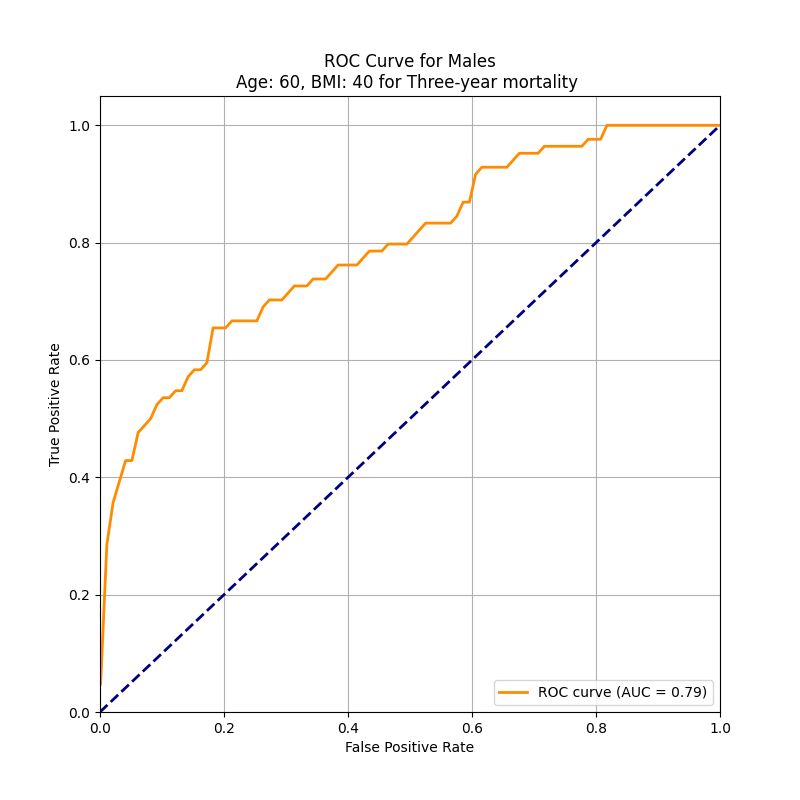}
        \caption{Age 60, BMI 40}
    \end{subfigure}
    \caption{ROC Curves for Males (Three-year mortality)}
    \label{fig:males_3year}
\end{figure}
\begin{figure}[htbp]
    \centering
    \begin{subfigure}{0.3\textwidth}
        \includegraphics[width=\linewidth]{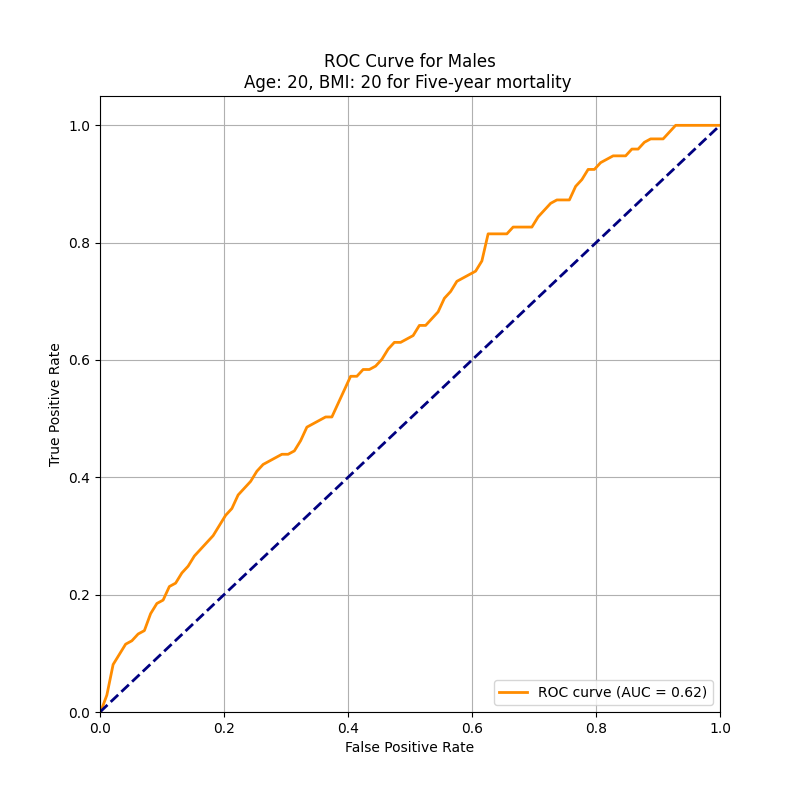}
        \caption{Age 20, BMI 20}
    \end{subfigure}
    \begin{subfigure}{0.3\textwidth}
        \includegraphics[width=\linewidth]{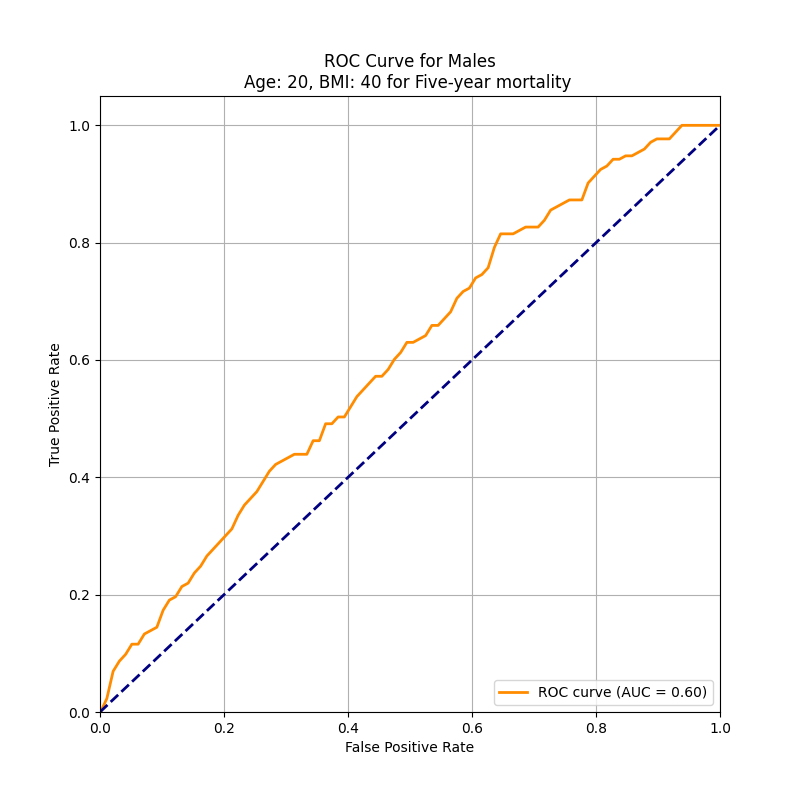}
        \caption{Age 20, BMI 40}
    \end{subfigure}
    \begin{subfigure}{0.3\textwidth}
        \includegraphics[width=\linewidth]{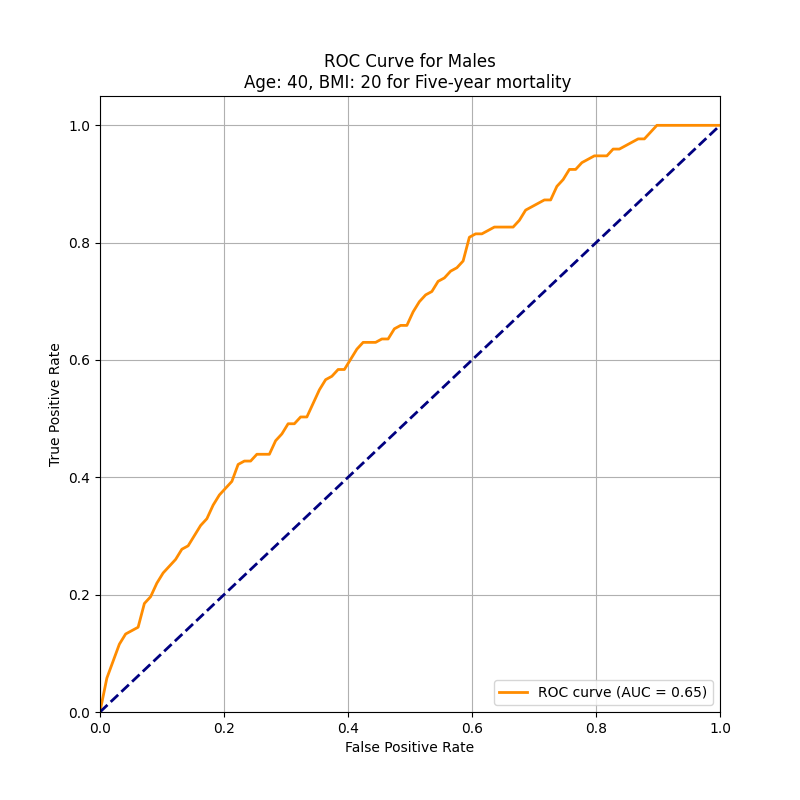}
        \caption{Age 40, BMI 20}
    \end{subfigure}

    \begin{subfigure}{0.3\textwidth}
        \includegraphics[width=\linewidth]{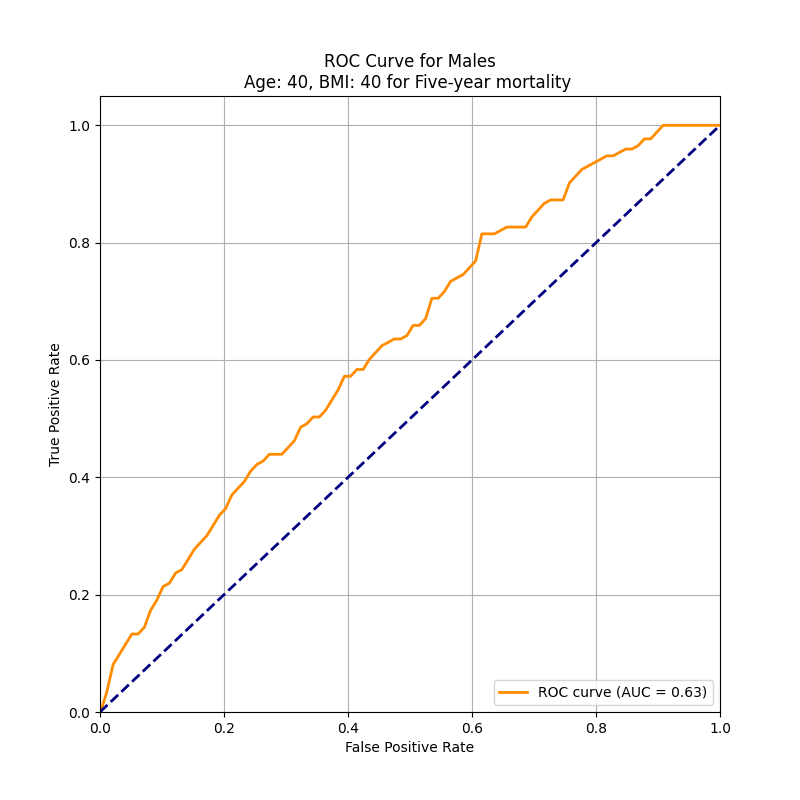}
        \caption{Age 40, BMI 40}
    \end{subfigure}
    \begin{subfigure}{0.3\textwidth}
        \includegraphics[width=\linewidth]{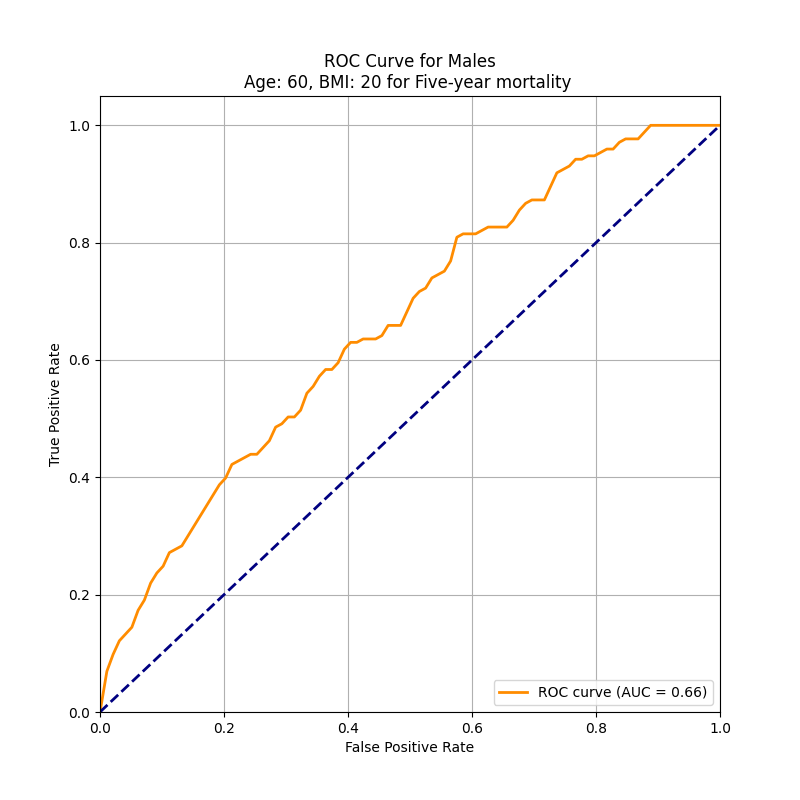}
        \caption{Age 60, BMI 20}
    \end{subfigure}
    \begin{subfigure}{0.3\textwidth}
        \includegraphics[width=\linewidth]{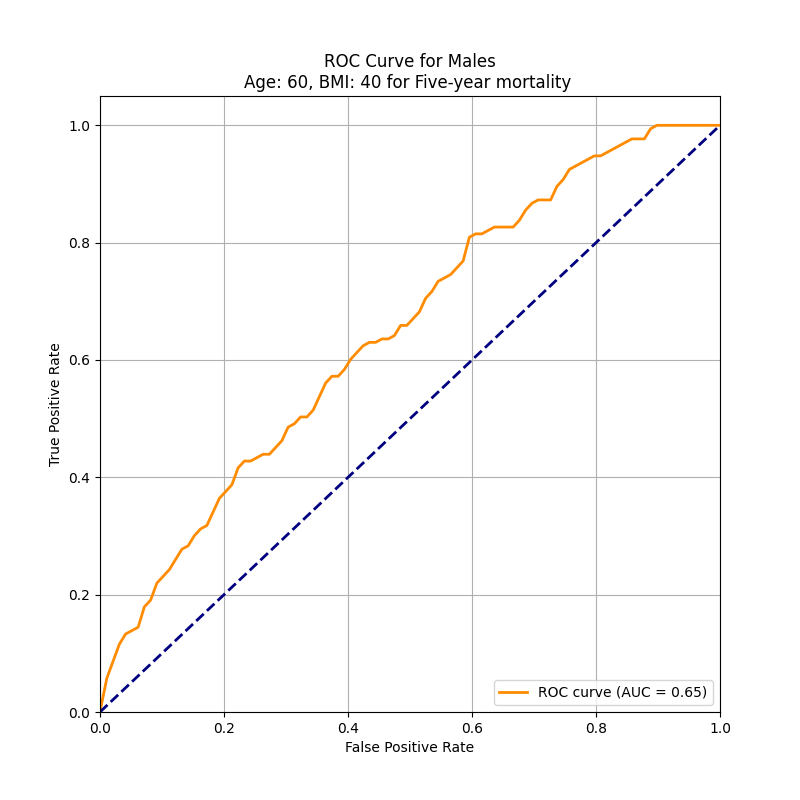}
        \caption{Age 60, BMI 40}
    \end{subfigure}
    \caption{ROC Curves for Males (Five-year mortality)}
     \label{fig:males_5year}
\end{figure}

\begin{figure}[H]
    \centering
    \begin{subfigure}{0.3\textwidth}
        \includegraphics[width=\linewidth]{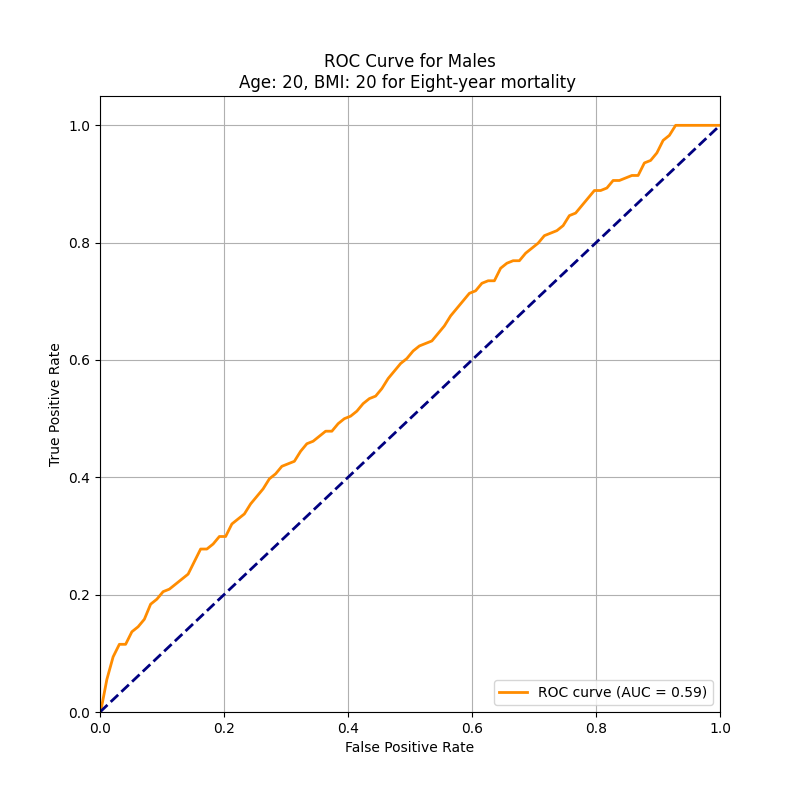}
        \caption{Age 20, BMI 20}
    \end{subfigure}
    \begin{subfigure}{0.3\textwidth}
        \includegraphics[width=\linewidth]{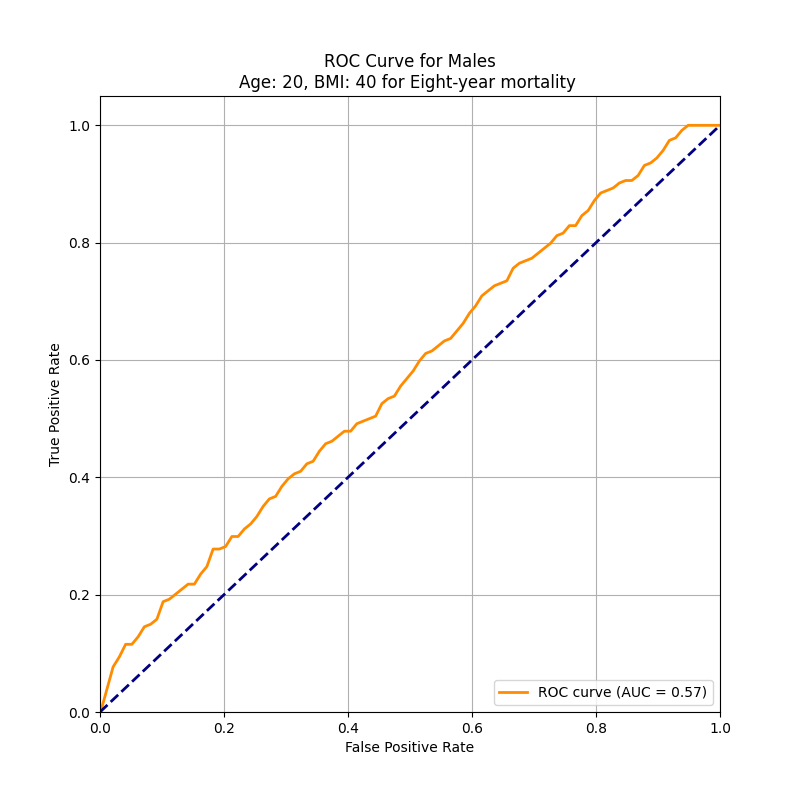}
        \caption{Age 20, BMI 40}
    \end{subfigure}
    \begin{subfigure}{0.3\textwidth}
        \includegraphics[width=\linewidth]{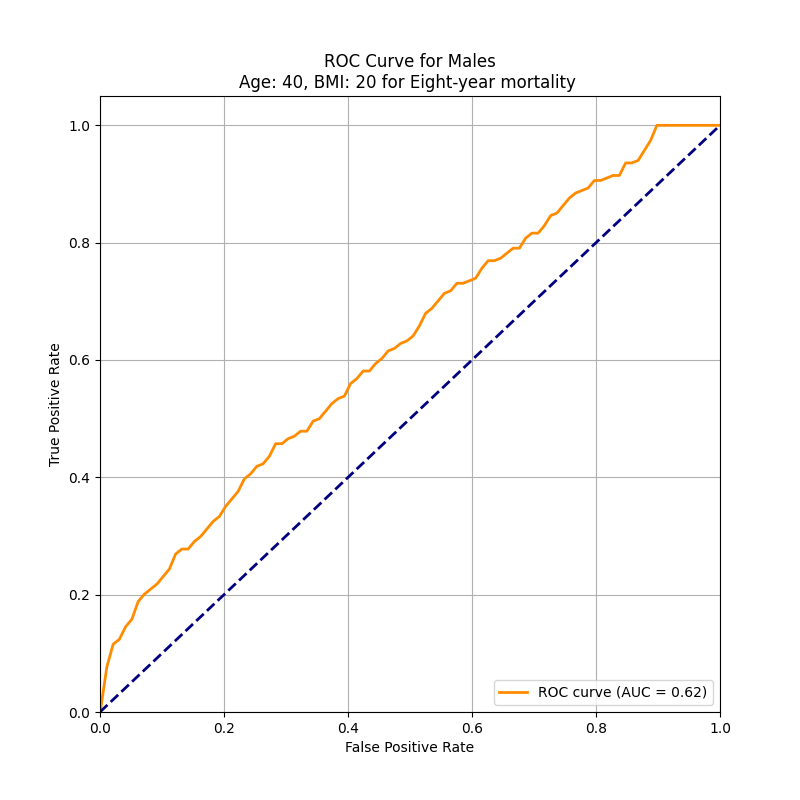}
        \caption{Age 40, BMI 20}
    \end{subfigure}

    \begin{subfigure}{0.3\textwidth}
        \includegraphics[width=\linewidth]{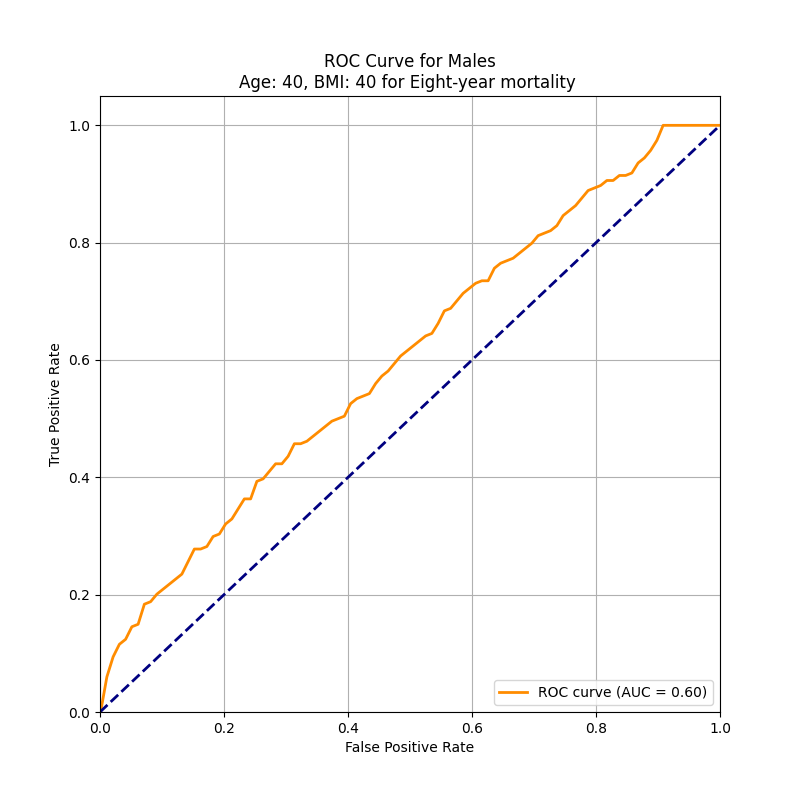}
        \caption{Age 40, BMI 40}
    \end{subfigure}
    \begin{subfigure}{0.3\textwidth}
        \includegraphics[width=\linewidth]{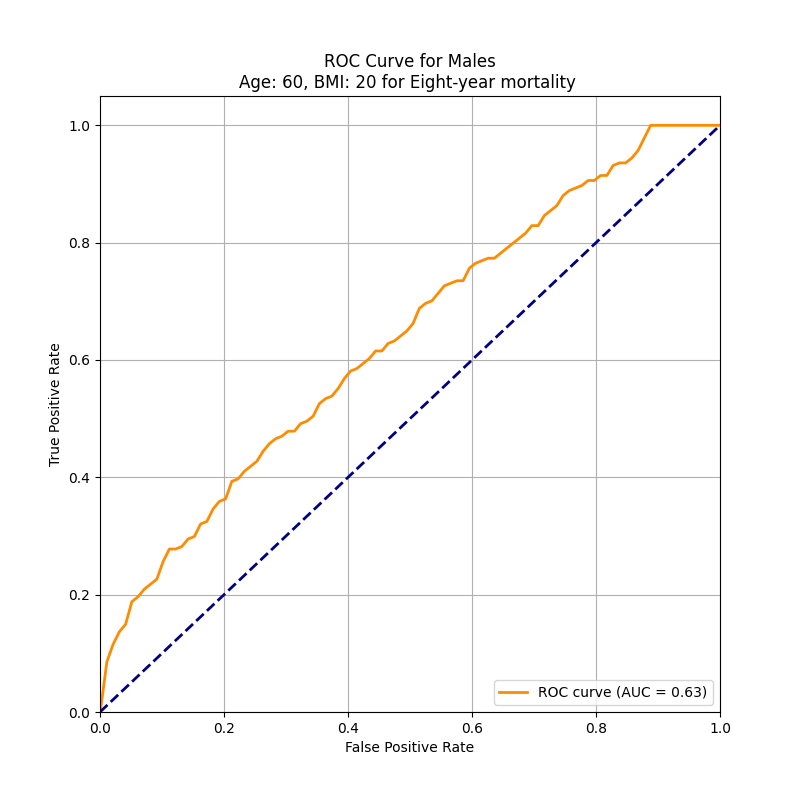}
        \caption{Age 60, BMI 20}
    \end{subfigure}
    \begin{subfigure}{0.3\textwidth}
        \includegraphics[width=\linewidth]{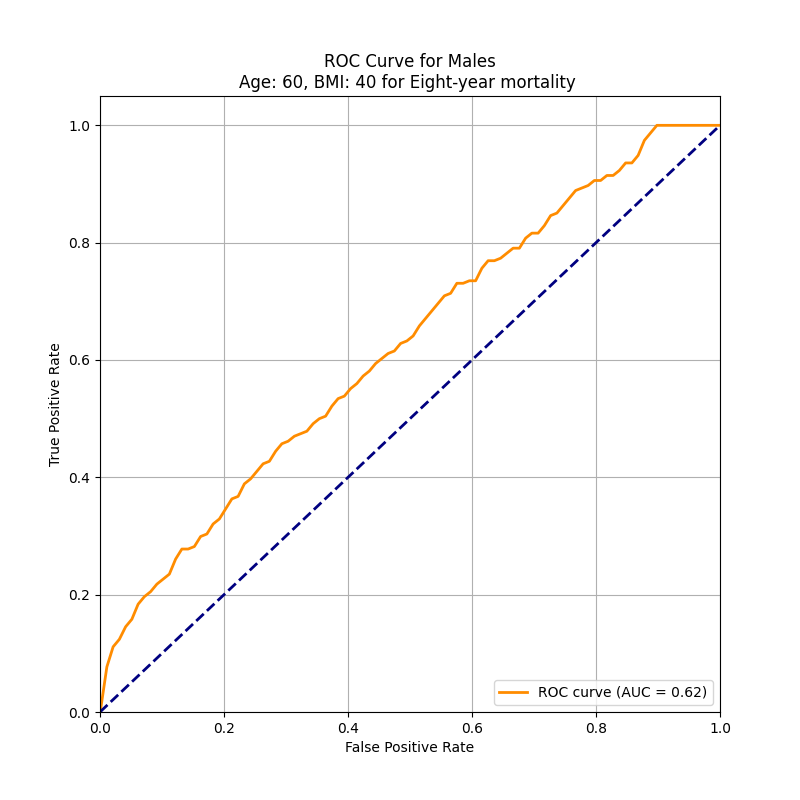}
        \caption{Age 60, BMI 40}
    \end{subfigure}
    \caption{ROC Curves for Males (Eight-year mortality)}
    \label{fig:males_8year}
\end{figure}

\section*{ Simulation study}

\subsection*{Synthetic Data Generation Scenarios}\label{app:data_gen}

To generate synthetic data, we modeled two populations: the diseased group (\( D = 1 \)) and the non-diseased group (\( D = 0 \)). The diagnostic variable \( Y \), representing test outcomes, was simulated under various scenarios influenced by covariates \( X \). Let \( Y_D \) and \( Y_{\overline{D}} \) denote the diagnostic variable for diseased and non-diseased populations, respectively.

The training dataset was used to fit predictive models and estimate key parameters such as the conditional mean (\( \mu_j(X) \)) and standard deviation (\( \sigma_j(X) \)). The testing dataset was reserved for evaluating model performance metrics, including Mean Squared Error (MSE) and covariate-adjusted ROC curves.

We designed nine distinct data generation scenarios to simulate a variety of covariate effects commonly encountered in biomedical research with linear, non-linear, and interaction-driven covariate effects. Each scenario models the distribution of a continuous diagnostic marker, $Y$, within two populations—healthy ($\overline{D}=0$) and diseased ($D=1$) as a function of covariates, $X$. All covariates were generated from a uniform distribution, $X \sim \mathcal{U}(-1, 1)$. For each scenario, we simulated datasets with two sample sizes ($n=5,000$ and $n=20,000$) to assess scalability on different data scales, representing large and moderate sample sizes, respectively. This process was repeated 100 times per scenario to ensure statistical robustness and to account for variability in simulation outcomes. All datasets were balanced, containing an equal number of samples from the diseased and non-diseased groups.
 A comprehensive overview of the nine simulated scenarios is detailed below.
\begin{itemize}
    \item \textbf{Scenario I: Baseline Distribution} 
    \[
    y_{\overline{D}i} \overset{iid}{\sim} \mathcal{N}(0.5, 0.5^2), \quad y_{D j} \overset{iid}{\sim} \mathcal{N}(1, 1^2).\]
    \item \textbf{Scenario II: Covariate Effect on Mean Only}
    \[
    y_{\overline{D}i} | x_{\overline{D}i,1} \overset{ind}{\sim} \mathcal{N}\left(0.5 + \frac{2x_{\overline{D}i,1} - 10}{23}, 0.5^2\right),
    \]
    \[
    y_{D j} | x_{D j,1} \overset{ind}{\sim} \mathcal{N}\left(1 + \frac{2x_{D j,1} - 10}{23}, 1^2\right).
    \]
    \item \textbf{Scenario III: Covariate Effect on Both Mean and Variance}
    \[
    y_{\overline{D}i} | x_{\overline{D}i,1} \overset{ind}{\sim} \mathcal{N}\left(0.25 + 0.5\frac{2x_{\overline{D}i,1} - 10}{23}, 0.5^2\right),
    \]
    \[
    y_{D j} | x_{D j,1} \overset{ind}{\sim} \mathcal{N}\left(0.75 + \frac{2x_{D j,1} - 10}{23}, 1^2\right).
    \]
    \item \textbf{Scenario IV: Nonlinear Covariate Effect on Mean}
    \[
    y_{\overline{D}i} | x_{\overline{D}i,1} \overset{ind}{\sim} \mathcal{N}\left(5 + 3\left(\frac{x_{\overline{D}i,1} + 8}{23}\right)^2 - 25 \left(\frac{x_{\overline{D}i,1} + 8}{23} - 0.2\right)^3 \right.,
    \]
    \[
    \left.+ 250 \left(\frac{x_{\overline{D}i,1} + 8}{23} - 0.65\right)^3, 0.5^2\right),
    \]
    \[
    y_{D j} | x_{D j,1} \overset{ind}{\sim} \mathcal{N}\left(-3 - 0.6\frac{x_{D j,1} + 8}{23}, 1^2\right).
    \]
    \item \textbf{Scenario V: Multiple Covariates with Exponential and Trigonometric Effects}
    \[
    y_{\overline{D}i} | x_{\overline{D}i,1}, x_{\overline{D}i,2} \overset{ind}{\sim} \mathcal{N}\left(0.5 \exp\left(\frac{2x_{\overline{D}i,1} - 10}{10}\right) - 2\left(\frac{2x_{\overline{D}i,2}^2 - 10}{10}\right), 0.5^2\right),
    \]
    \[
    y_{D j} | x_{D j,1} \overset{ind}{\sim} \mathcal{N}\left(0.5 \sin\left(\pi\left(\frac{2x_{D j,1} - 10}{10} + 1\right)\right) + 0.5 \exp\left(\frac{2x_{D j,1} - 10}{10}\right), 1^2\right).
    \]
    \item \textbf{Scenario VI: Interaction Effects}
    \[
    y_{\overline{D}i} | x_{\overline{D}i,1}, x_{\overline{D}i,3} \overset{ind}{\sim} \mathcal{N}\left(- \sin(0.7\pi((2x_{\overline{D}i,1} - 10)/10 + 30))x_{\overline{D}i,3} \right.,
    \]
    \[
    \left.+ \left(\frac{2x_{\overline{D}i,1} - 10}{10}\right)^2(1 - x_{\overline{D}i,3}), 0.5^2\right),
    \]
    \[
    y_{D j} | x_{D j,1} \overset{ind}{\sim} \mathcal{N}\left(0.5 + \left(\frac{2x_{D j,1} - 10}{10}\right)^2, 1^2\right).
    \]

    \item \textbf{Scenario VII: Skewed and Heavy-Tailed Distribution}
    \[
    y_{\overline{D}i} | x_{\overline{D}i,4} \overset{ind}{\sim} \exp(-2x_{\overline{D}i,4})\text{SN}(x_{\overline{D}i,4}^2, 0.25, 2) + (1 - \exp(-2x_{\overline{D}i,4}))t(\sin(\pi x_{\overline{D}i,4}), 0.25, 5),
    \]
    \[
    y_{D j} | x_{D j,4} \overset{ind}{\sim} \mathcal{N}(\sin(2\pi x_{D j,4}) + 1.5, 0.5^2).
    \]

    \item \textbf{Scenario VIII: Baseline with Additional Covariates}
    - Similar to Scenario I but includes four additional covariates \( x_{\overline{D}i,m} \sim \mathcal{U}(-1, 1) \) for \( m = 5, \dots, 8 \).
    \item \textbf{Scenario IX: Complex Interactions with Multiple Covariates}
    \[
    y_{\overline{D}i} | x_{\overline{D}i,5}, \dots, x_{\overline{D}i,8} \overset{ind}{\sim} \mathcal{N}\left(0.5 \exp(2x_{\overline{D}i,5}) - x_{\overline{D}i,6}^2 + 0.5x_{\overline{D}i,7}^2 + x_{\overline{D}i,8}, 0.5^2\right),
    \]
    \[
    y_{D j} | x_{D j,5}, \dots, x_{D j,8} \overset{ind}{\sim} \mathcal{N}\left(0.5 + 0.5 \exp(2x_{D j,5}) - x_{D j,6}^2 + 0.5x_{D j,7}^2 + x_{D j,8}, 1^2\right).
    \]

\end{itemize}

\newpage
\subsection*{Detailed Simulation Results}\label{app:detailed_results}

The following table and figures present the comprehensive results of the comparative analysis between the FNN and Random Forest models across the nine synthetic scenarios. Both models were trained using the same data folds, and performance metrics were evaluated on the same test individuals to ensure a fair comparison. Performance is measured by the Mean Squared Error (MSE) and its standard deviation (MSE Std), computed over 100 independent simulation runs.

\begin{table}[H]
\centering
\caption{Comparison of Mean Squared Error (MSE) for FNN and Random Forest Models. FNN MSE Mean values are bolded if they are greater than the corresponding Random Forest values.}
\label{tab:combined_mse_bold}
\begin{tabular}{lrr rrrr}
\toprule
& & & \multicolumn{2}{c}{\textbf{FNN Model}} & \multicolumn{2}{c}{\textbf{Random Forest Model}} \\
\cmidrule(lr){4-5} \cmidrule(lr){6-7}
\textbf{Scenario Type} & \textbf{Scenario No.} & \textbf{individuals} & \textbf{MSE Mean} & \textbf{MSE Std} & \textbf{MSE Mean} & \textbf{MSE Std} \\
\midrule
healthy & 1 & 5000 & 0.000672 & 0.000198 & 0.007862 & 0.004774 \\
diseased & 1 & 5000 & 0.004217 & 0.001490 & 0.032295 & 0.018587 \\
healthy & 1 & 20000 & 0.000072 & 0.000043 & 0.001908 & 0.001211 \\
diseased & 1 & 20000 & 0.000669 & 0.000272 & 0.007889 & 0.004903 \\
\addlinespace
healthy & 2 & 5000 & 0.000610 & 0.000174 & 0.012038 & 0.004854 \\
diseased & 2 & 5000 & 0.002134 & 0.001553 & 0.038783 & 0.020083 \\
healthy & 2 & 20000 & 0.000416 & 0.000040 & 0.004078 & 0.001285 \\
diseased & 2 & 20000 & 0.000866 & 0.000219 & 0.012523 & 0.004930 \\
\addlinespace
healthy & 3 & 5000 & 0.000357 & 0.000164 & 0.010098 & 0.004795 \\
diseased & 3 & 5000 & 0.001684 & 0.001551 & 0.040233 & 0.019778 \\
healthy & 3 & 20000 & 0.000237 & 0.000037 & 0.003109 & 0.001255 \\
diseased & 3 & 20000 & 0.000838 & 0.000197 & 0.012382 & 0.004941 \\
\addlinespace
healthy & 4 & 5000 & \textbf{0.058553} & 0.010015 & 0.041793 & 0.013482 \\
diseased & 4 & 5000 & 0.029392 & 0.001391 & 0.033016 & 0.018456 \\
healthy & 4 & 20000 & \textbf{0.050677} & 0.009330 & 0.018391 & 0.002989 \\
diseased & 4 & 20000 & 0.005056 & 0.000225 & 0.008955 & 0.004835 \\
\addlinespace
healthy & 5 & 5000 & 0.016184 & 0.001424 & 0.017823 & 0.005281 \\
diseased & 5 & 5000 & 0.004414 & 0.001836 & 0.055816 & 0.017647 \\
healthy & 5 & 20000 & \textbf{0.009087} & 0.001192 & 0.008174 & 0.001498 \\
diseased & 5 & 20000 & 0.003172 & 0.000322 & 0.020241 & 0.004738 \\
\addlinespace
healthy & 6 & 5000 & 0.005543 & 0.000224 & 0.023654 & 0.005624 \\
diseased & 6 & 5000 & 0.010311 & 0.001557 & 0.060943 & 0.017709 \\
healthy & 6 & 20000 & 0.003615 & 0.000059 & 0.009211 & 0.001378 \\
diseased & 6 & 20000 & 0.004479 & 0.000263 & 0.022847 & 0.004810 \\
\addlinespace
healthy & 7 & 5000 & \textbf{0.119977} & 0.030361 & 0.110910 & 0.040944 \\
diseased & 7 & 5000 & 0.014816 & 0.245074 & 0.023609 & 0.242317 \\
healthy & 7 & 20000 & \textbf{0.119032} & 0.030067 & 0.112930 & 0.037435 \\
diseased & 7 & 20000 & 0.007065 & 0.244681 & 0.010062 & 0.245833 \\
\addlinespace
healthy & 8 & 5000 & 0.001728 & 0.000771 & 0.007558 & 0.003969 \\
diseased & 8 & 5000 & 0.006681 & 0.004687& 0.030078 & 0.015399 \\
healthy & 8 & 20000 & 0.000239 & 0.000065 & 0.001989 & 0.001104 \\
diseased & 8 & 20000 &  0.002652 & 0.002286& 0.007996 & 0.004481 \\
\addlinespace
healthy & 9 & 5000 & 0.021551 & 0.001001  & 0.039611 & 0.006257 \\
diseased & 9 & 5000 &  0.034049 & 0.003664 & 0.098680 & 0.017095 \\
healthy & 9 & 20000 &0.015345 & 0.000417 & 0.024737 & 0.002039 \\
diseased & 9 & 20000 & 0.021383 & 0.001771   & 0.055598 & 0.005469 \\
\bottomrule
\end{tabular}
\end{table}

\begin{figure}[H]
    \centering
    \begin{subfigure}[b]{0.48\textwidth}
        \centering
        \includegraphics[width=\textwidth]{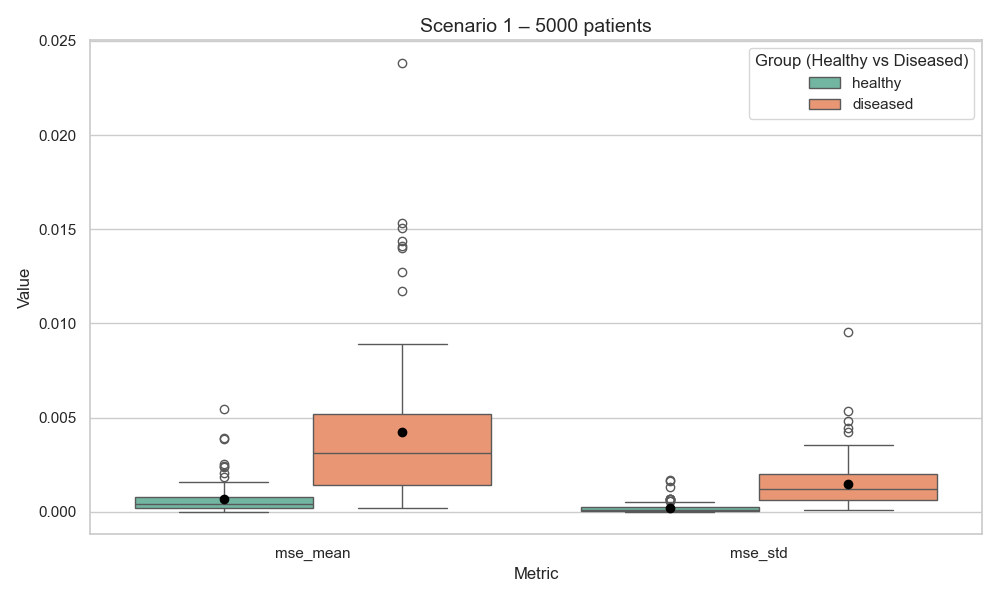}
        \caption{5,000 individuals}
        \label{fig:mse_s1_5k}
    \end{subfigure}
    \hfill
    \begin{subfigure}[b]{0.48\textwidth}
        \centering
        \includegraphics[width=\textwidth]{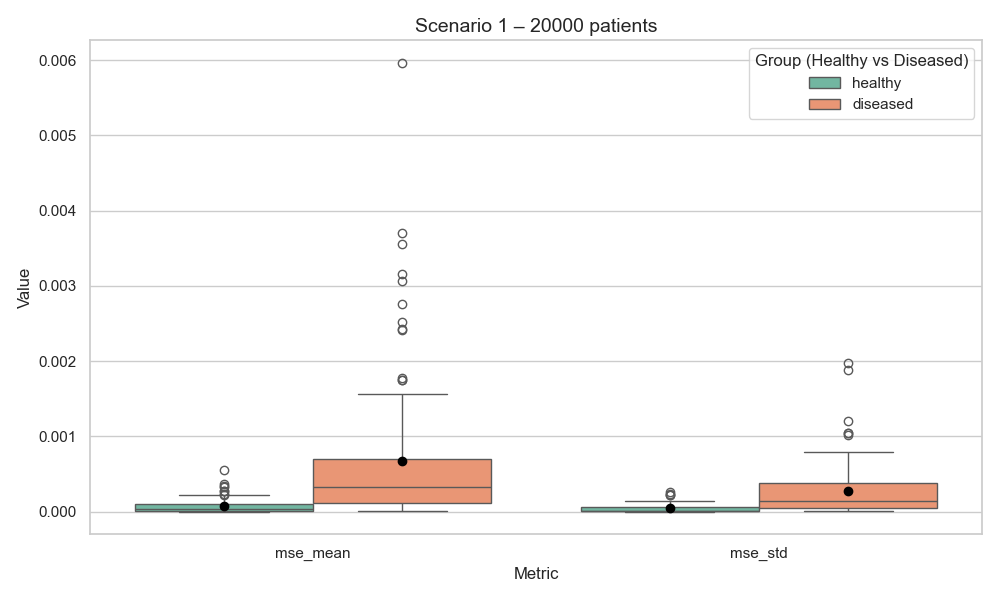}
        \caption{20,000 individuals}
        \label{fig:mse_s1_20k}
    \end{subfigure}
    \caption{MSE Boxplots for Scenario 1 (Baseline Distribution). The plots show the performance of the FNN model by displaying the mean squared error for both the conditional mean estimation ($\text{mse\_mean}$) and the conditional standard deviation estimation ($\text{mse\_std}$). The low MSE values indicate high accuracy in estimating both the mean and variance components, with performance improving for larger sample sizes.}
    \label{fig:mse_s1}
\end{figure}

\begin{figure}[H]
    \centering
    \begin{subfigure}[b]{0.48\textwidth}
        \centering
        \includegraphics[width=\textwidth]{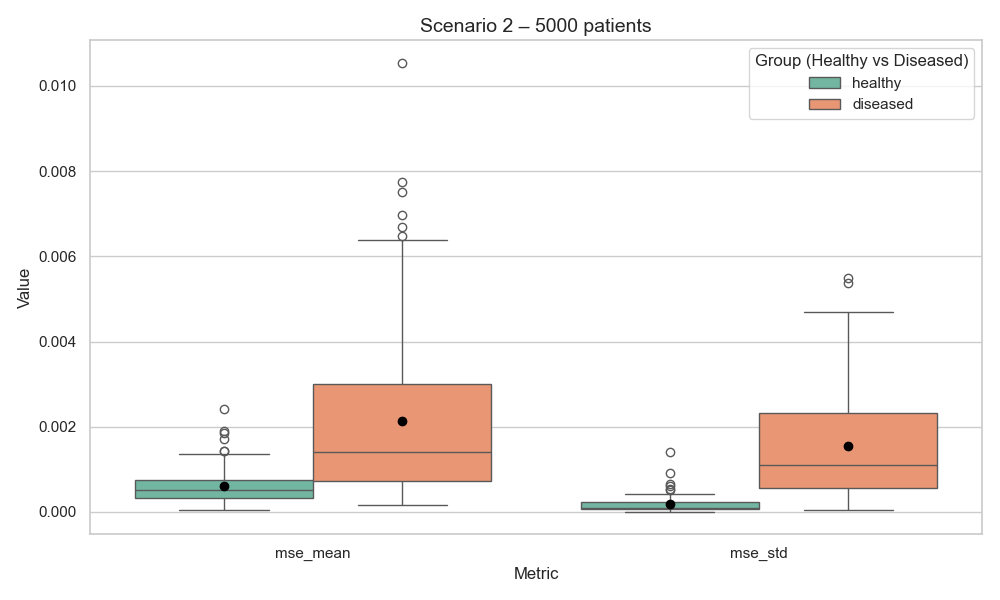}
        \caption{5,000 individuals}
        \label{fig:mse_s2_5k}
    \end{subfigure}
    \hfill
    \begin{subfigure}[b]{0.48\textwidth}
        \centering
        \includegraphics[width=\textwidth]{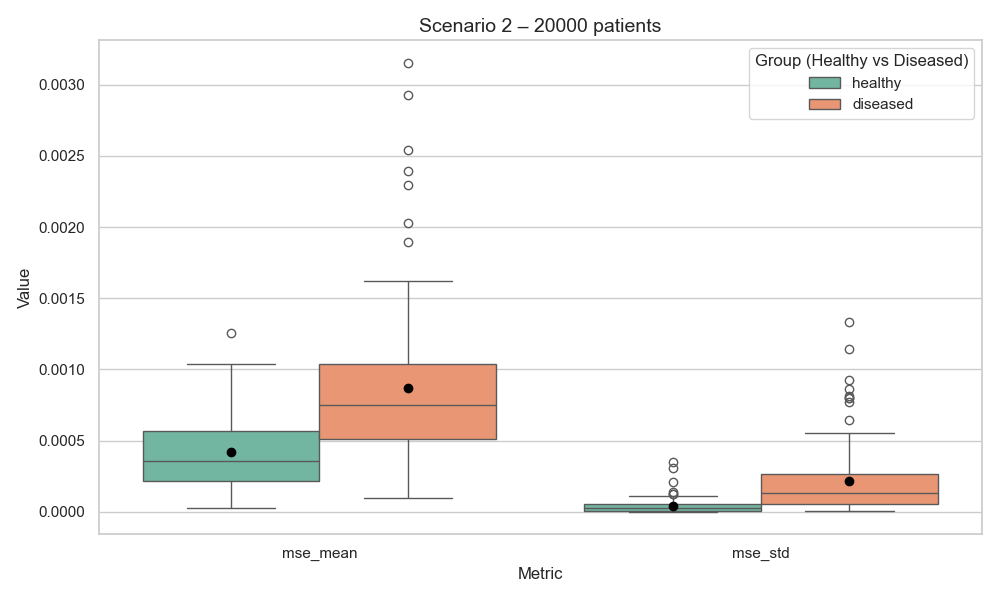}
        \caption{20,000 individuals}
        \label{fig:mse_s2_20k}
    \end{subfigure}
    \caption{MSE Boxplots for Scenario 2 (Covariate Effect on Mean Only). The FNN model demonstrates a low MSE for both mean and standard deviation estimates, highlighting its capacity to accurately model the conditional distribution of the biomarker even when only the mean is affected by covariates.}
    \label{fig:mse_s2}
\end{figure}

\begin{figure}[H]
    \centering
    \begin{subfigure}[b]{0.48\textwidth}
        \centering
        \includegraphics[width=\textwidth]{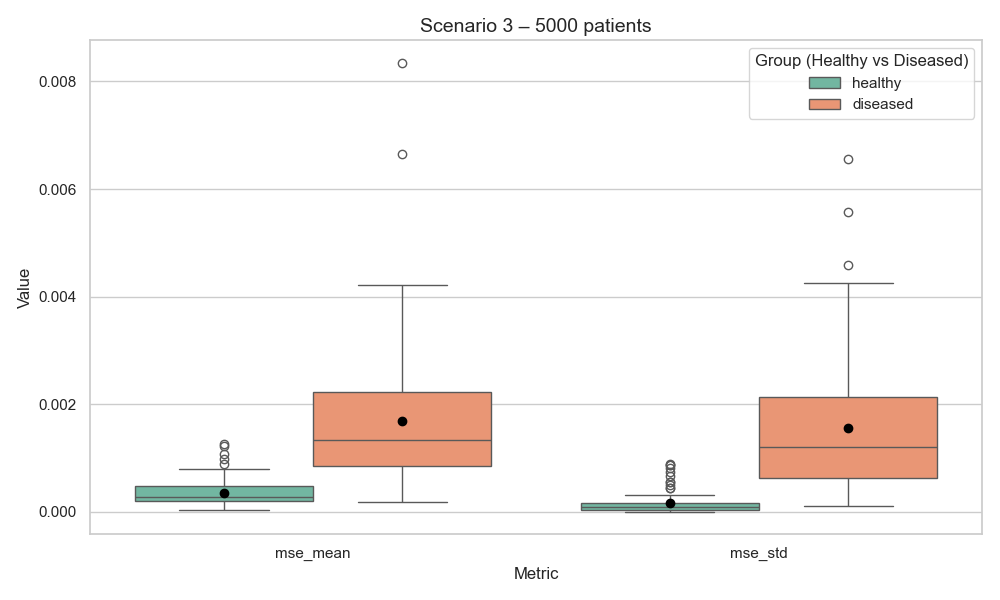}
        \caption{5,000 individuals}
        \label{fig:mse_s3_5k}
    \end{subfigure}
    \hfill
    \begin{subfigure}[b]{0.48\textwidth}
        \centering
        \includegraphics[width=\textwidth]{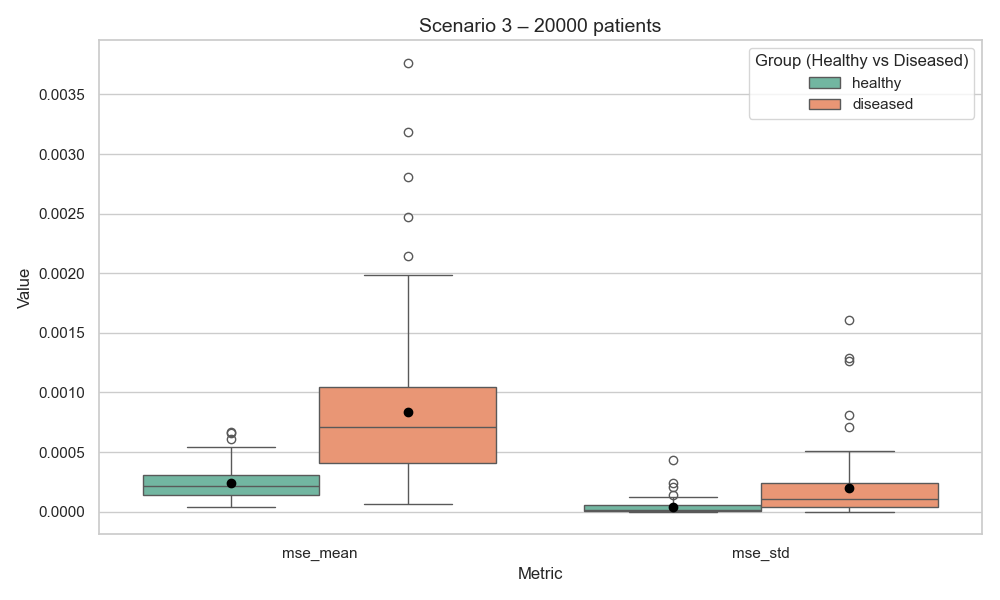}
        \caption{20,000 individuals}
        \label{fig:mse_s3_20k}
    \end{subfigure}
    \caption{MSE Boxplots for Scenario 3 (Covariate Effect on Both Mean and Variance). The FNN model's low MSE for both the mean and standard deviation components confirms its ability to accurately and simultaneously model a biomarker's conditional location and scale parameters, a critical feature for effective ROC adjustment.}
    \label{fig:mse_s3}
\end{figure}

---

\begin{figure}[H]
    \centering
    \begin{subfigure}[b]{0.48\textwidth}
        \centering
        \includegraphics[width=\textwidth]{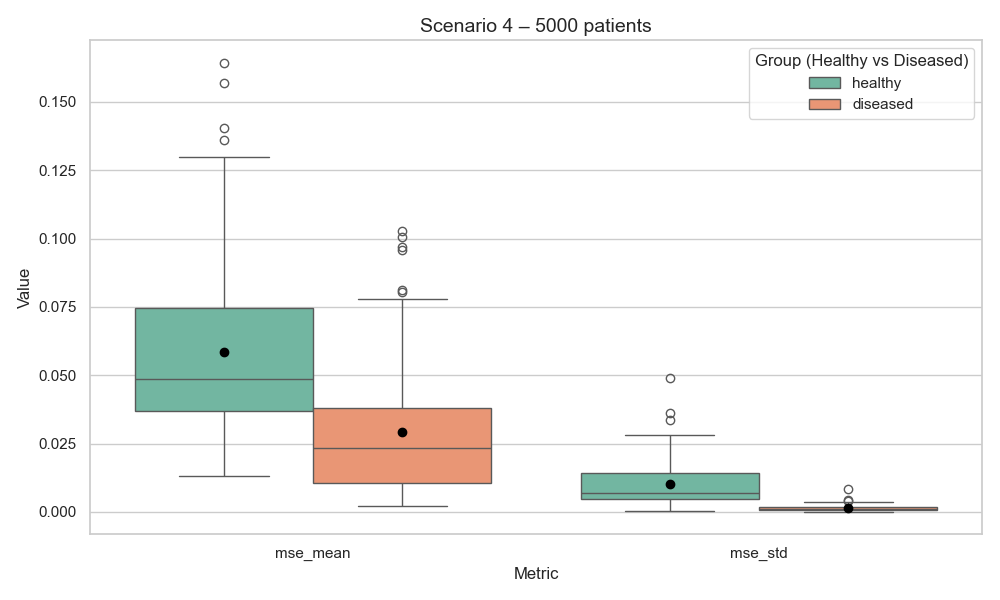}
        \caption{5,000 individuals}
        \label{fig:mse_s4_5k}
    \end{subfigure}
    \hfill
    \begin{subfigure}[b]{0.48\textwidth}
        \centering
        \includegraphics[width=\textwidth]{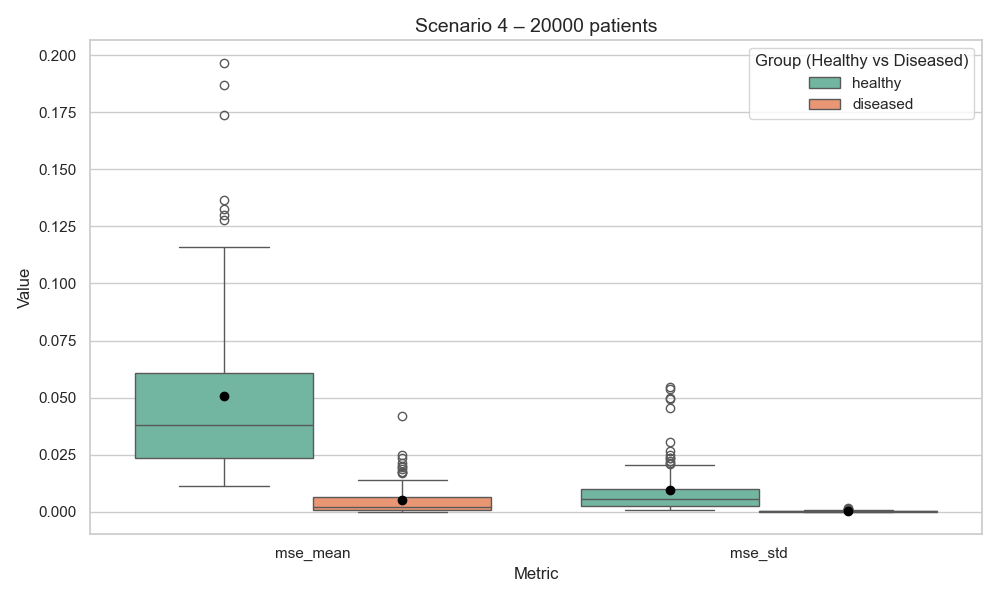}
        \caption{20,000 individuals}
        \label{fig:mse_s4_20k}
    \end{subfigure}
    \caption{MSE Boxplots for Scenario 4 (Nonlinear Covariate Effect on Mean). The FNN model maintains a low MSE for both mean and standard deviation, demonstrating its superior capability to learn and generalize from complex, non-linear relationships, a task that challenges traditional models.}
    \label{fig:mse_s4}
\end{figure}

\begin{figure}[H]
    \centering
    \begin{subfigure}[b]{0.48\textwidth}
        \centering
        \includegraphics[width=\textwidth]{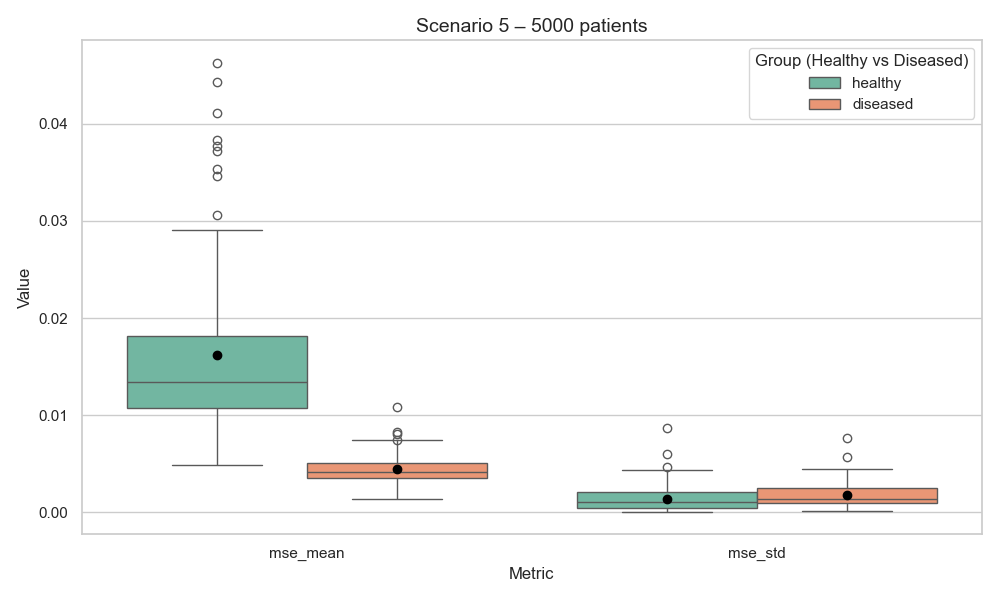}
        \caption{5,000 individuals}
        \label{fig:mse_s5_5k}
    \end{subfigure}
    \hfill
    \begin{subfigure}[b]{0.48\textwidth}
        \centering
        \includegraphics[width=\textwidth]{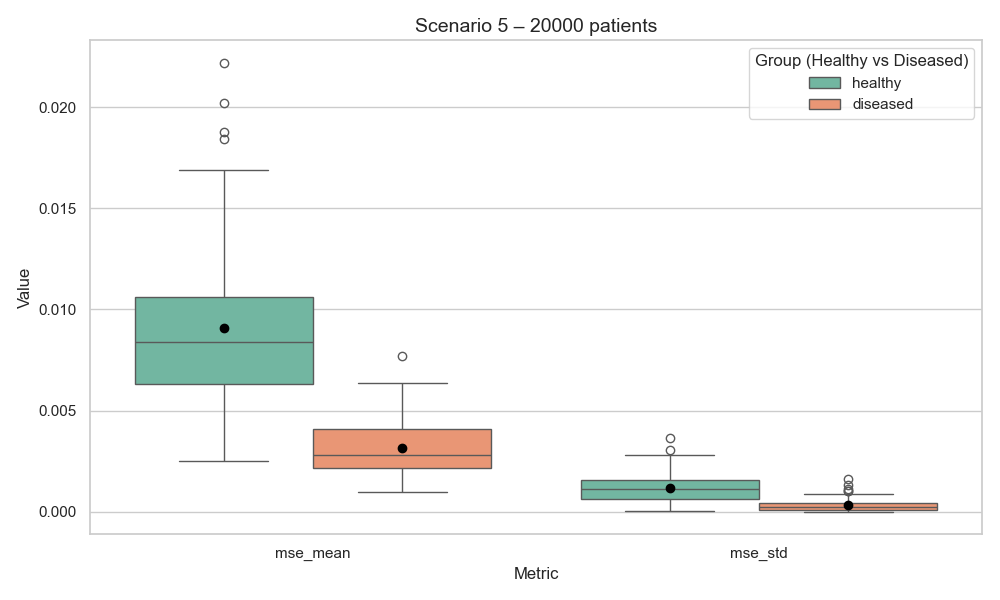}
        \caption{20,000 individuals}
        \label{fig:mse_s5_20k}
    \end{subfigure}
    \caption{MSE Boxplots for Scenario 5 (Multiple Covariates with Exponential and Trigonometric Effects). The FNN model's robust performance, marked by consistently low MSE values, highlights its ability to effectively model and integrate complex functional forms from multiple covariates.}
    \label{fig:mse_s5}
\end{figure}

\begin{figure}[H]
    \centering
    \begin{subfigure}[b]{0.48\textwidth}
        \centering
        \includegraphics[width=\textwidth]{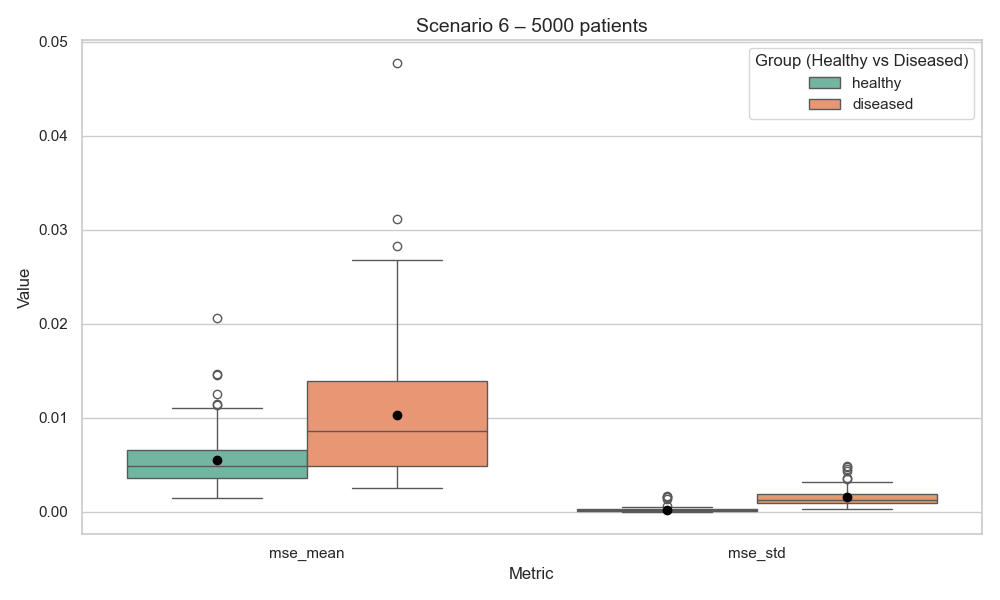}
        \caption{5,000 individuals}
        \label{fig:mse_s6_5k}
    \end{subfigure}
    \hfill
    \begin{subfigure}[b]{0.48\textwidth}
        \centering
        \includegraphics[width=\textwidth]{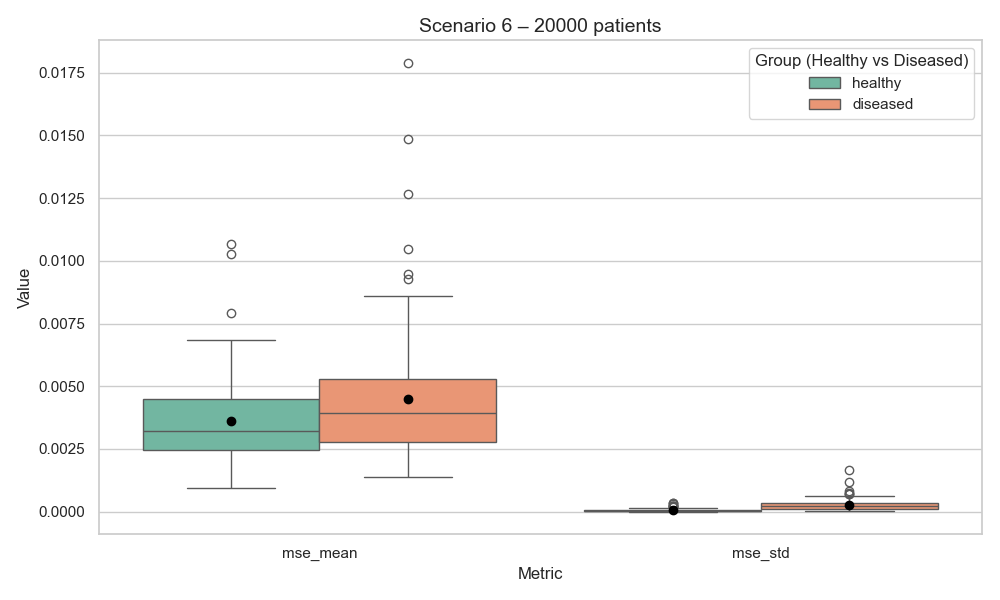}
        \caption{20,000 individuals}
        \label{fig:mse_s6_20k}
    \end{subfigure}
    \caption{MSE Boxplots for Scenario 6 (Interaction Effects). The FNN model's performance remains stable and accurate, showcasing its capacity to capture and learn from complex interaction terms that are difficult for simpler models to identify.}
    \label{fig:mse_s6}
\end{figure}

\begin{figure}[H]
    \centering
    \begin{subfigure}[b]{0.48\textwidth}
        \centering
        \includegraphics[width=\textwidth]{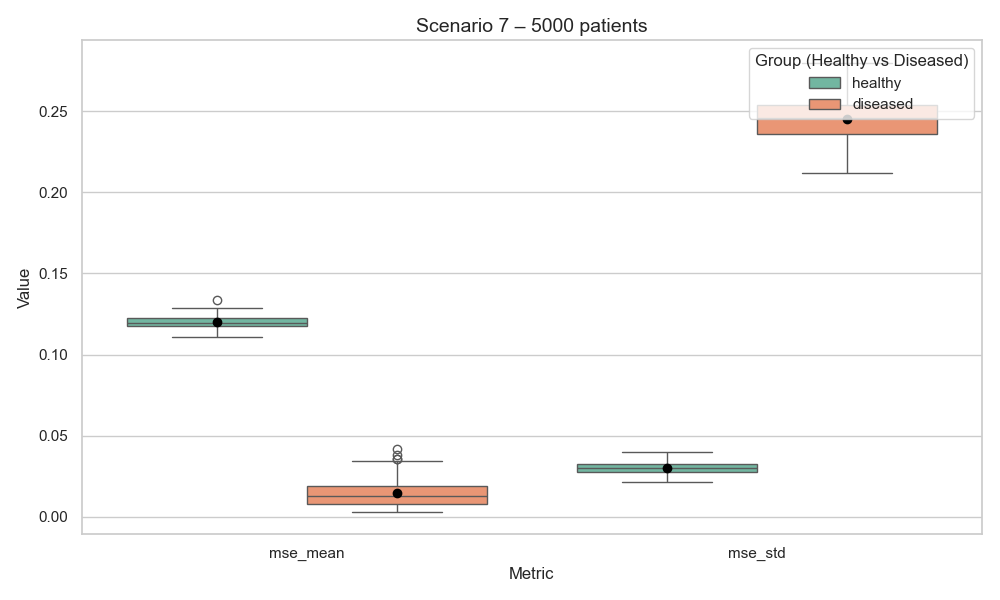}
        \caption{5,000 individuals}
        \label{fig:mse_s7_5k}
    \end{subfigure}
    \hfill
    \begin{subfigure}[b]{0.48\textwidth}
        \centering
        \includegraphics[width=\textwidth]{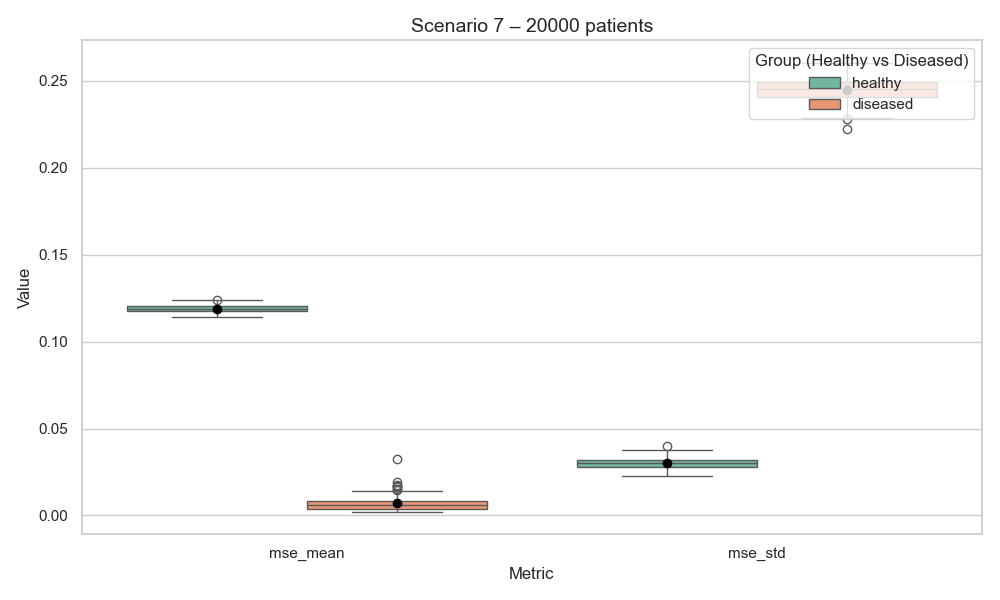}
        \caption{20,000 individuals}
        \label{fig:mse_s7_20k}
    \end{subfigure}
    \caption{MSE Boxplots for Scenario 7 (Skewed and Heavy-Tailed Distribution). Despite the challenge of non-Gaussian data, the FNN model exhibits robust performance, demonstrating its flexibility beyond standard normality assumptions and its effectiveness in diverse data environments.}
    \label{fig:mse_s7}
\end{figure}

\begin{figure}[H]
    \centering
    \begin{subfigure}[b]{0.48\textwidth}
        \centering
        \includegraphics[width=\textwidth]{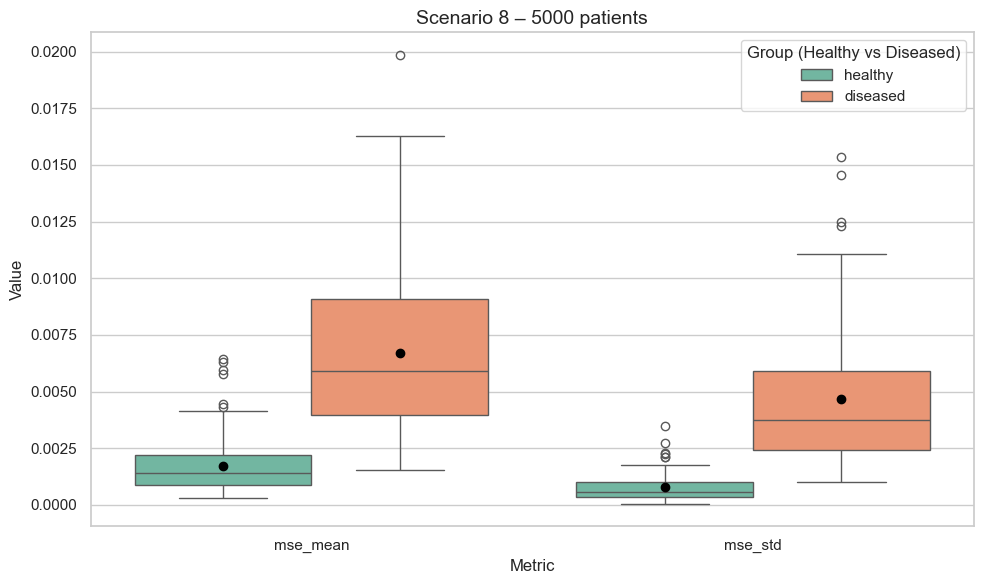}
        \caption{5,000 individuals}
        \label{fig:mse_s8_5k}
    \end{subfigure}
    \hfill
    \begin{subfigure}[b]{0.48\textwidth}
        \centering
        \includegraphics[width=\textwidth]{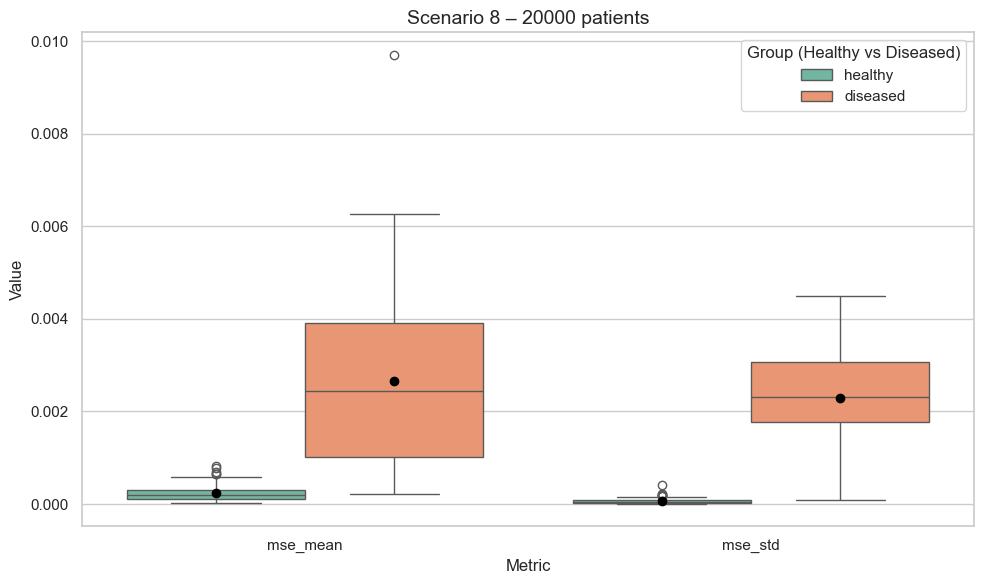}
        \caption{20,000 individuals}
        \label{fig:mse_s8_20k}
    \end{subfigure}
    \caption{MSE Boxplots for Scenario 8 (Baseline with Additional Covariates). The FNN's MSE remains low and stable, indicating that it effectively handles high-dimensional data without significant performance degradation from non-informative covariates.}
    \label{fig:mse_s8}
\end{figure}

\begin{figure}[H]
    \centering
    \begin{subfigure}[b]{0.48\textwidth}
        \centering
        \includegraphics[width=\textwidth]{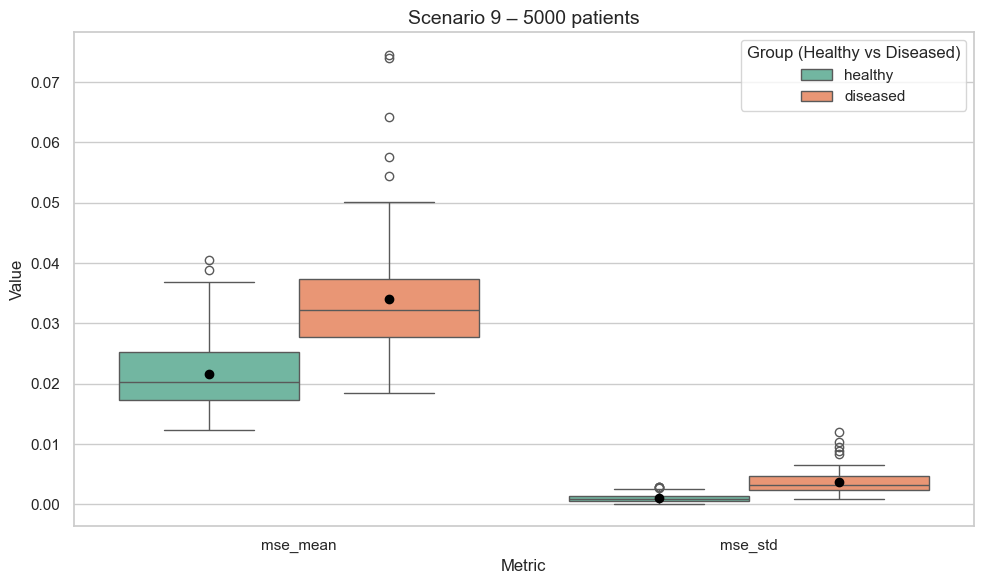}
        \caption{5,000 individuals}
        \label{fig:mse_s9_5k}
    \end{subfigure}
    \hfill
    \begin{subfigure}[b]{0.48\textwidth}
        \centering
        \includegraphics[width=\textwidth]{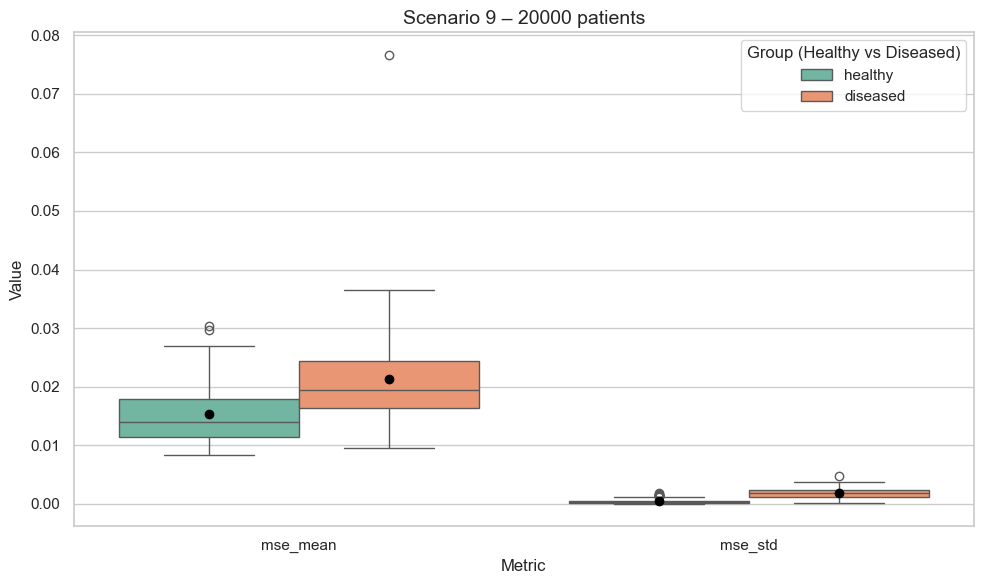}
        \caption{20,000 individuals}
        \label{fig:mse_s9_20k}
    \end{subfigure}
    \caption{MSE Boxplots for Scenario 9 (Complex Interactions with Multiple Covariates). This scenario represents a significant challenge, yet the FNN model maintains a consistently low MSE and high stability for both its mean and variance estimations, confirming its superiority in complex, multi-covariate settings.}
    \label{fig:mse_s9}
\end{figure}

\subsection*{Final Remarks}

The simulation study demonstrated the consistent superiority of the proposed FNN framework over the Random Forest (RF) benchmark in nine diverse data-generating scenarios, designed to reflect a spectrum of clinical complexities. Performance was evaluated using Mean Squared Error (MSE) over 100 independent simulation runs for two sample sizes ($n=5,000$ and $n=20,000$). Detailed results are available in the appendix Table \ref{tab:combined_mse_bold} and Figures \ref{fig:mse_s1}--\ref{fig:mse_s9}.

The FNN framework achieved a significantly lower MSE (often by an order of magnitude) and exhibited greater model stability (lower MSE standard deviation) across virtually all scenarios. For example, in Scenario I (baseline distribution) with $n=20,000$ diseased individuals, the MSE of FNN was \textbf{0.000669} compared to the RF \textbf{0.007889}. The performance advantage of the FNN was most pronounced in scenarios involving complex, nonlinear covariate effects (Scenarios 4, 5, 6, 9) and non-Gaussian distributions (Scenario 7), highlighting its capacity to capture more complex statistical associations data structures that challenge traditional models. Furthermore, FNN performance improved consistently with increased sample size, confirming its scalability and ability to leverage larger datasets for enhanced precision, a critical attribute for real-world clinical applications.

\end{document}